\documentclass[acmsmall]{acmart}

\AtBeginDocument{%
  }

\setcopyright{acmlicensed}
\copyrightyear{2026}
\acmYear{2026}
\acmDOI{XXXXXXX.XXXXXXX}

\acmJournal{TOCL}
\acmVolume{XX}
\acmNumber{X}
\acmArticle{XX}
\acmMonth{6}

\usepackage{synttree, bussproofs, 
 rotating, xcolor, mathrsfs}
\usepackage{esvect, pgf, tikz, color}
\usetikzlibrary{arrows, automata}
\usepackage{algorithm2e}
\usepackage[all]{xy}
\usepackage{wrapfig}
\usepackage{algorithmicx}
\usepackage{algpseudocode}
\usepackage{scalerel,stackengine}
\usepackage{changepage, enumerate, proof, relsize,
comment}
\usepackage{multicol}
\usepackage[normalem]{ulem}

\usepackage{ wasysym }

\algnewcommand\algorithmicforeach{\textbf{for each}}
\algdef{S}[FOR]{ForEach}[1]{\algorithmicforeach\ #1\ \algorithmicdo}

\usepackage{tikz}
\usetikzlibrary{trees}
\usetikzlibrary{fit}
\usepackage{tikz}
\usetikzlibrary{positioning,arrows,calc,decorations.markings}
\tikzset{
modal/.style={>=stealth',shorten >=1pt,shorten <=1pt,auto,node distance=1.5cm,semithick},
world/.style={circle,draw,minimum size=0.5cm,fill=gray!15},
point/.style={circle,draw,inner sep=0.5mm,fill=black},
reflexive above/.style={->,loop,looseness=7,in=120,out=60},
reflexive below/.style={->,loop,looseness=7,in=240,out=300},
reflexive left/.style={->,loop,looseness=7,in=150,out=210},
reflexive right/.style={->,loop,looseness=7,in=30,out=330}
}
\usetikzlibrary{decorations.pathmorphing}
\tikzset{snake it/.style={decorate, decoration=snake}}

\newtheorem{theorem}{Theorem}

\newtheorem{lemma}[theorem]{Lemma}
\newtheorem{observation}[theorem]{Observation}
\newtheorem{proposition}[theorem]{Proposition}
\theoremstyle{definition}
\newtheorem{definition}[theorem]{Definition}
\newtheorem{example}[theorem]{Example}
\theoremstyle{remark}
\newtheorem{remark}[theorem]{Remark}

\DeclareSymbolFont{extraup}{U}{zavm}{m}{n}
\DeclareMathSymbol{\vardiamond}{\mathalpha}{extraup}{87}

\newcommand{\probbox}[1]{\centerline{\framebox{\parbox{0.95\columnwidth}{#1}}}}

\def\phi{\varphi}

\definecolor{tim}{RGB}{0, 0, 250}
\definecolor{pio}{RGB}{200, 0, 200}
\definecolor{orange-red}{RGB}{210, 47, 0}
\definecolor{dark-green}{RGB}{51, 102, 0}

\newcommand{\iffi}{\textit{iff} }
\newcommand{\dfn}{Definition}
\newcommand{\fig}{Figure}
\newcommand{\lem}{Lemma}
\newcommand{\thm}{Theorem}

\newcommand{\sect}{Section}

\newcommand{\calc}{\mathfrak{A}}
\newcommand{\calcii}{\mathfrak{B}}
\newcommand{\calciii}{\mathfrak{C}}
\newcommand{\calciv}{\mathfrak{D}}
\newcommand{\gtint}{\mathsf{G3I}}
\newcommand{\gtr}{\mathsf{G3I}'}
\newcommand{\gtsfour}{\mathsf{G3S4}}
\newcommand{\sfour}{\mathsf{S4}}

\newcommand{\con}{C}
\newcommand{\confam}{\mathcal{C}}
\newcommand{\inc}[1]{\hat{#1}}
\newcommand{\dec}[1]{\check{#1}}
\newcommand{\csize}[1]{|#1|}

\newcommand{\rel}{\mathcal{R}}
\newcommand{\seq}{S}
\newcommand{\edgs}{\mathcal{E}}
\newcommand{\eset}{\edgs}

\newcommand{\ant}{\Gamma}
\newcommand{\antii}{\Sigma}
\newcommand{\suc}{\Delta}

\newcommand{\E}{\mathcal{E}}
\newcommand{\Ea}{\mathcal{E}_{a}}
\newcommand{\Eb}{\mathcal{E}_{b}}
\newcommand{\Ec}{\mathcal{E}_{c}}
\newcommand{\Es}{\mathcal{E}_{s}}
\newcommand{\gseq}{\mathcal{G}}
\newcommand{\gspace}[1]{\mathfrak{G}(#1)}
\newcommand{\gclass}{\mathfrak{G}}
\newcommand{\ops}{\mathfrak{R}}

\newcommand{\sar}{\vdash} 
\newcommand{\seqa}{\Rightarrow}

\newcommand{\invh}[1]{\overline{#1}}
\newcommand{\seqrel}{R}
\newcommand{\seqset}{\mathtt{S}}
\newcommand{\pseqset}{\mathtt{PS}}

\newcommand{\vset}{\mathcal{V}}
\newcommand{\lfunc}{\mathcal{L}}
\newcommand{\etypset}{\mathtt{E}}
\newcommand{\cetypset}{\mathtt{\conv{E}}}

\newcommand{\eslang}{(\etypset \cup \cetypset)^{*}}
\newcommand{\etypa}{a}
\newcommand{\etypb}{b}
\newcommand{\etypc}{c}

\newcommand{\univ}{\mathcal{U}}
\newcommand{\exby}{\subseteq}
\newcommand{\calcspace}[1]{\mathbb{S}(#1)}
\newcommand{\upspace}[1]{\mathbb{U}(#1)}
\newcommand{\downspace}[1]{\mathbb{D}(#1)}
\newcommand{\uprel}{\eqslantless}
\newcommand{\downrel}{\leqslant}
\newcommand{\cs}{\mathbf{S}}
\newcommand{\simplepath}[1]{\overset{#1}{\leadsto}}
\newcommand{\ppath}[1]{\hspace{3pt}\raise-3pt\hbox{$\simplepath{#1}$}\hspace{3pt}}

\newcommand{\charx}{x}
\newcommand{\chary}{y}

\newcommand{\abet}{\mathtt{A}}
\newcommand{\words}{\mathtt{A}^{*}}
\newcommand{\pto}{\longrightarrow}
\newcommand{\gram}{\mathbf{G}}
\newcommand{\conv}[1]{\overline{#1}}
\newcommand{\langgram}{\gram}
\newcommand{\lang}{\mathscr{L}}

\newcommand{\ptostep}{\longrightarrow_{\gram}}
\newcommand{\ptoder}{\longrightarrow_{\gram}^{*}}
\newcommand{\stra}{s}
\newcommand{\strb}{t}
\newcommand{\strc}{r}
\newcommand{\empstr}{\varepsilon}

\newcommand{\pru}{p}
\newcommand{\langsfour}{\mathscr{L}_{\Box}}

\newcommand{\id}{i(\con,\seqrel)}

\newcommand{\eru}{e(\confam,\seqrel)}

\newcommand{\rru}{r(\con,\seqrel)}
\newcommand{\lru}{l(\con,\seqrel)}
\newcommand{\tru}{t(\con,\seqrel)}
\newcommand{\fhru}{h_{f}}
\newcommand{\bhru}{h_{b}}
\newcommand{\rui}{\rho}
\newcommand{\ruii}{\sigma}
\newcommand{\ruiii}{\tau}
\newcommand{\hrus}{\mathrm{H}}

\newcommand{\rus}{\mathrm{R}}
\newcommand{\wk}{pw}

\newcommand{\init}{id}

\newcommand{\disr}{\lor}

\newcommand{\conr}{\land}

\newcommand{\boxr}{\Box}
\newcommand{\diar}{\dia}

\newcommand{\diarii}{\dia_{2}}

\newcommand{\refl}{r}
\newcommand{\trans}{t}

\newcommand{\deriv}{\mathcal{D}}
\newcommand{\prf}{\mathcal{P}}
\newcommand{\dsize}[1]{s(#1)}
\newcommand{\qsize}[1]{q(#1)}
\newcommand{\seqsize}[1]{s(#1)}

\newcommand{\ialg}{\textsc{Implicate}}

\newcommand{\ealg}{\textsc{Explicate}}

\newcommand{\halg}{\textsc{InvHorn}}

\newcommand{\pdeq}{\dashv \vdash_{p}}
\newcommand{\depgr}{\mathsf{DG}}
\newcommand{\depgrv}{P}
\newcommand{\depgre}{\sqsubseteq}

\newcommand{\ppatha}[1]{\mathcal{E}_{#1}}

\newcommand{\ug}{\oplus}
\newcommand{\dg}{\ominus}
\newcommand{\udg}{\odot}
\newcommand{\perm}{\rightleftharpoons}
\newcommand{\permabove}{\hspace{1mm}\text{\raise1pt\hbox{\rotatebox[origin=c]{45}{$\rightharpoonup$}}}\hspace{1mm}}
\newcommand{\simul}{\preceq}
\newcommand{\bisimul}{\simeq}
\newcommand{\fby}{\triangleright}

\newcommand{\set}[1]{\{#1\}}
\newcommand{\vspacer}{\vphantom{\big\vert}}

\newcommand{\iimp}{\supset}

\newcommand{\nsfour}{\mathsf{NS4}}

\newcommand{\conmap}{m} 
\newcommand{\dia}{\Diamond} 

\begin{document}


\title{Foundations for an Abstract Proof Theory in the Context of Horn Rules}

\author{Tim S. Lyon}
\email{timothy\_stephen.lyon@tu-dresden.de}
\orcid{0000-0003-3214-0828}
\affiliation{%
  \institution{Technische Universit{\"a}t Dresden}
  \streetaddress{N{\"o}thnitzer Stra{\ss}e 46}
  \city{Dresden}
  \state{Saxony}
  \country{Germany}
  \postcode{01069}
}
\author{Piotr Ostropolski-Nalewaja}
\orcid{0000-0002-8021-1638}
\email{postropolski@cs.uni.wroc.pl}
\affiliation{%
  \institution{University of Wrocław}
  \city{Wrocław}
  \country{Poland}
}


\begin{abstract}
We introduce a novel, logic-independent framework for the study of sequent-style proof systems, which covers a number of proof-theoretic formalisms and concrete proof systems that appear in the literature. In particular, we introduce a generalized form of sequents, dubbed \emph{g-sequents}, which are taken to be binary graphs of typical, Gentzen-style sequents. We then define a variety of \emph{inference rule types} as sets of operations that act over such objects, and define \emph{abstract (sequent) calculi} as pairs consisting of a set of g-sequents together with a finite set of operations. Our approach permits an analysis of how certain inference rule types interact in a general setting, demonstrating under what conditions rules of a specific type can be permuted with or simulated by others, and being applicable to any multisequent proof system that fits within our framework. We then leverage our permutation and simulation results to establish generic calculus and proof transformation algorithms, which show that every abstract calculus can be effectively transformed into a lattice of polynomially equivalent abstract calculi. We determine the complexity of computing this lattice and compute the relative sizes of proofs and sequents within distinct calculi of a lattice. We recognize that top and bottom elements in lattices correspond to many known deep-inference nested sequent systems and labeled sequent systems (respectively) for logics characterized by Horn properties.
\end{abstract}



\received{18 April 2024}
\received[revised]{28 March 2026}
\received[accepted]{1 June 2026}

\maketitle

\section{Introduction}\label{sec:intro}


Proof calculi are indispensable tools in the theory and application of logics, serving as engines that facilitate reasoning within a given logical paradigm. Of particular importance are \emph{sequent-style calculi}, which were first introduced by Gentzen in the 1930s~\cite{Gen35a,Gen35b}. Gentzen's sequent systems consist of \emph{inference rules}, which operate over expressions called \emph{(Gentzen) sequents}, i.e. expressions of the form $\phi_{1}, \ldots, \phi_{n} \seqa \psi_{1}, \ldots, \psi_{k}$ with $\phi_{i}$ and $\psi_{j}$ logical formulae, used to derive theorems of a specified logic. Gentzen's sequent formalism has become one of the preferred formalisms for constructing calculi exhibiting the so-called \emph{sub-formula property}, meaning every formula occurring in the premise of an inference rule is a sub-formula of one occurring in the conclusion of the rule. 
This feature, and the sequent formalism more generally, have proven to be fruitful from both a theoretical and practical standpoint, being used to supply proof systems for a wide array of logics \cite{Cor89,Cur52,Ten87}, to discover new logics~\cite{Gir87}, to establish non-trivial properties of logics (e.g. interpolation~\cite{Mae60}), and 
to develop automated reasoning methods for logics~\cite{Dyc92,Sla97}. 
 
Yet, the discovery of new, expressive logics (e.g. the tense logic $\mathsf{Kt}$ and bi-intuitionistic logic) led to the realization that the sequent formalism was \emph{too strict} as sequent calculi exhibiting the sub-formula property remained elusive; cf.~\cite{BuiGor07,Wan94}. In response, a variety of formalisms extending Gentzen's traditional sequent formalism were introduced to recapture the sub-formula property. Such formalisms are referred to as \emph{multisequents} and are generalizations of Gentzen sequents, obtained by embedding Gentzen sequents into more complex data structures. 

Multisequent formalisms include \emph{hypersequents}, which are multisets of Gentzen sequents~\cite{Avr96,Pot83}, $2$-sequents/linear nested sequents, which are lines of Gentzen sequents~\cite{Mas92,Lel15}, nested sequents, which are trees of Gentzen sequents~\cite{Bul92,Kas94}, and labeled sequents~\cite{Sim94,Vig00}, which are binary graphs of sequents. Such proof systems have found a broad range of applications for diverse classes of logics, being used in the design of interpolant construction algorithms~\cite{FitKuz15,LyoTiuGorClo20,LyoKar24}, in writing decision algorithms (with counter-model extraction)~\cite{Bru09,LyoBer19,TiuIanGor12}, and in knowledge integration scenarios~\cite{LyoGom22}. Nevertheless, it was found that differing formalisms possessed distinct advantages over one another; e.g. nested calculi were found to be suitable for writing proof-search and decision algorithms~\cite{TiuIanGor12,LyoGom22}, whereas labeled calculi were found to admit algorithmic construction for large classes of logics~\cite{CiaMafSpe13}.

\begin{figure}[t]
\begin{center}
\scalebox{1.0}{
\bgroup
\def\arraystretch{1.1}
\begin{tabular}{|c|c|c|}
\hline
Logics & Labeled & Nested\\
\hline
Normal Modal Logics & \cite{NegPla11} & \cite{Bru09,Pog09b}\\
\hline
Bi-intuitionistic Logic & \cite{PinUus09} & \cite{PinUus18}\\
\hline
STIT Logics & \cite{BerLyo19a} & \cite{LyoBer19}\\
\hline 
FO Modal Logics & \cite{NegPla11,Lyo22a} & \cite{Lyo22a,LyoOrl23}\\
\hline 
FO Intuitionistic Logics & \cite{Lyo21,Lyo21thesis} & \cite{Fit14,Lyo21thesis}\\
\hline 
Provability Logic & \cite{NegPla11} & \cite{Pog09}\\
\hline 
Intuitionistic (Multi-)Modal Logics & \cite{Sim94,Lyo22b} & \cite{Str13,Lyo21a,Lyo25}\\
\hline 
Tense/Grammar Logics & \cite{Bor08,Lyo21thesis} & \cite{GorPosTiu11,TiuIanGor12}\\
\hline
\end{tabular}
\egroup
}
\end{center}
\caption{Examples of logics and associated multisequent calculi covered by our abstract framework. Citations to papers containing relevant (cut-free) labeled sequent and (variants of) reachability nested systems are provided in the second and third columns.\label{fig:logics-and-calculi}}
\end{figure}
 
Naturally, the arrival of new sequent-based formalisms gave rise to questions concerning their relationships: How are calculi in one formalism transformed into `deductively equivalent' calculi in another? What are the relative sizes of proofs and sequents in one formalism compared to another? Under what conditions are proofs transformable between formalisms and what are the complexity bounds thereof? Such questions have typically been investigated in restricted concrete settings, focusing on specific multisequent calculi for known classes of logics~\cite{CiaLyoRamTiu21,Gal19a,GorRam12,Lel15,PinUus18}.  Our work deviates from these approaches. Instead of studying calculi tailored to particular logics, we propose a novel and unifying \emph{abstract framework} that captures a broad family of multisequent calculi and logics. Many systems studied in the literature can therefore be seen as concrete instances (i.e. specific parameter choices) of the general machinery we develop.

In particular, we formulate calculi as pairs consisting of (1) a set of objects called \emph{generalized sequents} accompanied by (2) a finite set of inference rules. We therefore shift our attention from proof systems for logics, and instead, focus on proof systems in and of themselves, yielding a \emph{logic-independent} approach for studying the properties of, and relationships between, multisequent systems. Due to its generality, our results hold for any logic or multisequent system that can be viewed as an object in our framework. Specifically, we accomplish the following:

$\bullet$ We generalize the notion of \emph{sequent} to a labeled graph (where nodes are assumed to be labeled by Gentzen sequents). We refer to these objects as \emph{g-sequents}, and they cover various kinds of sequents (e.g. labeled, nested, linear nested) that commonly appear in proof-theoretic works.

$\bullet$ We generalize inference rules to select \emph{inference rule types} that operate over g-sequents, revealing the critical components that constitute an inference rule. These inference rule types subsume standard inference rules for multisequent systems and include parameters (which we call \emph{sequent constraints} and \emph{structural constraints}) whose instantiation yields concrete inference rules. 

$\bullet$ We define \emph{abstract (sequent) calculi} as pairs, whose first component is a set of g-sequents and whose second component is a finite collection of inference rules. This generic notion of calculus ensures that our results hold for any multisequent calculus that can be viewed as an object within our framework.
 
$\bullet$ Our abstract calculi include structural rules that facilitate reasoning with Horn properties. Therefore, a sizable number of (non-)classical logics semantically characterized by `Horn' frame conditions, and their accompanying multisequent systems, are subsumed by our work. Examples of logics with proof systems covered by our framework can be viewed in Figure~\ref{fig:logics-and-calculi}, though we remark that this list is not exhaustive.

$\bullet$ We define proof transformation notions (e.g. \emph{permutation} and \emph{simulation}) as well as explain how to strengthen or weaken 
certain rules (via new operations called \emph{absorption} and \emph{fracturing}), which are used to provide generic calculus and derivation transformation algorithms and to compute complexity bounds thereof. This work contributes to a better understanding of how structural rules are eliminated from proofs, and how reachability and propagation rules~\cite{GorPosTiu11,Lyo21thesis} arise from this process. This is important as proof systems with such rules typically incorporate less bureaucracy, utilize more economical data structures, and 
are amenable to automated reasoning tasks~\cite{Bru09,LyoTiuGorClo20,LyoGom22,TiuIanGor12}.

$\bullet$ We discover that every abstract calculus exists within a finite lattice of `polynomially equivalent' abstract calculi, which we show how to compute. We observe that the top and bottom of a lattice is one of two calculus types, which we call \emph{implicit} and \emph{explicit} calculi, respectively. When we instantiate a lattice with known multisequent systems, we find that various deep-inference nested sequent calculi with reachability rules (e.g.~\cite{TiuIanGor12,Fit14,Lyo21a}) serve as top elements, whereas various standard labeled sequent calculi (e.g.~\cite{Sim94,Vig00,NegPla11}) serve as bottom elements, establishing a \emph{duality} between many known nested and labeled sequent systems, which includes pairs of systems mentioned in \fig~\ref{fig:logics-and-calculi}.

Our abstract approach has explanatory value, yielding deep 
insights into the nature of, and connection between, multisequent systems. Also, we provide a widely applicable toolkit for the manipulation of proofs and proof systems.

\smallskip

\noindent
\textbf{Related Work.} Various works have been put forth exploring proof transformation algorithms within or between distinct multisequent systems. Examples of such works include studying the relationship between labeled sequent systems and tree-hypersequent systems for provability logic~\cite{GorRam12}, defining proof transformations between linear nested sequent, 2-sequent, and hypersequent systems for modal and intuitionistic logics~\cite{Lel15}, and exploring simulations between display calculi, nested sequent calculi, and labeled calculi for tense/temporal logics~\cite{CiaLyoRamTiu21}, though the literature abounds with studies of a similar nature~\cite{Fit14,Gal19a,GorPosTiu11,Lyo20b,Lyo21a,Lyo25b,PinUus18}. Typically, such works define proof transformations by means of \emph{permutations} and \emph{simulations}, that is, such works demonstrate how proofs can be re-written by swapping applications of inference rules (permutations), or by replacing applications of inference rules with alternative applications (simulations), deriving the same conclusion. As mentioned above, these projects have always been carried out in \emph{concrete settings}, analyzing specific proof systems for specific logics. In the current paper, we generalize the methodology of such works, studying permutations and simulations for inference rule types (as opposed to specific inference rules), giving rise to generic proof transformations 
that hold for any calculus which can be viewed as an object within our framework.

\textit{Structural Refinement.} Of particular relevance to this article are works detailing the relationship between labeled sequent systems and nested sequent systems for non-classical logics. The formalism of labeled sequents was initiated by Kanger in the 1950’s~\cite{Kan57}, though it was arguably Simpson~\cite{Sim94} who provided the contemporary form of such systems. Labeled sequent systems extend traditional sequent systems by incorporating semantic information directly into the syntax of sequents. Consequently, labeled sequents take the form of graphs of Gentzen sequents. For logics admitting a relational semantics (e.g. those mentioned in \fig~\ref{fig:logics-and-calculi}), the edges in their labeled sequents encode the accessibility relation. A characteristic feature of labeled sequent systems is the inclusion of \emph{structural rules} that manipulate the edges in labeled sequents and semantically correspond to frame conditions imposed on the accessibility relation. As a consequence, distinct non-classical logics may be supplied a labeled sequent system by transforming the semantics of the logic and its frame conditions into inference rules. This has the effect that the labeled sequent formalism is quite general and modular, allowing for the specification of numerous non-classical logics (e.g.~\cite{Sim94,Vig00,NegPla11}).

The formalism of nested sequents arose out of the work of Kashima~\cite{Kas94} and Bull~\cite{Bul92}, and is distinct from the labeled sequent formalism both in terms of the sequents used and types of rules used.\footnote{Although it should be noted that Leivant~\cite[p. 361]{Lei81} introduced a notational variant of nested sequents in 1981 (which prefixes formulae with so-called \emph{execution sequences}) in his proof-theoretic work on propositional dynamic logic.} Nested sequents are another extension of Gentzen's sequent formalism, which take the form of trees of Gentzen sequents. There are at least three prominent kinds of nested sequent systems that commonly appear in the literature. 
The first kind, \emph{shallow nested sequent systems} (e.g.~\cite{Kas94,GorPosTiu11,TiuIanGor12}), are essentially one-sided display calculi~\cite{Bel82}. In such systems, rules are only applicable to the root of a nested sequent and require \emph{residuation rules} (also called \emph{display rules})---which change the root of a nested sequent to an internal node---for cut-elimination and completeness. 
The second and third kinds of nested sequent systems are often lumped together as \emph{deep-inference systems} (e.g.~\cite{Bru09,GorPosTiu11,Str13}). These latter two kinds of systems exhibit deep-inference, meaning, rules are applicable to any node in a nested sequent; this has the effect that residuation rules are admissible (cf.~\cite{CiaLyoRamTiu21,GorPosTiu08,GorPosTiu11}). 
Despite that fact that the second and third kinds both fall within the deep-inference paradigm, they may be distinguished based on their inclusion of either structural rules or reachability rules. In the nested sequent setting, structural rules change the tree structure of a nested sequent without affecting the Gentzen sequents associated with nodes. Conversely, reachability rules operate by propagating or consuming data along paths in a nested sequent and only affect the Gentzen sequents associated with nodes. To separate these two kinds of systems, we refer to deep-inference nested sequent systems with structural rules as \emph{structural nested systems} and deep-inference nested sequent systems with reachability rules as \emph{reachability nested systems}. Numerous works define both kinds of deep-inference systems and even establish syntactic correspondences between them; e.g.~\cite{Bru09,Pog09,GorPosTiu11,Str13}. Last, we note that the above tripartite classification of nested sequent systems is not intended to suggest that \emph{every} nested system fits within one of the three classes. Rather, these classes highlight three different kinds of systems that commonly appear in the literature. There are exceptions and nested sequent systems exist that include aspects of differing classes; e.g. nested systems that incorporate both structural rules and reachability rules~\cite{Bru09,Lyo23,MarStr14}.

The relationship between labeled sequent systems and reachability nested systems has been explored in a series of papers~\cite{CiaLyoRamTiu21,GorRam12,Lyo21,Lyo21thesis,Lyo21a}. These numerous case studies led to the realization that, in many settings, structural rule elimination can be used to transform labeled sequent proofs into nested sequent proofs and that structural rule introduction can be used for the converse transformation. 
These observations led to the formulation of a methodology, referred to as \emph{structural refinement}, for extracting reachability nested systems from labeled sequent systems. The central idea was to leverage the automatic construction procedures of labeled systems along with structural rule elimination to transform labeled systems into reachability nested systems (see~\cite{Lyo21thesis}). This methodology did, in fact, lead to the identification of numerous new reachability nested systems~\cite{LyoBer19,Lyo21thesis,Lyo21a,Lyo22a,Lyo25}. 


The current paper can be seen as the end result of the first author's work on structural refinement, yet goes significantly beyond this work. While structural refinement offers guidelines on how one might extract a reachability nested system from a given labeled sequent system, this work presents concrete calculus transformation algorithms that have the ability to definitively transform labeled sequent systems into reachability nested systems and vice-versa. Furthermore, this paper establishes general permutation and simulation relationships between inference rule types, along with novel operations (absorption and fracturing) for synthesizing and analyzing inference rules. This allows for a number of proof transformations to be defined, including structural rule elimination and introduction.

\smallskip

\noindent
\textbf{Organization of Paper.} In \sect~\ref{sec:overview}, we explain how our framework was designed by abstracting general underlying patterns appearing in calculi, considering various inference rule types and proof manipulation techniques. In \sect~\ref{sec:abstract-calc}, we define our framework, and then put it to use in \sect~\ref{sec:perm-sim} to establish a large number of permutation and simulation relationships between various inference rule types. In \sect~\ref{sec:generic-algs}, we demonstrate how abstract calculi can be converted into lattices of polynomially equivalent calculi and specify our generic calculus and proof transformation algorithms. We also discuss how reachability nested and labeled systems can be identified with top and bottom elements of these lattices. Subsequently, in \sect~\ref{sec:example}, we give an example showing how our abstract framework can be instantiated and used to generate a lattice of polynomially equivalent calculi for the modal logic $\sfour$. Finally, in \sect~\ref{sec:conclusion}, we conclude and discuss avenues for future research.

\section{Overview of our Approach}\label{sec:overview}


\begin{figure*}[t]

\begin{center}
\begin{tabular}{c c}
\AxiomC{}
\RightLabel{$(id)$}
\UnaryInfC{$\rel,\, wEu,\, \Gamma,\, w : p \sar u : p,\, \Delta$}
\DisplayProof\hspace{5mm}

&

\AxiomC{$\rel,\, \Gamma,\, w : \phi \sar \Delta$}
\AxiomC{$\rel,\, \Gamma,\, w : \psi \sar \Delta$}
\RightLabel{$(\lor_{L})$}
\BinaryInfC{$\rel,\, \Gamma,\, w : \phi \lor \psi \sar \Delta$}
\DisplayProof
\end{tabular}
\end{center}
\begin{center}
\begin{tabular}{c}
\AxiomC{$\rel,\, w E u,\, \Gamma,\, w : \phi \supset \psi \sar u : \phi,\, \Delta$}
\AxiomC{$\rel,\, w E u,\, \Gamma,\, w : \phi \supset \psi,\, u : \psi \sar \Delta$}
\RightLabel{$(\supset_{L})$}
\BinaryInfC{$\rel,\, wEu,\, \Gamma,\, w : \phi \supset \psi \sar \Delta$}
\DisplayProof
\end{tabular}
\end{center}
\begin{center}
\begin{tabular}{c @{\hskip .1em} c @{\hskip .1em} c}
\AxiomC{$\rel,\, wEw,\, \Gamma \sar \Delta$}
\RightLabel{$(ref)$}
\UnaryInfC{$\rel,\, \Gamma \sar \Delta$}
\DisplayProof\hspace{2mm}

&

\AxiomC{$\rel,\, wEu,\, uEv,\, wEv,\, \Gamma \sar \Delta$}
\RightLabel{$(tra)$}
\UnaryInfC{$\rel,\, wEu,\, uEv,\, \Gamma \sar \Delta$}
\DisplayProof\hspace{2mm}

&

\AxiomC{$\rel,\, w E u,\, \Gamma,\, u : \phi \sar u : \psi,\, \Delta$}
\RightLabel{$(\supset_{R})^{\dag}$}
\UnaryInfC{$\rel,\, \Gamma \sar w : \phi \supset \psi,\, \Delta$}
\DisplayProof
\end{tabular}
\end{center}

\caption{Some inference rules from the labeled calculus $\gtint$ for propositional intuitionistic logic~\cite{NegPla11}. We let $\gtr$ denote the collection of the above rules. The side condition $\dag$ stipulates that the rule is applicable only if $u$ is fresh, i.e. $u$ does not occur in the surrounding context $\rel, \Gamma, \Delta$. 
\label{fig:int-calc}}
\end{figure*}

We now turn our attention toward explaining our abstract proof-theoretic framework, which arose out of analyzing a range of calculi and formalizing underlying patterns. Rather than attempting to recount this entire development, we use a single labeled sequent calculus as a running example. The goal of this section is to provide the intuition that motivates the general framework introduced in Section~\ref{sec:abstract-calc}. While the present discussion offers only a guided example, Section~\ref{sec:abstract-calc} presents the full formal account.

We have chosen a fragment of the labeled sequent calculus $\gtint$~\cite{NegPla11} for propositional intuitionistic logic to use as our running example. We denote this fragment by $\gtr$, and define it to be the set 
 of rules shown in \fig~\ref{fig:int-calc}.\footnote{We employ a slight variation of the notation used for labeled sequents in~\cite{NegPla11} to better fit within the notation of our framework.} This calculus avoids the unnecessary complexities of other multisequent systems, while also possessing revelatory attributes that justify concepts later defined within our~framework.
%

\subsection{The Structure of Multisequents and Inference Rules}

As mentioned in \sect~\ref{sec:intro}, multisequents take various forms, typically being types of graphs with sequents as vertices. 
These may take the form of binary graphs~\cite{Sim94}, polytrees~\cite{CiaLyoRamTiu21}, trees~\cite{Kas94}, lines~\cite{Mas92}, or single points, yielding standard sequents~\cite{Gen35a,Gen35b}. For instance, the labeled sequents employed in $\gtr$ take the form of graphs, as we will now explain. 

We define the language $\mathcal{L}$ of propositional intuitionistic logic to be the set of formulae generated via the following grammar in BNF: $\phi ::= p \ | \ \bot \ | \ \phi \lor \phi \ | \ \phi \land \phi \ | \ \phi \supset \phi$, where $p$ ranges over a set of propositional variables. We let $\{w, u, v, \ldots\}$ be a set of \emph{labels}. A \emph{relational atom} (or, \emph{edge}) is an expression of the form $wEu$ and a \emph{labeled formula} is an expression of the form $w : \phi$ such that $w$ and $u$ are labels and $\phi \in \mathcal{L}$. A \emph{labeled sequent} in $\gtr$ is an expression of the form $\rel, \Gamma \sar \Delta$ such that $\rel$ is a set of relational atoms, and $\Gamma$ and $\Delta$ are multisets of labeled formulae. 

\begin{wrapfigure}{r}{0.4\textwidth}
\vspace{-1em} 
\centering
\begin{tikzpicture}[
    world/.style={circle, draw, minimum size=6mm},
    active/.style={very thick},
    node distance=1.4cm,
    >=stealth
]


\node (wlabel) at (0,0.7) {$(\phi \seqa \psi)$};
\node[world, active] (w) at (0,0) {$w$};

\node (zlabel) at (3,0.7) {$(\psi \seqa \chi, \xi)$};
\node[world, active] (z) at (3,0) {$z$};

\node[world, active] (u) at (0,-2) {$u$};
\node (ulabel) at (0,-2.7) {$(\chi \seqa \emptyset)$};

\node[world, active] (v) at (3,-2) {$v$};
\node (vlabel) at (3,-2.7) {$(\emptyset \seqa \emptyset)$};

\draw[->, thick] (w) -- (z);
\draw[->, thick] (w) -- (u);
\draw[->, thick] (w) -- (v);
\draw[->, thick] (u) -- (v);

\end{tikzpicture}
\vspace{-1em} 
\end{wrapfigure}
Each labeled sequent can be viewed as a binary graph of sequents, obtained by depicting all labels as vertices, all relational atoms as edges, and all labeled formulae as sequents labeling nodes (cf.~\cite{Lyo21thesis}). For instance, the graph shown 
to the right corresponds to $\rel, \Gamma \sar \Delta$ with $\rel := wEw, wEz, wEv, wEu, uEv$, $\Gamma := w : \phi, u : \chi, z : \psi$, and $\Delta := w : \psi, z : \chi, z : \xi$. Every labeled sequent can be rewritten in an equivalent form $\rel \sar \Sigma$, where $\rel$ is a set of relational atoms as before, but $\Sigma$ is a set of prefixed sequents. As an example, the labeled sequent $\rel, \Gamma \sar \Delta$ can also be written as $\rel \sar w : (\phi \seqa \psi), z : (\psi \seqa \chi, \xi), u : (\chi \seqa \emptyset), v : (\emptyset \seqa \emptyset).$ We view this perspective of labeled (or, graphical) sequents as beneficial for a couple reasons. First, the internal structure of a sequent is \emph{logic-dependent}; e.g. certain intuitionistic logics may restrict the succedent to at most one formula~\cite{Gen35a,Gen35b} or certain sub-structural logics may employ sequences of formulae as opposed to (multi)sets in the antecedent or succedent~\cite{Gir87}. As we are interested in providing a generic framework that studies the \emph{graphical properties} of `generalized or graphical sequents' and their associated proof systems, we may simply view sequents as \emph{types of labels} (or, \emph{colors}). We will therefore define the notion of a \emph{generalized sequent} (\emph{g-sequent} for short) as a graph of `sequents' without specifying the internal structure of such sequents, yielding a logic-independent study of multisequent systems, as presented in \sect~\ref{sec:abstract-calc}. Second, as mentioned above, various multisequent formalisms beget proof systems that operate over certain types of graphs of sequents. Thus, our notion of g-sequent captures all such formalisms uniformly as restricting the g-sequents used yields a certain formalism. For example, restricting to linear structures corresponds to the linear nested sequent formalism~\cite{Lel15}.


Since we view labeled sequents as graphs of sequents, we will reformulate the labeled sequents and inference rules of 
$\gtr$ accordingly. We now take a labeled sequent to be an expression of the form $\rel \sar \Sigma$ such that $\rel$ is a set of edges (i.e. relational atoms) as before and $\Sigma$ is a set of \emph{prefixed sequents}, which are of the form $w : (X \seqa Y)$ with $X = \phi_{1}, \ldots, \phi_{n}$ and $Y = \psi_{1}, \ldots, \psi_{k}$ multisets of intuitionistic formulae. We rewrite the inference rules of $\gtr$ in this notation and discuss the \emph{inference rule type} that each rule falls within. Informally, an inference rule type is a set of inference rules that all exhibit a similar behavior. The inference rule types we consider in this paper are \emph{initial rules}, \emph{local rules}, \emph{expansion rules}, \emph{transmission rules}, \emph{Horn rules}, and \emph{reachability rules}. We will use the rules of $\gtr$ as concrete examples of specific rule types with the goal of providing intuition, and will provide formal definitions of each rule type in Section~\ref{sec:abstract-calc}.

\begin{figure}[t]
\centering
\scalebox{0.95}{
\begin{tabular}{@{\hskip -1cm} c}
\begin{minipage}[t]{0.5\columnwidth}
\centering
\begin{tikzpicture}[baseline={(0,0.6)}, node distance=1.6cm,
    world/.style={circle, draw, minimum size=6mm},
    active/.style={very thick},
    fitbox/.style={draw, dashed, inner sep=8pt, rounded corners},
    >=stealth
]

\draw[thick] (-2.2,0.975) -- (2.2,0.975);
\node[right=3pt] at (2.2,0.975) {$(id)$};

\coordinate (mid) at (0,0.35);

\node[world, active] (w2) at ($(mid)+(-1, -0.4)$) {$w$};
\node[world, active] (u2) at ($(mid)+( 1, -0.4)$) {$u$};

\draw[->, thick] (w2) -- (u2);

\node (w2lab) at ($(w2)+(0,-0.7)$) {$(X,p \seqa Y)$};
\node (u2lab) at ($(u2)+(0,-0.7)$) {$(X' \seqa p, Y')$};

\node[fitbox, fit=(w2)(u2)(w2lab)(u2lab)] (ConBox) {};
\node[left=6pt of ConBox.west] {$\mathcal{G}$};

\end{tikzpicture}
\end{minipage}
\begin{minipage}[t]{0.5\columnwidth}
\centering
\begin{tikzpicture}[baseline={(0,0.6)}, node distance=1.6cm,
    world/.style={circle, draw, minimum size=6mm},
    active/.style={very thick},
    fitbox/.style={draw, dashed, inner sep=6pt, rounded corners}
]

\node[world, active] (w1) at (-2,2.5) {$w$};
\node (G1label) at (-2,1.75) {$(X,~\phi~\seqa~Y)$};

\node[fitbox, fit=(w1) (G1label)] (G1box) {};
\node[right=6pt of G1box.east] {$\mathcal{G}_1$};

\node[world, active] (w2) at (2,2.5) {$w$};
\node (G2label) at (2,1.75) {$(X,~\psi~\seqa~Y)$};

\node[fitbox, fit=(w2) (G2label)] (G2box) {};
\node[right=6pt of G2box.east] {$\mathcal{G}_2$};

\draw[thick] ($(G1box.south west)+(0,-0.3)$)
             -- ($(G2box.south east)+(0,-0.3)$);
\node[right=3pt] at ($(G2box.south east)+(0,-0.3)$) {$(\lor_L)$};

\node[world, active] (wC) at (0,0) {$w$};
\node (Glabel) at (0,-.75) {$(X,~\phi\lor\psi~\seqa~Y)$};

\node[fitbox, fit=(wC) (Glabel)] (Gbox) {};
\node[right=6pt of Gbox.east] {$\mathcal{G}$};

\end{tikzpicture}
\end{minipage}
\end{tabular}
}
\caption{Graphical presentations of $(id)$ and $(\lor_L)$.\label{fig:intial-local-eg}}
\end{figure}

\smallskip
\noindent
\textbf{Initial Rules.} In our new notation, the $(id)$ rule takes the form shown below. This rule is depicted graphically to the left in Figure~\ref{fig:intial-local-eg} with the relational atom $wEu$ and prefixed sequents $w : (X, p \seqa Y)$ and $u : (X' \seqa p, Y')$ explicitly displayed. The dashed rectangle is taken to represent the remaining structure (i.e., the context) of the conclusion $\gseq$, viz. the other vertices, edges, and labels.
\begin{center}
\AxiomC{$\phantom{\rel}$}
\RightLabel{$(id)$}
\UnaryInfC{$\underbrace{\rel,\, wEu \sar \Sigma,\, w : (X, p \seqa Y),\, u : (X' \seqa p, Y')}_{\mathcal{G}}$}
\DisplayProof
\end{center}
We ask: what are the features of an inference rule that make it initial? Obviously, they are free of premises, and dictate what is taken to be axiomatic. Second, we observe that such rules may rely on the existence of relational data; e.g. in the $(id)$ rule above, an edge $wEu$ connecting one sequent of a certain type to another sequent of a certain type must be present. This gives rise to what we call a \emph{structural constraint}, which in general is a constraint on the shape of paths that must exist in the g-sequents comprising a rule application and which we encode using labeled graphs. 

We formalize the structural constraint of the $(id)$ rule as the labeled graph $\con = (\{w,u\},\{(w,u)\},L)$ such that $L(w,u) = \{E\}$. One may verify that $(id)$ satisfies such a constraint in the sense that any instance of $(id)$ can be `pattern matched' to such a constraint (with the edge $(w,u)$ being associated with $wEu$). That is to say, the structural constraint $\con$ encodes that fact that there must be a single edge present between the distinguished vertices $w$ and $u$ in any instance of $(id)$. Third, although structural constraints appear to be critical features of initial rules (and of inference rules more broadly), such objects are not enough to clearly express the operation of $(id)$. We also require that prefixed sequents satisfy a certain relation, which we refer to as a \emph{sequent constraint}; e.g. in the $(id)$ rule above, a relation $R$ must hold between the sequent $S_{1} = X_{1} \seqa Y_{1}$ prefixed with $w$ and the sequent $S_{2} = X_{2} \seqa Y_{2}$ prefixed with $u$, where $R(\seq_{1},\seq_{2})$ holds 
\iffi $p \in X_{1}$ and $p \in Y_{2}$.

\smallskip
\noindent
\textbf{Local Rules.} We define a \emph{local rule} to be an inference rule that operates only on sequents at a single label. The terminology reflects the fact that such rules act locally at a specific vertex of the graph and are independent of the edge structure. In the context of modal logics, local rules typically correspond to the (de)composition of Boolean connectives. As an example, if we rewrite $(\lor_{L})$ in our notation, the rule has the form shown below. The rule is depicted graphically on the right in Figure~\ref{fig:intial-local-eg}.

\smallskip
\begin{center}
\AxiomC{$\rel  \sar \Sigma, w : (X, \phi \seqa Y)$\hspace{-3mm}}
\AxiomC{$\rel  \sar \Sigma, w : (X, \psi \seqa Y)$}
\RightLabel{$(\lor_{L})$}
\BinaryInfC{$\rel  \sar \Sigma, w : (X, \phi \lor \psi \seqa Y)$}
\DisplayProof
\end{center}
\smallskip
Observe that this rule only manipulates data occurring in sequents at the label $w$. (NB. Below, we let $w : \seq_{1}$, $w : \seq_{2}$, and $w : \seq_{3}$ denote the displayed prefixed sequents occurring in the left premise, right premise, and conclusion of the rule, respectively.) As in the initial rule case above, we recognize that a sequent constraint is required to fully specify the operation of the $(\lor_{L})$ rule: for sequents $\seq_{1} = X_{1} \seqa Y_{1}$, $\seq_{2} = X_{2} \seqa Y_{2}$, and $\seq_{3} = X_{3} \seqa Y_{3}$, we define $R(\seq_{1},\seq_{2},\seq_{3})$ \iffi $\phi \in X_{1}$, $\psi \in X_{2}$, $\phi \lor \psi \in X_{3}$, $X_{1} \setminus \phi = X_{2} \setminus \psi = X_{3} \setminus \phi \lor \psi$, and $Y_{1} = Y_{2} = Y_{3}$. Such a relation must hold in any application of this rule for it to qualify as a valid rule application. 

\medskip
\noindent
\textbf{Expansion Rules.} In contrast to local rules, which operate on sequents at a single vertex without modifying the relational structure, \emph{expansion rules} extend the underlying graph by bottom-up adding an edge (in $\rel$) to a new label. The freshness of the introduced label---enforced by a side condition requiring that it does not appear elsewhere in the conclusion---is essential, as it ensures that the new vertex carries no prior assumptions and thus represents an arbitrary accessible world. The $(\supset_{R})$ rule serves as an example of an expansion rule, taking the form shown below when rewritten in our notation. A pictorial representation of $(\supset_{R})$ is given to the left in Figure~\ref{fig:example-impR-tra}.
\begin{center}
\AxiomC{$\rel,\, wEu \sar \Sigma,\, w : (X \seqa Y), u : (\phi \seqa \psi)$}
\RightLabel{$(\supset_{R})$}
\UnaryInfC{$\rel \sar \Sigma,\, w : (X \seqa Y, \phi \supset \psi)$}
\DisplayProof
\end{center}
Similar to the case of the $(\lor_{L})$ rule above, we observe that a sequent constraint must hold, specifying how the sequents at $w$ and $u$ in the premise relate to each other and the sequent at $w$ in the conclusion. This sequent constraint can be defined as follows: for sequents $\seq_{1} = X_{1} \seqa Y_{1}$, $\seq_{2} = X_{2} \seqa Y_{2}$, and $\seq_{3} = X_{3} \seqa Y_{3}$, we define $R(\seq_{1},\seq_{2},\seq_{3})$ \iffi $X_{2} = \phi$, $Y_{2} = \psi$, $X_{3} = X_{1}$, and $Y_{3} = Y_{2}, \phi \iimp \psi$. One can readily verify that $R(\seq_{1},\seq_{2},\seq_{3})$ holds if we take $S_{1}$ to be the sequent at $w$ in the premise, $S_{2}$ to be the sequent at $u$ in the premise, and $S_{3}$ to be the displayed sequent in the conclusion.

\begin{figure}[t]
\centering

\begin{tikzpicture}[node distance=1.8cm,
    world/.style={circle, draw, minimum size=6mm},
    active/.style={very thick},
    fitbox/.style={draw, dashed, inner sep=6pt, rounded corners},
    >=stealth
]


\node[world, active] (w1) at (-3,2.5) {$w$};
\node (w1lab)      at (-3,1.7) {$(X,\,\phi\supset\psi \seqa Y)$};

\node[world, active] (u1) at (-0.5,2.5) {$u$};
\node (u1lab)      at (-0.5,1.7) {$(X' \seqa \phi,\ Y')$};

\draw[->, thick] (w1) -- (u1);

\node[fitbox, fit=(w1)(u1)(w1lab)(u1lab)] (G1box) {};
\node[left=6pt of G1box.west] {$\mathcal{G}_1$};


\node[world, active] (w2) at (3.5,2.5) {$w$};
\node (w2lab)      at (3.5,1.7) {$(X,\,\phi\supset\psi \seqa Y)$};

\node[world, active] (u2) at (6.3,2.5) {$u$};
\node (u2lab)      at (6.3,1.7) {$(X',\,\psi \seqa Y')$};

\draw[->, thick] (w2) -- (u2);

\node[fitbox, fit=(w2)(u2)(w2lab)(u2lab)] (G2box) {};
\node[left=6pt of G2box.west] {$\mathcal{G}_2$};


\coordinate (ruleSW) at ($(G1box.south west)+(0,-0.3)$);
\coordinate (ruleSE) at ($(G2box.south east)+(0,-0.3)$);

\draw[thick] (ruleSW) -- (ruleSE);
\node[right=4pt] at (ruleSE) {$(\supset_L)$};


\path let \p1 = (ruleSW), \p2 = (ruleSE)
      in coordinate (mid) at ({(\x1+\x2)/2}, {(\y1+\y2)/2});

\node[world, active] (wc) at ($(mid)+(-1.25,-1.0)$) {$w$};
\node[world, active] (uc) at ($(mid)+( 1.25,-1.0)$) {$u$};

\node (wclab) at ($(wc)+(0,-0.75)$) {$(X,\,\phi\supset\psi \seqa Y)$};
\node (uclab) at ($(uc)+(0,-0.75)$) {$(X' \seqa Y')$};

\draw[->, thick] (wc) -- (uc);

\node[fitbox, fit=(wc)(uc)(wclab)(uclab)] (Gbox) {};
\node[left=6pt of Gbox.west] {$\mathcal{G}$};

\end{tikzpicture}

\caption{Graphical presentation of $(\supset_L)$.\label{fig:example-imp-left}}
\end{figure}

\medskip
\noindent
\textbf{Transmission Rules.} We take a \emph{transmission rule} to be an inference rule that updates two sequents connected by a single edge. A transmission rule bottom-up transmits data from one vertex to another vertex along an edge. Unlike expansion rules, transmission rules do not introduce fresh labels or new edges; rather, they operate over relational structure that is already present in the conclusion. The $(\supset_{L})$ rule serves as an example of a transmission rule, which takes the form shown below in our notation. A graphical depiction of this rule is given in Figure~\ref{fig:example-imp-left}.
\begin{center}
$\gseq_{1} = \rel,\, w E u \sar \Sigma, w : (X, \phi \supset \psi \seqa Y), u :(X' \seqa \phi, Y')$
\end{center}
\begin{center}
$\gseq_{2} = \rel,\, w E u \sar \Sigma, w : (X, \phi \supset \psi \seqa Y), u :(X', \psi \seqa Y')$
\end{center}
\begin{center}
\AxiomC{$\gseq_{1}$}
\AxiomC{$\gseq_{2}$}
\RightLabel{$(\supset_{L})$}
\BinaryInfC{$\rel,\, w E u \sar \Sigma,\, w : (X, \phi \supset \psi \seqa Y),\, u :(X' \seqa Y')$}
\DisplayProof
\end{center}
Similar to the case of the $(id)$ rule, the use of an edge (viz. $wEu$) in updating the sequents at $w$ and $u$ implies that a structural constraint must be enforced. 
We can formalize this constraint as the graph $\con = (\{w,u\},\{(w,u)\},L)$ such that $L(w,u) = \{E\}$. We observe that $(\supset_{L})$ satisfies this constraint in the sense that it captures what kind of edge structure must be present in the premises and conclusion of the rule. In addition, a sequent constraint is required to fully specify the operation of the above rule, relating the sequents at $w$ and $u$ in the left premise, right premise, and conclusion. This can be formalized in a manner similar to the sequent constraints defined for $(\lor_L)$ and $(\iimp_R)$.

\begin{figure}[t]
\centering

\begin{minipage}[t]{0.47\textwidth}
\centering
\begin{tikzpicture}[baseline={(0,0)}, node distance=1.6cm,
    world/.style={circle, draw, minimum size=6mm},
    active/.style={very thick},
    fitbox/.style={draw, dashed, inner sep=6pt, rounded corners},
    >=stealth
]

\node[world, active] (w1) at (0,2.5) {$w$};
\node (w1label) at (0,1.75) {$(X \seqa Y)$};

\node[world, active] (u1) at (3,2.5) {$u$};
\node (u1label) at (3,1.75) {$(\phi \seqa \psi)$};

\draw[->, thick] (w1) -- (u1);

\node[fitbox, fit=(w1)(u1)(w1label)(u1label)] (G1box) {};
\node[left=6pt of G1box.west] {$\mathcal{G}'$};

\coordinate (ruleSW) at ($(G1box.south west)+(0,-0.3)$);
\coordinate (ruleSE) at ($(G1box.south east)+(0,-0.3)$);

\draw[thick] (ruleSW) -- (ruleSE);
\node[right=3pt] at (ruleSE) {$(\iimp_R)$};

\path let \p1 = (ruleSW), \p2 = (ruleSE)
    in coordinate (mid) at ({(\x1+\x2)/2}, {(\y1+\y2)/2});

\node[world, active] (wC) at ($(mid)+(0,-1)$) {$w$};
\node (Glabel) at ($(mid)+(0,-1.7)$) {$(X \seqa Y,\ \phi \iimp \psi)$};

\node[fitbox, fit=(wC)(Glabel)] (Gbox) {};
\node[left=6pt of Gbox.west] {$\mathcal{G}$};

\end{tikzpicture}
\end{minipage}
\hspace{1.5em}
\begin{minipage}[t]{0.47\textwidth}
\centering
\begin{tikzpicture}[baseline={(0,0)}, node distance=1.6cm,
    world/.style={circle, draw, minimum size=6mm},
    active/.style={very thick},
    fitbox/.style={draw, dashed, inner sep=8pt, rounded corners},
    >=stealth
]

\node[world, active] (w) at (0,1.8) {$w$};
\node[world, active] (u) at (2,1.8) {$u$};
\node[world, active] (v) at (4,1.8) {$v$};

\draw[->, thick] (w) -- (u);
\draw[->, thick] (u) -- (v);
\draw[->, thick] (w) .. controls (2,2.7) .. (v);

\node (curvepoint) at (2,2.7) {};

\node[fitbox, fit=(w)(u)(v)(curvepoint)] (PremBox) {};
\node[right=6pt of PremBox.east] {$\mathcal{G}'$};

\coordinate (pSW) at ($(PremBox.south west)+(0,-0.3)$);
\coordinate (pSE) at ($(PremBox.south east)+(0,-0.3)$);

\draw[thick] (pSW) -- (pSE);
\node[right=3pt] at (pSE) {$(tra)$};


\path let \p1 = (pSW), \p2 = (pSE)
  in coordinate (trueMid) at ({(\x1+\x2)/2}, {(\y1+\y2)/2});

\node[world, active] (w2) at ($(trueMid)+(-2, -1.3)$) {$w$};
\node[world, active] (u2) at ($(trueMid)+( 0, -1.3)$) {$u$};
\node[world, active] (v2) at ($(trueMid)+( 2, -1.3)$) {$v$};

\draw[->, thick] (w2) -- (u2);
\draw[->, thick] (u2) -- (v2);

\node (toppad)    at ($(trueMid)+(0, -0.8)$) {};
\node (bottompad) at ($(trueMid)+(0, -1.8)$) {};

\node[fitbox, fit=(w2)(u2)(v2)(toppad)(bottompad)] (ConBox) {};
\node[right=6pt of ConBox.east] {$\mathcal{G}$};

\end{tikzpicture}
\end{minipage}

\caption{Graphical presentations of $(\supset_L)$ and $(tra)$.\label{fig:example-impR-tra}}
\end{figure}

\medskip
\noindent
\textbf{Horn Rules.}\label{def:horn-rule-example} In our setting, a \emph{Horn property} is any property that can be expressed as a universally closed, first-order formula of the form $\forall x_{1} \cdots x_{n} (x_{1} E_{1} x_{2} \land x_{2} E_{1} x_{3} \land \cdots \land x_{n-1} E_{n-1} x_{n} \rightarrow x_{1} E_{n} x_{n})$ with each $E_{i}$ a binary predicate. Such properties cover a variety of frame conditions encountered in the proof theory of modal, tense, intuitionistic, and related logics. 
A \emph{Horn rule} is an inference rule that encodes a Horn property and stipulates that if a certain sequence of edges exist in the conclusion of the rule, then a single type of edge must occur in the premise (cf.~\cite{Vig00,Lyo21a}). Such rules serve as types of \emph{structural rules}~\cite{CiaLyoRamTiu21} or \emph{relational rules}~\cite{Vig00} existing in the literature. The significance of Horn rules lies in the modularity they afford: rather than designing a calculus for each logic in a given class, one may take a base calculus and extend it with Horn rules to obtain calculi for other logics.

The $(ref)$ and $(tra)$ rules stand as examples of Horn rules, which take the form shown below in our notation. We have provided a graphical depiction of $(tra)$ in Figure~\ref{fig:example-impR-tra} to emphasize its functionality.
\begin{center}
\begin{tabular}{c c}
\AxiomC{$\rel,\, wEw \sar \Sigma$}
\RightLabel{$(ref)$}
\UnaryInfC{$\rel\sar \Sigma$}
\DisplayProof\hspace{5mm}

&

\AxiomC{$\rel,\, wEu,\, uEv,\, wEv \sar \Sigma$}
\RightLabel{$(tra)$}
\UnaryInfC{$\rel,\, wEu,\, uEv \sar \Sigma$}
\DisplayProof
\end{tabular}
\end{center}
The $(ref)$ rule encodes reflexivity, by adding a single `loop' (i.e. $wEw$) to the premise, whereas the $(tra)$ rule encodes transitivity, requiring a sequence of two edges (i.e. $wEu, uEv$), and connecting $w$ to $v$ via a single edge (i.e. $wEv$) in the premise. Both rules encode types of Horn properties, and we note that such rules can be specified without the use of constraints. As we discuss below, Horn rules can be `absorbed' into the constraints associated with initial and transmission rules, producing new inference rules.

\subsection{Calculus Transformation and Rule Trading}

Permutations arguments are at the heart of proof theory; e.g. Gentzen's celebrated cut-elimination theorem shows how the cut rule can be eliminated via permutations, yielding a proof exhibiting the sub-formula property~\cite{Gen35a,Gen35b}. Likewise, simulations between sets of inference rules are of critical importance as they can be used to establish the `relative strength' of proof systems and to establish the relative sizes of proofs. In \sect~\ref{sec:perm-sim}, we will define these notions, using them to confirm a broad set of general relationships between rule types within our framework, and assisting us in writing generic algorithms (with complexity bounds) that transform calculi and their associated proofs. 
 
 We now exemplify simulations and permutations in the context of $\gtr$. In particular, we look at how initial, transmission, and Horn rules relate to one another. This investigation will demonstrate the connection between structural constraints and Horn rules, justifying their presence in our framework. 
 
We begin by studying simulations between the initial rule $(id)$ and the Horn rules $(ref)$ and $(tra)$, and look at the cases where the explicit edge $wEu$ in $(id)$ is `active' in applications of $(ref)$ and $(tra)$. The first case yields a proof of the following form:
\begin{center}
\AxiomC{}
\RightLabel{$(id)$}
\UnaryInfC{$\rel,\, wEw \sar \Sigma,\, w : (X, p \seqa p, Y)\vspacer$}
\RightLabel{$(ref)$}
\UnaryInfC{$\rel \sar \Sigma,\, w : (X, p \seqa p, Y)$}
\DisplayProof
\end{center}
 while the second case yields a proof of the form:
\begin{center}
\AxiomC{}
\RightLabel{$(id)$}
\UnaryInfC{$\rel,\, wEu,\, uEv,\, wEv \sar \Sigma, w : (X, p \seqa Y), v : (X' \seqa p, Y')$} 
\RightLabel{$(tra)$}
\UnaryInfC{$\rel,\, wEu,\, uEv \sar \Sigma, w : (X, p \seqa Y), v : (X' \seqa p, Y')$}
\DisplayProof
\end{center}
 We observe that the conclusion in the $(ref)$ case is similar to an instance of $(id)$. However, whereas $(id)$ requires the existence of prefixed sequents $w : (X, p \seqa Y)$ and $u : (X' \seqa p, Y')$ connected by a single edge $wEu$, the conclusion of $(ref)$ \emph{identifies} these two sequents as $w : (X, p \seqa p, Y)$ and omits the occurrence of an edge. In the $(tra)$ case, the conclusion of $(tra)$ contains two prefixed sequents like $(id)$, but with these two prefixed sequents connected by a path of edges $wEu, uEv$. Taking this into account, we recognize that we could simulate such proofs with a stronger form of $(id)$ that absorbs the functionality of the $(ref)$ and $(tra)$ rules:
 \begin{center}
\AxiomC{}
\RightLabel{$(id)'$}
\UnaryInfC{$\rel \sar \Sigma,\, w : (X, p \seqa Y),\, u : (X' \seqa p, Y')$}
\DisplayProof
\end{center}
where $(id)'$ is subject to the side condition that a path $wEv_{1}, \ldots, v_{n-1}Eu$ of relational atoms of length $0$ (meaning $w = u$) or greater exists between $w$ and $u$ in $\rel$. We can formalize this requirement as a structural constraint of the form $\con = (\{w,u\}, \{(w,u)\}, L)$ with $L(w,u) = \{\empstr, E, EE, \ldots\}$, where $\empstr$ is the empty string (meaning $w = u$), $E$ is treated as a character, and each $EE \cdots E$ is a word. Moreover, we require the same sequent relation to be enforced on $(id)'$ just as it was with $(id)$. We can take the conclusion of $(ref)$ (in the proof above) to be an instance of $(id)'$ where $L(w,u) = \{\empstr\}$, the conclusion of a typical $(id)$ rule to be an instance of $(id)'$ where $L(w,u) = \{E\}$, and the conclusion of $(tra)$ (in the proof above) to be an instance of $(id)'$ where $L(w,u) = \{EE\}$.

One can indeed show that any labeled sequent derivable by $(id)$ followed by applications of $(ref)$ or $(tra)$ can be simulated by $(id)'$ and vice-versa~\cite{Lyo21,Lyo21thesis}. Furthermore, this example justifies the inclusion of constraints in our framework as it shows that constraints can be modified, generating stronger inference rules, and forging new derivations that simulate others, effectively yielding new types of calculi.

We also observe a similar behavior when applying $(ref)$ and $(tra)$ to the transmission rule $(\supset_{L})$. Let us consider applying the $(ref)$ rule after an instance of $(\supset_{L})$ such that the relational atom `active' in the latter is removed by $(ref)$. We then have a derivation of the following form:
\begin{center}
\AxiomC{$\rel,\, w E w \sar \Sigma,\, w : (X, \phi \supset \psi \seqa \phi, Y)$}
\AxiomC{$\rel,\, w E w \sar \Sigma,\, w : (X, \phi \supset \psi, \psi \seqa Y)$}
\RightLabel{$(\supset_{L})$}
\BinaryInfC{$\rel,\, w E w \sar \Sigma,\, w : (X, \phi \supset \psi \seqa Y)\vspacer$}
\RightLabel{$(ref)$}
\UnaryInfC{$\rel \sar \Sigma,\, w : (X, \phi \supset \psi \seqa Y)$}
\DisplayProof
\end{center}
 Whereas $(\supset_{L})$ acts on prefixed sequents at $w$ and $u$, separated by a single edge $wEu$, $(ref)$ requires the identification of these two prefixed sequents, similar to what happens in the $(id)$ case above. 
 An investigation of applying $(tra)$ to an instance of $(\supset_{L})$ would exhibit behavior as in the $(id)$ case as well, where the two prefixed sequents are connected via a chain of relational atoms greater than one. We could therefore modify the constraint imposed on $(\supset_{L})$, enforcing a new constraint $\inc{\con} = (\{w,u\},\{(w,u)\},L)$ such that $L(w,u) = \{\empstr, E, EE, \ldots\}$. This constraint can be imposed to define a new rule $(\supset_{L})'$, which operates like $(\supset_{L})$, but applies between sequents connected via a chain of relational atoms of length zero or greater. Using this modified rule, we find that the above derivation can be simulated by applications of $(ref)$ followed by an application of $(\supset_{L})'$, yielding a type of permutation, as shown below.
\begin{center}
\AxiomC{$\rel,\, wEw \sar \Sigma,\, w : (X, \phi \supset \psi \seqa \phi, Y)$}
\RightLabel{$(ref)$}
\UnaryInfC{$\rel\sar \Sigma,\, w : (X, \phi \supset \psi \seqa \phi, Y)$}
\AxiomC{$\rel,\, wEw \sar \Sigma,\, w : (X, \phi \supset \psi, \psi \seqa Y)$}
\RightLabel{$(ref)$}
\UnaryInfC{$\rel\sar \Sigma,\, w : (X, \phi \supset \psi, \psi \seqa Y)$}
\RightLabel{$(\supset_{L})'$}
\BinaryInfC{$\rel \sar \Sigma, w : (X, \phi \supset \psi \seqa Y)$}
\DisplayProof
\end{center}

If we replace $(id)$ and $(\supset_{L})$ by $(id)'$ and $(\supset_{L})'$ in $\gtr$, we find that $(ref)$ and $(tra)$ can be permuted upward in any given proof and ultimately eliminated~\cite{Pim18,Lyo21}. Rules such as $(id)'$ and $(\supset_{L})'$ have been referred to as \emph{reachability rules}~\cite{Lyo21thesis} or \emph{propagation rules}~\cite{GorPosTiu11}, and form a crucial component of our framework. Such rules witness the importance of structural constraints, and as we will show in \sect~\ref{sec:perm-sim}, the interplay between constraints, reachability rules, and Horn rules uncover a number of permutation and simulation relationships between classes of inference rule types. Ultimately, in \sect~\ref{sec:generic-algs}, such rules will play a vital role, helping us identify spaces of polynomially equivalent calculi.

Finally, we comment on the relationship between Horn rules and \emph{semi‑Thue systems}~\cite{Pos47}. A semi‑Thue system is a finite set of string‑rewriting rules (or \emph{production rules}) of the form
$\stra \pto \strb$, where $\stra$ and $\strb$ are strings over a fixed alphabet. (NB. A formal definition is provided in the next section.) In the structural constraints introduced above, edge labels are languages---for example, $L(w,u)= \{\empstr,E,EE, \ldots\}$. Such a language can be generated by a semi‑Thue system of the form $\{E \pto\empstr, E \pto EE\}$, and this correspondence is not accidental.

Horn rules can be encoded as particular kinds of semi‑Thue systems, and conversely, certain semi‑Thue systems naturally give rise to Horn rules. For instance, the production $E\pto\empstr$ captures reflexivity: it states that an $E$‑edge may be obtained from the empty word, mirroring the bottom‑up reading of the $(ref)$ rule, where a loop at a label may always be added. Likewise, the production $E \pto EE$ corresponds to transitivity: it asserts that an $E$‑edge from $w$ to $v$ is justified whenever a path of two consecutive $E$‑edges exists, matching the behavior of the $(tra)$ rule.

This connection motivates the introduction of semi‑Thue systems in the next section and provides insight into the connection between constraints and Horn rules.

\section{Abstract Sequent Calculi}\label{sec:abstract-calc}

This section introduces the formal framework that underpins all subsequent results. The material is organized into four subsections, each building systematically on the previous one. Section~\ref{subsec:gen-seq} begins by defining \emph{generalized sequents (g‑sequents)}, the basic syntactic objects of our approach. These are edge‑labeled graphs whose vertices carry sequents, understood here simply as atomic labels, since our focus lies on how inference rules manipulate graphical structure, rather than on the internal form of sequents. The use of such objects is motivated by more expressive sequent systems that operate over graphs of sequents (e.g. labeled calculi~\cite{Sim94,Vig00} and nested calculi~\cite{Bul92,Kas94}).

In Section~\ref{subsec:E-sys}, we introduce a restricted class of semi-Thue systems~\cite{Pos47} , called \emph{$\etypset$-systems}, which rewrite individual edge types into strings of edge types. These systems generate the languages needed to express path-based conditions on g-sequents, and are motivated by their natural correspondence with Horn properties. Section~\ref{subsec:constraints} then defines the two kinds of constraints that govern the applicability of inference rules: \emph{structural constraints}, which specify the edge structure that must be present in g-sequents, and \emph{sequent constraints}, which specify how the sequents at designated vertices are related. 

Finally, in Section~\ref{subsec:rules-abstract-sys}, we assemble these components to define several classes of inference rules, i.e. \emph{inference rule types}---initial, local, expansion, reachability, and Horn rules---each parameterized by structural and sequent constraints. These rule classes are then collected into \emph{abstract (sequent) calculi}, and we introduce the accompanying notions of derivation, proof, and polynomial equivalence that will be used throughout the remainder of the paper.


\subsection{Generalized Sequents}\label{subsec:gen-seq}

We let $\seqset = \{\seq_{1}, \seq_{2}, \seq_{3}, \ldots\}$ be a countably infinite set of \emph{sequents}, which are denoted by $\seq$ and annotated versions thereof. As sequents are taken to be atomic entities in our framework, we do not describe their internal structure. We let $\univ = \{w, u, v, \ldots\}$ be the \emph{universe}, whose entities are denoted by $w$, $u$, $v$, $\ldots$ (potentially annotated), and which serve as vertices in the various graphs we define. Below, we define \emph{g-sequents} relative to a non-empty, finite set $\etypset = \{\etypa, \etypb, \etypc, \ldots\}$ of \emph{edge types}, which are used to index the edges of a g-sequent.

\begin{definition}[Generalized Sequent]\label{def:g-sequent} A \emph{generalized sequent (g-sequent)} is defined to be a tuple $\gseq = (\vset, \edgs, \lfunc)$ such that
\begin{itemize}

\item $\vset \subseteq\, \univ$ is a (potentially empty) finite set of vertices;

\item $\edgs = \{\Ea \ | \ a \in \etypset\}$ with $\Ea \subseteq \vset \times \vset$ for each $a \in \etypset$;

\item $\lfunc : \vset \to \seqset$.

\end{itemize}
 We use $\gseq$ (possibly annotated) to denote g-sequents, and let $\gspace{\etypset}$ be the set of all g-sequents defined relative to a set $\etypset$ of edge types. For a g-sequent $\gseq = (\vset, \edgs, \lfunc)$, we let $\univ(\gseq) := \vset$.
\end{definition}

As proof systems are concerned with the manipulation of syntactic entities via inference rules, we employ a more standard `sequent-style' notation for g-sequents in our technical work. In particular, we use the equivalent notation $\ant \sar \suc$ to denote a g-sequent $\gseq = (\vset, \edgs, \lfunc)$, where the \emph{antecedent} $\ant$ is a set of \emph{edge atoms} of the form $w \Ea u$ and the \emph{succedent} $\suc$ is a set of \emph{prefixed sequents} of the form $w : \seq$ such that (1) for each $a \in \etypset$, $w \Ea u \in \ant$ \iffi $(w,u) \in \Ea$, and (2) $w : \seq \in \suc$ \iffi $\lfunc(w) = \seq$. We define the \emph{size} of a g-sequent $\gseq = \ant \sar \suc = (\vset,\edgs,\lfunc)$ to be $\seqsize{\gseq} = |\ant| + |\suc| = |\bigcup \edgs| + |\vset|$. Also, we let $\pseqset = \univ \times \seq$ denote the set of prefixed sequents and let $\univ(\ant)$ and $\univ(\suc)$ denote the set of all vertices occurring in $\ant$ and $\suc$, respectively.
  
To improve intuition concerning g-sequents and their representations, we provide examples in \fig~\ref{fig:g-seq}.  We also specify a special subclass of g-sequents (whose importance will be discussed in Sections~\ref{sec:generic-algs} and~\ref{sec:conclusion}) referred to as \emph{polytree g-sequents}. A polytree g-sequent is a g-sequent $\gseq = (\vset, \edgs, \lfunc)$ such that $(\vset, \edgs)$ is a \emph{polytree}, i.e. a graph that is (1) connected and (2) free of (un)directed cycles. Observe that the g-sequent shown right in \fig~\ref{fig:g-seq} is a polytree g-sequent.

\begin{figure}[t]
\begin{center}
\begin{tikzpicture}[node distance=1.6cm,
    world/.style={circle, draw, minimum size=6mm, fill=white},
    active/.style={very thick},
    >=stealth
]

\node[world, active] (w0) [label=above:$\seq_{1}$] {$w_{1}$};
\node[world, active] (w1) [below left=of w0, label=above:$\seq_{2}$] {$w_{2}$};
\node[world, active] (w2) [below right=of w0, label=above:$\seq_{3}$] {$w_{3}$};
\node[world, active] (w4) [below=of w1, label=below:$\seq_{4}$] {$w_{4}$};
\node[world, active] (w5) [right=of w4, label=below:$\seq_{5}$] {$w_{5}$};

\draw[->, thick] (w0) -- (w1) node [midway, above] {$c$};
\draw[->, thick] (w0) -- (w2) node [midway, above] {$b$};
\draw[->, thick] (w4) -- (w1) node [midway, left] {$b$};
\draw[->, thick] (w5) -- (w1) node [midway, right] {$a$};

\node[world, active] (w6) [left=of w0, xshift=-3cm, label=above:$\seq_{1}$] {$u_{1}$};
\node[world, active] (w7) [below left=of w6, label=below:$\seq_{2}$] {$u_{2}$};
\node[world, active] (w8) [below right=of w6, label=below:$\seq_{3}$] {$u_{3}$};
\node[world, active] (w9) [below right=of w7, label=below:$\seq_{4}$] {$u_{4}$};

\draw[->, thick] (w6) -- (w7) node [midway, above] {$b$};
\draw[->, thick] (w6) -- (w8) node [midway, above] {$a$};
\draw[->, thick] (w7) -- (w9) node [midway, below] {$a$};
\draw[->, thick] (w8) -- (w9) node [midway, below] {$c$};
\draw[->, thick] (w7) -- (w8) node [midway, above] {$a$};
\draw[->, thick] (w9) to [out=225, in=315, looseness=6] node[below] {$c$} (w9);

\node[] [left=of w6, xshift=-3cm] (x0) {$w$};
\node[] [below=of x0, yshift=-0.25cm] (x1) {$u$};
\node[] [below=of x1, yshift=-0.25cm] (x2) {$v$};

\draw[->, thick] (x0) -- (x1) node [midway, left] {$\gram(b)$};
\draw[->, thick] (x2) -- (x1) node [midway, left] {$\gram'\!(c)$};

\draw[->, dotted, decorate, decoration={snake, segment length=2.25cm, amplitude=0.75mm}] (x0) -- (w6) node [midway, above] {$\conmap$};
\draw[->, dotted, decorate, decoration={snake, segment length=2.25cm, amplitude=0.75mm}] (x1) -- (w9) node [midway, above] {$\conmap$};
\draw[->, dotted, decorate, decoration={snake, segment length=2.25cm, amplitude=0.75mm}] (x2) -- (w9) node [midway, above] {$\conmap$};

\end{tikzpicture}
\end{center}

\caption{We give two examples: $\gseq = (\vset, \{\Ea,\Eb,\Ec\}, \lfunc)$ is shown middle with $\vset = \{u_{1}, u_{2}, u_{3}, u_{4}\}$, $\Eb = \{(u_{1},u_{2})\}$, $\Ea = \{(u_{1},u_{3}),(u_{2},u_{3}),(u_{2},u_{4})\}$, $\Ec = \{(u_{3},u_{4}),(u_{4},u_{4})\}$, and for $i \in [4]$, $\lfunc(u_{i}) = \seq_{i}$. $\ant \sar \suc$ is shown right with $\ant = w_{1} \Ec w_{2}, w_{1} \Eb w_{3}, w_{4} \Eb w_{2}, w_{5} \Ea w_{2}$ and $\suc = w_{1} : \seq_{1}, w_{2} : \seq_{2}, w_{3} : \seq_{3}, w_{4} : \seq_{4}, w_{5} : \seq_{5}$. A graphical representation of the structural constraint $\con$ and constraint map $\conmap$ from Example~\ref{eg:struc-con-con-map} is shown above left; one can see that $\gseq$ satisfies $\con$ with $\conmap$.\label{fig:g-seq}}
\end{figure}
 

\subsection{$\etypset$-Systems and Propagation}\label{subsec:E-sys}

To control the functionality of certain inference rules, we make use of a restricted version of semi-Thue systems~\cite{Pos47} that rewrite single edge types into strings of edge types. Our use of semi-Thue systems is motivated by the fact that such systems are well-suited for expressing and encoding Horn properties. Recall that a Horn property states that a path of edges necessitates the existence of an edge between the initial/terminal and terminal/initial point of the path (see p.~\pageref{def:horn-rule-example}). By taking strings of edge types to represent paths and single edge types to represent single edges, we can encode such properties as semi-Thue systems. We will make the correspondence between semi-Thue systems and Horn rules (which encode Horn properties) explicit in the sequel.

Given a set $\abet$ of characters, we define the set $\words$ of \emph{strings} over $\abet$ to be the set of finite sequences of elements of $\abet$ including the \emph{empty string} $\empstr$. We denote strings with (possibly annotated) letters $\stra$, $\strb$, $\strc$. A \emph{production rule} is defined to be an object of the form $\stra \pto \stra'$ such that $\stra,\stra' \in \words$. We often use $\pru$ and annotated versions thereof to denote production rules. A \emph{semi-Thue system} is defined to be a (potentially empty) finite set $\gram$ of production rules. Semi-Thue systems permit us to derive strings via repeated applications of production rules. Given a semi-Thue system $\gram$ over $\abet$, and a pair of strings $\strb, \strb' \in \words$ we write $\strb \ptostep \strb'$ \iffi there exists a rule $\stra \pto \stra' \in \gram$ such that $\stra$ is a sub-string of $\strb$, and $\strb'$ can be obtained from $\strb$ by replacing some occurrence of $\stra$ in $\strb$ by $\stra'$. A \emph{$\gram$-derivation} of a string $\strb \in \words$ from a string $\stra \in \words$, denoted $\stra \ptoder \strb$, is defined accordingly: (1) $\stra \ptoder \stra$, (2) if $\stra \ptoder \stra_{1} \strc \stra_{2}$ and $\strc \ptostep \strb$, then $\stra \ptoder \stra_{1} \strb \stra_{2}$. We define the \emph{length} of a $\gram$-derivation of a string $\strb \in \words$ from a string $\stra \in \words$ to be the minimal number of rule applications used to derive $\strb$ from $\stra$. The \emph{language} of a string $\stra \in \words$ relative to a semi-Thue system $\gram$ is defined as: $\gram(\stra) = \{\strb  \mid  \stra \ptoder \strb\}$. 


Let $\cetypset$ be the following set $\set{\conv{\etypa} \mid \etypa \in \etypset}$. For a production rule of the form $\pru = x \pto y_1 \cdots y_{n}$ with $x,y_{1},\ldots,y_{n} \in \etypset \cup \cetypset$, we define $\conv{\pru} = \conv{x} \pto \conv{y}_n\cdots\conv{y}_1$, where $\conv{\conv{z}} = z$ for $z \in \etypset \cup \cetypset$. We define an \emph{$\etypset$-system} to be a semi-Thue system $\gram$ over $\etypset \cup \cetypset$ satisfying: (1) for every rule $\stra \pto \strb \in \gram$ we have $|\stra| = 1$, and (2) $\stra \pto \strb \in \gram$ \iffi $\conv{\stra} \pto \conv{\strb} \in \gram$. A \emph{production pair} from $\gram$ is defined to be a pair $(\pru, \conv{\pru})$ such that $\pru, \conv{\pru} \in \gram$. We define $P(\gram)$ to be the set of all production pairs in $\gram$. For a set $P$ of production pairs, we let $\gram(P)$ be the set of all production rules found in a production pair of $P$.


Given a g-sequent $\gseq = (\vset, \edgs, \lfunc)$, two vertices $u,w \in \vset$, and an element $\etypa \in \etypset$ we write $\gseq \models u \ppath{\etypa} w$ \iffi $(u,w) \in \eset_{\etypa}$, and $\gseq \models u \ppath{\conv{\etypa}} w$ \iffi $(w,u) \in \eset_{\etypa}$. Moreover, given a string $x\stra \in \eslang$ where $x \in \etypset \cup \cetypset$, we inductively define $\gseq \models u \ppath{x\stra} w$ as `$\exists_{v \in \vset}\ \gseq \models u\ppath{x} v$ and $\gseq \models v \ppath{\stra}w$', and $\gseq \models u \ppath{\conv{(\stra x)}} w$ as `$\exists_{v \in \vset}\ \gseq \models u\ppath{\conv{x}}v$ and $\gseq \models v \ppath{\conv{\stra}}w$'. Additionally, when $\gseq$ is clear from the context we may simply write $u \ppath{\stra} w$ to express $\gseq \models u \ppath{\stra} w$. Finally, given a language $\lang$ (of some $\etypset$-system) we use $u \ppath{\lang} w$ \iffi  there is a string $\stra \in \lang$ such that $u \ppath{\stra} w$.

\begin{example} Let $\gram = \set{\etypa \pto \etypb \etypa , \conv{\etypa} \pto \conv{\etypa} \conv{\etypb} , \etypa \pto \etypa \etypc , \conv{\etypa} \pto \conv{\etypc} \conv{\etypa}}$ and $\gseq$ be the g-sequent shown middle in Figure~\ref{fig:g-seq}. One can verify that $\gseq \models u_{1} \ppath{\gram(\etypa)} u_{4}$ because $\gseq \models u_{1} \ppath{\etypb \etypa \etypc} u_{4}$. It is evident that $\gseq \not\models u_{1} \ppath{\gram(\etypb)} u_{4}$ because $\gram(\etypb) = \set{\etypb}$.
\end{example}

\subsection{Constraints}\label{subsec:constraints}


Having introduced g‑sequents and the $\etypset$‑systems that generate the languages used to describe paths within them, we now turn to the notion of constraints, which govern when an inference rule may be applied to a given g‑sequent. The first kind of constraint we introduce is a structural constraint, which specifies the edge configurations and path conditions that must be present in any g‑sequent serving as a premise (or conclusion) of a rule application.

 
\begin{definition}[Structural Constraint]\label{def:constraint} Let $\etypset$ be a set of edge types. We define a \emph{structural constraint} $\con$ to be a finite labeled polytree $(V, E, L)$ such that $V \subseteq \univ$, $E \subseteq V \times V$, and if $(w,u) \in E$, then $L(w,u) = \langgram(\etypa)$ for $\etypa \in \etypset$ and $\gram$ an $\etypset$-system. We define a \emph{constraint family} to be a finite sequence $\confam = (\con_{1}, \ldots, \con_{n})$ of constraints, and we say that an $\etypset$-system $\gram$ \emph{participates} in a constraint $\con = (V,E,L)$ \iffi there exists an edge $(w,u) \in E$ and $\etypa \in \etypset$ such that $L(w,u) = \langgram(\etypa)$. Likewise, we say that an $\etypset$-system $\gram$ \emph{participates} in a constraint family $\confam$ \iffi there exists a constraint $\con$ in $\confam$ such that $\gram$ participates in $\con$. We let $\gram(\con) = \gram_{1} \cup \cdots \cup \gram_{n}$ such that $\gram_{1}, \ldots, \gram_{n}$ are all $\etypset$-systems participating in $\con$, and define the \emph{size} of a constraint $\con$ as: $|\con| = |\gram(\con)|$. 
\end{definition}

As shown in the definition above, we represent structural constraints using finite labeled polytrees of the form $(V, E, L)$. We use labeled polytrees as opposed to other kinds of graphs in structural constraints as they minimally generalize the various kinds of structural constraints that are normally imposed on inference rules in the literature; e.g. the $Prop(\mathbf{P})$ rules in Gor{\'{e}} et al.~\cite{GorPosTiu11} or the $(id^{n}_{q})$ rule in~\cite{Lyo21}. A structural constraint determines which paths must be present in the g-sequents of a rule application by being `homomorphically mappable' into each g-sequent. 
These mappings take place by means of \emph{constraint maps} $\conmap \colon V \to \vset$ from the structural constraint $\con = (V, E, L)$ to each g-sequent $\gseq = (\vset, \edgs, \lfunc)$ such that for all $w,u \in V$, if $L(w,u) = \gram(a)$, then there exists a string $\stra \in \gram(a)$ such that $\gseq \models \conmap(w) \ppath{\stra} \conmap(u)$ (see \dfn~\ref{def:constraint-sat} below). To make these notions more concrete, we have provided an example of a structural constraint and constraint map in Example~\ref{eg:struc-con-con-map}.

\begin{definition}[Constraint Satisfaction]\label{def:constraint-sat} Let $\con = (V,E,L)$ be a structural constraint and $\gseq = (\vset, \edgs, \lfunc)$ be a g-sequent. We define a \emph{constraint map} to be a function $\conmap \colon V \to \vset$. We say that $\gseq$ \emph{satisfies} $\con$ with constraint map $\conmap$ \iffi for all $w,u \in V$, if $L(w,u) = \gram(\etypa)$, then $\gseq \models  \conmap(w) \ppath{\gram(\etypa)} \conmap(u)$. We say that $\gseq$ \emph{satisfies} $\con$ \iffi there exists a constraint map $\conmap$ such that $\gseq$ \emph{satisfies} $\con$ with $\conmap$.
\end{definition}

\begin{example}\label{eg:struc-con-con-map} The structural constraint $\con = (\{w,u,v\},\{(w,u),(v,u)\},L)$ such that $L(w,u) = \gram(b)$, $\gram = \{b \pto ba, \conv{b} \pto \conv{a}\conv{b}\}$, $L(v,u) = \gram'\!(c)$, and $\gram' = \emptyset$ can be visualized as the labeled polytree shown left in \fig~\ref{fig:g-seq}. We let $\conmap \colon V \to \vset$ be a constraint map from $\con$ to $\gseq = (\vset, \{\Ea,\Eb,\Ec\}, \lfunc)$ shown middle in \fig~\ref{fig:g-seq} such that $\conmap(w) = u_{1}$ and $\conmap(u) = \conmap(v) = u_{4}$. Observe that $\gseq$ satisfies $\con$ with $\conmap$ since a path corresponding to the string $ba \in \gram(b)$ exists between $u_{1}$ and $u_{4}$ and a path corresponding to the string $c \in \gram'\!(c) = \{c\}$ exists between $u_{4}$ and $u_{4}$.
\end{example}

\begin{definition}[Sequent Constraint]\label{def:correctness-rel} We define a \emph{sequent constraint} $\seqrel$ to be an $(n+1)$-ary relation such that:
$$
\seqrel \subseteq \underbrace{\seqset \times \cdots \times \seqset}_{n} \times \ 2^{\pseqset}.
$$ 
We say that $\seq_{1}, \ldots, \seq_{n} \in \seqset$ and $\suc \subseteq \pseqset$ \emph{satisfy} $\seqrel$ \iffi $(\seq_{1}, \ldots, \seq_{n},\suc) \in \seqrel$. 
\end{definition}

As certain inference rules in the literature are \emph{context dependent}, e.g. the $L\exists$ rule of Fitting~\cite{Fit14}, sequent constraints must take the entire succedent $\suc$ of a g-sequent into account in inference rule applications. This explains the presence of $\suc$ in sequent constraints. Note that we will hitherto refer to structural constraints as \emph{constraints} more simply, while referring to sequent constraints as \emph{sequent constraints}.

\subsection{Rules and Abstract Systems}\label{subsec:rules-abstract-sys}

We now specify certain classes of inference rules, which will be collected together into finite sets to define our abstract calculi later on. 
For inference rules with multiple premises, we use $i \in [n]$ to mean $1 \leq i \leq n$.

\smallskip
\noindent
\textbf{Initial Rule.} We define an \emph{initial rule} to be an operation of the following form:
\begin{center}
\AxiomC{ } 
\RightLabel{$\id$}
\UnaryInfC{$\ant \sar \suc$}
\DisplayProof
\end{center}
with $\con := (V,E,L)$, $V = \{w_{1}, \ldots, w_{n}\}$, and there exists a constraint map $\conmap$ such that 
\begin{itemize}

\item[(1)] the g-sequent $\ant \sar \suc := (\vset,\edgs,\lfunc)$ satisfies the constraint $\con$ with $\conmap$, and 

\item[(2)] $\lfunc(\conmap(w_{1})), \ldots, \lfunc(\conmap(w_{n}))$, and $\suc' = (\suc \setminus \{\conmap(w_{i}) : \lfunc(\conmap(w_{i})) \ | \  i \in [n] \})$ satisfy $\seqrel$.
\end{itemize}
Examples of initial rules include $\mathsf{init}_{2}$ in~\cite{KuzLel18} and $(\bot L)$ in~\cite{Sim94}.

 
\smallskip
\noindent
\textbf{Local Rule.} We define a \emph{local rule} to be an operation of the following form:
\begin{center}
\AxiomC{$\{\,\ant \sar \suc,\, w : \seq_{i}\,\}_{i \in [n]}$}
\RightLabel{$\lru$}
\UnaryInfC{$\ant \sar \suc,\, w : \seq_{n{+}1}$}
\DisplayProof
\end{center}
such that $\con := (\{w'\},\emptyset,\emptyset)$, $(\ant \sar \suc,\, w : \seq_{i}) := (\vset_{i},\edgs_{i},\lfunc_{i})$ for each $i \in [n{+}1]$, and constraint maps $\conmap_{i} \colon \{w'\} \to \vset_{i}$ exist such that
\begin{itemize}

\item[(1)] $\conmap_{i}(w') = \conmap_{j}(w') = w$ for $i \neq j \in [n{+}1]$, and 

\item[(2)] $\lfunc_{1}(\conmap_{1}(w')) = \seq_{1}, \ldots, \lfunc_{n{+}1}(\conmap_{n{+}1}(w')) = \seq_{n{+}1}$, and $\suc$ satisfy $\seqrel$.

\end{itemize}
Examples of local rules include $(\neg {\rightarrow})$ in~\cite{Bul92} and CUT~in~\cite{Gir87}.

 
\smallskip
\noindent
\textbf{Expansion Rule.}\label{def:expansion-rule} We define an \emph{expansion rule} to be an operation of the following form:
\begin{center}
\AxiomC{$\ant, \antii \sar \suc, w : \seq_{1}, u : \seq_{2}$}
\RightLabel{$\eru$}
\UnaryInfC{$\ant \sar \suc,\, w : \seq$}
\DisplayProof
\end{center}
such that $\confam := (\con_{1},\con_{2})$ is a constraint family, $\con_{1} := (\{w',u'\},\emptyset,\emptyset)$, and $\con_{2} := (\{w'\},\emptyset,\emptyset)$. Also, $(\ant, \antii \sar \suc, w : \seq_{1}, u : \seq_{2}) := (\vset_{1},\edgs_{1},\lfunc_{1})$ and $(\ant \sar \suc,\, w : \seq) := (\vset_{2},\edgs_{2},\lfunc_{2})$. For such a rule to be applied, there must exist constraint maps $\conmap_{1} \colon \{w',u'\} \to \vset_{1}$ and $\conmap_{2} \colon \{w'\} \to \vset_{2}$ such that 
\begin{itemize}

\item[(1)] $\conmap_{1}(w') = \conmap_{2}(w') = w$ and $\conmap_{1}(u') = u$,

\item[(2)] $\lfunc_{1}(\conmap_{1}(w')) = \seq_{1}, \lfunc_{1}(\conmap_{1}(u')) = \seq_{2}, \lfunc_{2}(\conmap_{2}(w')) = \seq$, and $\suc$ satisfy $\seqrel$, and

\item[(3)] $\univ(\ant \sar \suc,  w : \seq) \cap \univ(\antii) = \{w\}$ with $\antii \in \{w \Ea u, u \Ea w \ | \ \etypa \in \etypset\}$.

\end{itemize}
Examples of such rules are $\Box R$ in~\cite{Vig00} and $[a]$ in \cite{TiuIanGor12}.



\smallskip
\noindent
\textbf{Forward Horn Rule.} If $\stra = \etypa_{1} \cdots \etypa_{n} \in (\etypset \cup \conv{\etypset})^{*}$, then we define $w \Es u = w \E_{\etypa_{1}} v_{1}, \ldots, v_{n-1} \E_{\etypa_{n}} u$, where $w \edgs_{\conv{a}} u := u \Ea w$ and $w \E_{\empstr} u = (w = u)$. We define a \emph{forward Horn rule} to be an operation of the form shown below left, which takes the form shown below right when $\stra = \empstr$.
\begin{center}
\begin{tabular}{c c}
\AxiomC{$\ant,\, w \Es u,\, w \Ea u \sar \suc$}
\RightLabel{$\fhru$}
\UnaryInfC{$\ant,\, w \Es u \sar \suc$}
\DisplayProof\hspace{5mm}

&

\AxiomC{$\ant,\, w \Ea w \sar \suc$}
\RightLabel{$\fhru$}
\UnaryInfC{$\ant \sar \suc$}
\DisplayProof
\end{tabular}
\end{center}
For a production rule $\pru = \etypa \pto \stra$, we define the singleton set $\hrus(\pru,\conv{\pru})$ to be the set containing the forward Horn rule above left, which takes the form above right when $\stra = \empstr$.

\smallskip
\noindent
\textbf{Backward Horn Rule.} We define a \emph{backward Horn rule} to be an operation of the form shown below left, which takes the form shown below right when $\stra = \empstr$.
\begin{center}
\begin{tabular}{c c}
\AxiomC{$\ant,\, w \Es u,\, u \Ea w \sar \suc$}
\RightLabel{$\bhru$}
\UnaryInfC{$\ant,\, w \Es u \sar \suc$}
\DisplayProof\hspace{5mm}

&

\AxiomC{$\ant,\, w \Ea w \sar \suc$}
\RightLabel{$\bhru$}
\UnaryInfC{$\ant \sar \suc$}
\DisplayProof
\end{tabular}
\end{center}
For a production rule $\pru = \conv{\etypa} \pto \stra$, we define the singleton set $\hrus(\pru,\conv{\pru})$ to be the set containing the backward Horn rule shown above left, which takes the form shown above right when $\stra = \empstr$. We define a \emph{Horn rule} to be either a forward or backward Horn rule, and for a set $P$ of production pairs, we let $\hrus(P) = \bigcup_{(\pru,\conv{\pru}) \in P} \hrus(\pru,\conv{\pru})$. Examples of Horn rules include $\chi_{B}$ in \cite{Sim94} and $(\mathrm{Path})$ in~\cite{CiaLyoRamTiu21}. We remark that Horn rules encode (universally closed) relational properties of the form $w \Es u \rightarrow w \edgs_{x} u$ with $x \in \etypset \cup \conv{\etypset}$, covering standard frame conditions, e.g. for tense logics~\cite{GorPosTiu11} and first-order intuitionistic logics~\cite{FitMen98}. 

\smallskip
\noindent
\textbf{Reachability and Transmission Rules.} We define a \emph{reachability rule} to be an operation:
\begin{center}
\AxiomC{$\{\,\ant \sar \suc,\, w : \seq_{i},\, u : \seq_{i}'\,\}_{i \in [n]}$}
\RightLabel{$\rru$}
\UnaryInfC{$\ant \sar \suc,\, w : \seq_{n{+}1},\, u : \seq_{n{+}1}'$}
\DisplayProof
\end{center}
with $\con = (\{w',u'\},E,L)$, $(\ant \sar \suc,\, w : \seq_{i}, u : \seq_{i}') := (\vset_{i},\edgs_{i},\lfunc_{i})$ for $i \in [n{+}1]$, and where constraint maps $\conmap_{i} \colon \{w',u'\} \to \vset_{i}$ exist such that 
\begin{itemize}

\item[(1)] $\conmap_{i}(w') = \conmap_{j}(w') = w$ and $\conmap_{i}(u') = \conmap_{j}(u') = u$ for $i \neq j \in [n{+}1]$,

\item[(2)] $\ant \sar \suc,\, w : \seq_{i}, u : \seq_{i}'$ satisfies $\con$ with $\conmap_{i}$, and

\item[(3)] $\lfunc_{1}(\conmap_{1}(w')) = \seq_{1}, \lfunc_{1}(\conmap_{1}(u'))  = \seq_{1}', \ldots, \lfunc_{n{+}1}(\conmap_{n{+}1}(w')) = \seq_{n{+}1}, \lfunc_{n{+}1}(\conmap_{n{+}1}(u'))  = \seq_{n{+}1}'$, and $\suc$ satisfy $\seqrel$.

\end{itemize}
Examples of reachability rules include $\mathrm{Prop}(\mathbf{P})$ in \cite{GorPosTiu11} and $(\forall_{l}^{n})$ in \cite{Lyo21}.

 
We define a \emph{transmission rule} $\tru$ (as discussed in the previous section) to be a special instance of a reachability rule where the constraint $\con = (\{w',u'\},\{(w',u')\},L)$ with $L(w',u') = \set{a}$ for some $\etypa \in \etypset$. Examples of transmission rules include $\lozenge^{\circ}$ in \cite{Str13} and $\mathsf{Lift}$ in~\cite{KuzLel18}.
 
\smallskip
We refer to any inference rule of the above form as either an \emph{inference rule} or \emph{rule}, more generally, and use $\rui$, $\ruii$, $\ruiii$, $\ldots$ (potentially annotated) to denote them. For those inference rules parameterized by a constraint $\con$ or constraint family $\confam$, we say that an $\etypset$-system \emph{participates in the rule} \iffi the $\etypset$-system participates in the constraint $\con$ or constraint family $\confam$. Let us now define the notion of an \emph{abstract calculus}.

\begin{definition}[Abstract Calculus] Let $\etypset$ be a set of edge types. We define an \emph{abstract (sequent) calculus} (over $\etypset$) to be an ordered pair $\calc = (\gclass, \ops)$ such that (1) $\gclass \subseteq \gspace{\etypset}$ is a set of g-sequents closed under applications of the rules in $\ops$ and (2) $\ops$ is a finite collection of inference rules. We use $\calc$, $\calcii$, $\calciii$, $\ldots$ (occasionally annotated) to denote abstract calculi and define $\calcspace{\etypset}$ to be the collection of all abstract calculi over $\etypset$. Furthermore, for an abstract calculus $\calc = (\gclass_{1}, \ops_{1})$ and $\calcii = (\gclass_{2}, \ops_{2})$, we say that $\calcii$ is an \emph{extension} of $\calc$, and write $\calc \subseteq \calcii$, \iffi $\etypset_{1} \subseteq \etypset_{2}$, $\gclass_{1} \subseteq \gclass_{2}$, and $\ops_{1} \subseteq \ops_{2}$.
\end{definition}

\begin{remark} In this paper, we confine our study to abstract calculi of the form $\calc = (\gspace{\etypset}, \ops)$, i.e. where the set of all g-sequents $\gspace{\etypset}$ defined relative to $\etypset$ is used by the calculus.
\end{remark}

Given a set $\rus$ of rules, we define a \emph{derivation} $\deriv$ to be a finite tree of g-sequents from $\gspace{\etypset}$ such that every parent node is the conclusion of an application of a rule from $\rus$ with all children nodes the corresponding premises. If a g-sequent $\gseq$ occurs in a derivation $\deriv$, then we write $\gseq \in \deriv$ to indicate this. The \emph{quantity} of a derivation $\deriv$ is defined as $\qsize{\deriv} = |\{\gseq \in \gspace{\etypset} \ | \ \gseq \in \deriv\}|$ and the \emph{size} of a derivation $\deriv$ is defined to be $\dsize{\deriv} = \max\{\seqsize{\gseq} \ | \ \gseq \in \deriv\} \times \qsize{\deriv}$. 
 
A \emph{proof} $\prf$ is defined to be a derivation beginning with applications of initial rules, and a \emph{complete proof} is any proof ending with a g-sequent of the form $\sar w : \seq$. Finally, a \emph{polytree proof} is defined to be a proof such that every g-sequent occurring in the proof is a polytree g-sequent.

\begin{example} We illustrate a proof in an abstract calculus. For readability, we present the example at a high level and omit the explicit specification of the structural and sequent constraints associated with each rule instance; giving these in full would require substantial additional space. A complete instantiation of our framework---including fully spelled‑out constraints and the construction of an entire multisequent calculus---appears later in Section~\ref{sec:example}.
\begin{center}
\AxiomC{$\phantom{\gseq}$}
\RightLabel{$i(\confam_{3},\seqrel_{3})$}
\UnaryInfC{$w \Ea u, u \Ea v , w \Ec v \sar w : \seq_{4}, u : \seq_{5}$}
\RightLabel{$h_{1}$}
\UnaryInfC{$w \Ea u, u \Ea v \sar w : \seq_{4}, u : \seq_{5}$}
\RightLabel{$e(\confam_{0},\seqrel_{0})$}
\UnaryInfC{$w \Ea u \sar w : \seq_{1}, u : \seq_{3}$}

\AxiomC{$\phantom{\gseq}$}
\RightLabel{$i(\confam_{3},\seqrel_{3})$}
\UnaryInfC{$w \Ea u, u \Eb u \sar w : \seq_{6}, u : \seq_{7}$}
\RightLabel{$r(\confam_{2},\seqrel_{2})$}
\UnaryInfC{$w \Ea u, u \Eb u \sar w : \seq_{1}, u : \seq_{3}$}
\RightLabel{$h_{2}$}
\UnaryInfC{$w \Ea u \sar w : \seq_{1}, u : \seq_{3}$}

\RightLabel{$l(\confam_{1},\seqrel_{1})$}
\BinaryInfC{$w \Ea u \sar w : \seq_{1}, u : \seq_{2}$}
\RightLabel{$e(\confam_{0},\seqrel_{0})$}
\UnaryInfC{$\sar w : \seq_{0}$}
\DisplayProof
\end{center}
\end{example}
 
Two abstract calculi $\calc,\calcii \in \calcspace{\etypset}$ are defined to be \emph{polynomially equivalent}, written $\calc \pdeq \calcii$, when a proof of a g-sequent $\gseq$ exists in $\calc$ \iffi a proof $\prf'$ of $\gseq$ exists in $\calcii$, and there exist $\mathrm{PTIME}$ functions $f$ and $g$ such that $f(\prf) = \prf'$ and $g(\prf') = \prf$.\label{def:poly-equiv} We also lift specific set-theoretic operations to abstract calculi: for a set $\rus$ of rules and an abstract calculus $\calc = (\gspace{\etypset}, \ops)$, we define $\calc \setminus \rus := (\gspace{\etypset}, \ops \setminus \rus)$ and $\calc \cup \rus := (\gspace{\etypset}, \ops \cup \rus)$. Last, we let $\hrus(\calc)$ denote the set of Horn rules in an abstract calculus $\calc$, and remark that $\hrus$ and annotated versions thereof will be exclusively used to denote sets of Horn rules throughout the remainder of the paper.

\begin{definition}\label{def:gram-rule-relation} Let $\rui \in \{\id, \rru\}$. We define the \emph{grammar} $\gram(\rui)$ of $\rui$ as: $\pru, \conv{\pru} \in \gram(\rui)$ \iffi there exists an $\etypset$-system $\gram$ that participates in $\rui$ such that $\pru, \conv{\pru} \in \gram$. For $\rui \in \{\lru, \eru\}$, we define $\gram(\rui) = \emptyset$. For a Horn rule $\fhru$ or $\bhru$, we define the grammar $\gram(\fhru) := \{\etypa \pto \stra, \conv{\etypa} \pto \conv{\stra}\}$ and $\gram(\bhru) := \{\conv{\etypa} \pto \stra, \etypa \pto \conv{\stra}\}$, respectively. Given a set of rules $\rus = \{\rui_{1}, \ldots, \rui_{n}\}$, we let $\gram(\rus) := \gram(\rui_{1}) \cup \cdots \cup \gram(\rui_{n})$. For an abstract calculus $\calc = (\gspace{\etypset}, \ops)$, $\gram(\calc) := \gram(\ops)$.

Similarly, for a set $\rus$ of rules, we define the set of production pairs of $\rus$ as $P(\rus) := P(\gram(\rus))$, and for an abstract calculus $\calc = (\gspace{\etypset}, \ops)$, we let $P(\calc) := P(\ops)$.
\end{definition}

\begin{example} Let $\rus = \set{h_{1},h_{2}}$ such that $h_{1}$ is the forward Horn rule shown below left and $h_{2}$ is the backward Horn rule shown below right.
\begin{center}
\begin{tabular}{c c c c}
\AxiomC{$\ant, w \Eb u, u \Ea v, w \Ea v \sar \suc$}
\RightLabel{$h_{1}$}
\UnaryInfC{$\ant, w \Eb u, u \Ea v \sar \suc$}
\DisplayProof

&

\AxiomC{$\ant, w \Eb u, u \Ec w \sar \suc$}
\RightLabel{$h_{2}$}
\UnaryInfC{$\ant, w \Eb u  \sar \suc$\vspacer}
\DisplayProof
\end{tabular}
\end{center}
By Definition~\ref{def:gram-rule-relation}, we can compute the grammar of $\rus$ accordingly:
$$
\gram(\rus) = \gram(h_{1}) \cup \gram(h_{2}) 
= \set{ \etypa \pto \etypb \etypa , \conv{\etypa} \pto \conv{\etypa} \conv{\etypb} , \conv{\etypc} \pto \etypb , \etypc \pto \conv{\etypb} }.
$$
Therefore, the set of production pairs is $P(\rus) = \set{ (\etypa \pto \etypb \etypa , \conv{\etypa} \pto \conv{\etypa} \conv{\etypb}) , (\conv{\etypc} \pto \etypb , \etypc \pto \conv{\etypb}) }$. Observe that $\hrus(P(\rus)) = \set{h_{1},h_{2}}$, that is, $\hrus(\cdot)$ acts as a left inverse returning the original set of rules.
\end{example}

\section{Permutations and Simulations}\label{sec:perm-sim}

This section develops the theoretical basis for our main results and systematically investigates the interplay between different inference rule types---most notably, the relationship between Horn rules and constraints. These results lay the foundation for two central questions that are explored and answered in the sequel:

\begin{itemize}

\item How can Horn rules be embedded into constraints to yield new, yet provably equivalent, abstract calculi? 

\item How can the resulting theory of abstract calculi be applied to concrete logics and existing proof systems? 

\end{itemize}
These questions are answered in Sections~\ref{sec:generic-algs} and~\ref{sec:example}, respectively.

A central feature of our approach is the introduction of two dual operations on constraints: the \emph{absorb} operation~($\ug$), which increases the expressiveness of a constraint by incorporating an $\etypset$-system, and the \emph{fracture} operation~($\dg$), which acts as its inverse by decreasing expressiveness. Intuitively, absorbing a set of Horn rules~$\hrus$ into the constraint of a rule internalizes, within that constraint, the inferential content that $\hrus$ would otherwise contribute; fracturing reverses this process, externalizing part of the constraint's content back into explicit Horn rules. This dynamic interplay between absorption and fracture is what enables us to define generic proof transformations and to compute deductively equivalent calculi from one another.

\subsection{Permutation and Absorption}

We begin by defining the absorb operation, which `adds' an $\etypset$-system to the constraint of a rule. We note that this operation only affects rules parameterized with constraints that associate $\etypset$-systems with the edges of a constraint, namely, the $\id$ and $\rru$ rules (see \sect~\ref{sec:abstract-calc}). As local, expansion, and Horn rules omit the use of such constraints, such inference rules are unaffected by the absorb operation, and thus, we disregard the absorb operation in these cases.

\begin{definition}[Absorb]\label{def:absorb} 
We define the \emph{absorb} operation between a constraint $\con = (V, E, L)$ and an $\etypset$-system $\gram$ denoted $\con \oplus \gram$, as the constraint $(V, E, L')$ such that for each $(w,u) \in E$, $L'(w,u) = (\gram' \! \cup \gram)(\etypa)$ \iffi $L(w,u) = \gram'(\etypa)$. 
We lift the absorb operation from constraints to initial and reachability rules as follows: $\id \oplus \gram = i(\con \oplus \gram, \seqrel)$ and $r(\con,\seqrel) \ug \gram = r(\con \ug \gram, \seqrel)$.
\end{definition}

We now define the notion of \emph{permutation} in our setting, clarifying what it means for two rule sets to be permutable with one another.

\begin{definition}[Permutation]\label{def:permutation} Let $\etypset$ be a set of edge types, and $\rus_{1}$ and $\rus_{2}$ be two sets of rules. We say that $\rus_{1}$ \emph{permutes above} $\rus_{2}$, written $\rus_{1} \permabove \rus_{2}$, \iffi for any g-sequents $\gseq, \gseq_{1}, \ldots, \gseq_{n} \in \gspace{\etypset}$, if $\gseq$ can be derived via an application of a rule $\ruii \in \rus_{2}$ followed by an application of a rule $\rui \in \rus_{1}$ from $\gseq_{1}, \ldots, \gseq_{n}$, then $\gseq$ can be derived via an application of $\rui$ followed by an application of $\ruii$ from $\gseq_{1}, \ldots, \gseq_{n}$. If $\rus_{1} \permabove \rus_{2}$ and $\rus_{2} \permabove \rus_{1}$, then we say that $\rus_{1}$ and $\rus_{2}$ are \emph{permutable} with one another, and write $\rus_{1} \perm \rus_{2}$. We note that when $\rus_{1}$ or $\rus_{2}$ is a singleton (i.e. a single rule $\rui$), we simply write the rule name $\rui$ in the notation defined above. 
\end{definition}




\subsubsection{Permuting Horn rules}

We now present a sequence of permutation results regarding Horn rules. These results rest on a close correspondence between Horn rules and the production rules that determine constraint languages. Specifically, for any set of Horn rules $\hrus$, we show:
\begin{itemize}
\item $\hrus$ always permutes with local rules (Theorem~\ref{thm:local-horn-perm}), 

\item $\hrus$ can always be permuted above expansion rules (Theorem~\ref{thm:expansion-horn-perm}), and 

\item $\hrus$ permutes with reachability rules given that 
$\hrus$ has been absorbed into their constraints (Theorem~\ref{thm:reach-horn-perm}).

\end{itemize}
%
%
For the remainder of the section, we fix a set $\etypset$ of edge types, and consider relationships between rules that participate in an abstract calculus $\calc = (\gspace{\etypset}, \ops) \in \calcspace{\etypset}$, unless specified otherwise.

\begin{theorem}\label{thm:local-horn-perm} If $\lru$ is a local rule and $\hrus$ is a set of Horn rules, then $\lru \perm \hrus$.
\end{theorem}

\begin{proof}
Consider any $\ant$, $\ant'$, $\suc$, $\suc'$, and $\suc_i$ with $i \in [n]$. Observe that the applicability of any local rule $\lru$ to a set of g-sequents does not depend on the edges thereof, that is, the derivation shown below left is a valid application of $\lru$ \iffi the derivation shown below right is:
\begin{center}
\begin{tabular}{c c}
\AxiomC{$\set{\ant \sar \suc_i}_{i \in [n]}$}
\RightLabel{$\lru$}
\UnaryInfC{$\ant \sar \suc$}
\DisplayProof\hspace{5mm}

&

\AxiomC{$\set{\ant' \sar \suc_i}_{i \in [n]}$}
\RightLabel{$\lru$}
\UnaryInfC{$\ant' \sar \suc$}
\DisplayProof
\end{tabular}
\end{center}
Analogously, the application of any rule $\rui \in \hrus$ does not depend on the labeling of vertices in g-sequents, that is, the derivation shown below left is a valid application of $\rui$ \iffi the derivation shown below right is:
\begin{center}
\begin{tabular}{c c}
\AxiomC{$\ant \sar \suc$}
\RightLabel{$\rui$}
\UnaryInfC{$\ant' \sar \suc$}
\DisplayProof\hspace{10mm}

&

\AxiomC{$\ant \sar \suc'$}
\RightLabel{$\rui$}
\UnaryInfC{$\ant' \sar \suc'$}
\DisplayProof
\end{tabular}
\end{center}
Thus, for any $\rui \in \hrus$, we have a derivation of the form shown below left \iffi if we have a derivation of the form shown below right, where $\rui \times n$ indicates that $\rui$ is applied $n$ times.
\begin{center}
\begin{tabular}{c c}
\AxiomC{$\set{\ant \sar \suc_i}_{i \in [n]}$}
\RightLabel{$\lru$}
\UnaryInfC{$\ant \sar \suc$\vspacer}
\RightLabel{$\rui$}
\UnaryInfC{$\ant'\sar \suc$}
\DisplayProof\hspace{5mm}

&

\AxiomC{$\set{\ant \sar \suc_i}_{i \in [n]}$}
\RightLabel{$\rui \times n$}
\UnaryInfC{$\set{\ant' \sar \suc_i}_{i \in [n]}$\vspacer}
\RightLabel{$\lru$}
\UnaryInfC{$\ant'\sar \suc$}
\DisplayProof
\end{tabular}
\end{center}
This concludes the proof.
\end{proof}

\begin{theorem}\label{thm:expansion-horn-perm} If $\eru$ is an expansion rule and $\hrus$ is a set of Horn rules, then $\hrus \permabove \eru$.
\end{theorem}

\begin{proof} We want to prove that for any rule $\rui \in \hrus$, if we are given a derivation of the form shown below left, then we can swap the rule applications, yielding a derivation as shown below right, where $\antii \in \{w \Ea u, u \Ea w \ | \ \etypa \in \etypset\}$.
\begin{center}
\begin{tabular}{c @{\hskip 0em} c c @{\hskip 0em} c}
\AxiomC{$\gseq_{1} =$}
\noLine
\UnaryInfC{$\gseq_{2} =$}
\noLine
\UnaryInfC{\phantom{$\gseq$}}
\DisplayProof

&

\AxiomC{$\ant, \antii \sar \suc, w : \seq_{1}, u : \seq_{2}$}
\RightLabel{$\eru$}
\UnaryInfC{$\ant \sar \suc, w : \seq_{3}$\vspacer}
\RightLabel{$\rui$}
\UnaryInfC{$\ant' \sar \suc, w : \seq_{3}$}
\DisplayProof\hspace{5mm}

&

\AxiomC{\phantom{$\gseq$}}
\noLine
\UnaryInfC{$\gseq_{3} =$}
\noLine
\UnaryInfC{$\gseq_{4} =$}
\DisplayProof

&

\AxiomC{$\ant, \antii \sar \suc, w : \seq_{1}, u : \seq_{2}$}
\RightLabel{$\rui$}
\UnaryInfC{$\ant', \antii \sar \suc, w : \seq_{1}, u : \seq_{2}$\vspacer}
\RightLabel{$\eru$}
\UnaryInfC{$\ant' \sar \suc, w : \seq_{3}$}
\DisplayProof
\end{tabular}
\end{center}
To prove the claim, we must show that the right application of $\eru$ is indeed a valid application of the rule. Let the constraint family $\confam$ be as in the definition of an expansion rule (see p.~\pageref{def:expansion-rule}) and let $\gseq_{i} = (\vset_{i},\edgs_{i},\lfunc_{i})$ for $i \in [4]$. By assumption, we know that constraint maps $\conmap_{1} \colon \{w',u'\} \to \vset_{1}$ and $\conmap_{2} \colon \{w'\} \to \vset_{2}$ exist such that 
(1) $\conmap_{1}(w') = \conmap_{2}(w') = w$ and $\conmap_{1}(u') = u$, (2) $\lfunc_{1}(\conmap_{1}(w')), \lfunc_{1}(\conmap_{1}(u')), \lfunc_{2}(\conmap_{2}(w'))$, and $\suc$ satisfy $\seqrel$, and (3) $\univ(\ant \sar \suc, w : \seq_{3}) \cap \univ(\antii) = \{w\}$. We now argue that conditions (1)--(3) of an expansion rule hold in the right derivation as well.

Let us define the constraint maps $\conmap_{3} \colon \{w',u'\} \to \vset_{3}$ and $\conmap_{4} \colon \{w'\} \to \vset_{4}$ such that $\conmap_{3}(w') := w$, $\conmap_{3}(u') := u$, and $\conmap_{4}(w') := w$. By definition then, (1) $\conmap_{3}(w') = \conmap_{4}(w')$ and $\conmap_{3}(u') = u$. Since, $\lfunc_{1}(\conmap_{1}(w')) = \lfunc_{3}(\conmap_{3}(w'))$, $\lfunc_{1}(\conmap_{1}(u')) = \lfunc_{3}(\conmap_{3}(u'))$, and $\lfunc_{2}(\conmap_{2}(w')) = \lfunc_{4}(\conmap_{4}(w'))$, we have that (2) $\lfunc_{3}(\conmap_{3}(w')), \lfunc_{3}(\conmap_{3}(u')), \lfunc_{4}(\conmap_{4}(w'))$, and $\suc$ satisfy $\seqrel$. Last, since $\rui$ will only remove an edge between vertices $v,z \in \univ(\ant)$, we have that (3) $\univ(\ant' \sar \suc, w : \seq_{3}) \cap \univ(\antii) = \{w\}$. Therefore, $\rui$ may be permuted above $\eru$, showing that the derivation above right exists.
\end{proof}

\begin{theorem}\label{thm:reach-horn-perm} If $\rru$ is a reachability rule, and $\hrus$ is a set of Horn rules, then $\rru \ug \gram(\hrus) \perm \hrus.$
\end{theorem}

\begin{proof} 
Let $\rui \in \hrus$ and $\gram(\rui) = \{a \pto \stra, \conv{a} \pto \conv{\stra}\}$. We will only consider the case of $\fhru$, as the case for $\bhru$ is analogous. We let $\inc{\con} = \con \ug \gram(\hrus)$ be the constraint of $\ruii = \rru \ug \gram(\hrus)$. Our aim is to prove that we have a derivation of the form shown below left \iffi we have a derivation of the form shown below right.
\begin{center}
\AxiomC{$\{\,\ant,\, w \Es u,\, w \Ea u \sar \suc_{i}\,\}_{i \in [n]}$}
\RightLabel{$\ruii$}
\UnaryInfC{$\ant,\, w \Es u,\, w \Ea u \sar \suc_{n{+}1}$\vspacer}
\RightLabel{$\fhru$}
\UnaryInfC{$\ant,\, w \Es u \sar \suc_{n{+}1}$\vspacer}
\DisplayProof\hspace{5mm}
\AxiomC{$\{\,\ant,\, w \Es u,\, w \Ea u \sar \suc_{i}\,\}_{i \in [n]}$}
\RightLabel{$\fhru \times n$}
\UnaryInfC{$\{\,\ant,\, w \Es u \sar \suc_{i}\,\}_{i \in [n]}$\vspacer}
\RightLabel{$\ruii$}
\UnaryInfC{$\ant,\, w \Es u \sar \suc_{n{+}1}$\vspacer}
\DisplayProof
\end{center}
We note that the right-to-left direction is trivial, and thus, we focus on the left-to-right direction. In particular, we want to argue that if a derivation of the form shown below left is a valid application of $\ruii$, then the derivation below right is a valid application of $\ruii$, where the premises are the g-sequents $\gseq_{i}$ and $\gseq_{i}'$, respectively. 
\begin{center}
\begin{tabular}{@{\hskip 1mm} c @{\hskip .5em} c}
\AxiomC{$\{\overbrace{\rule{0pt}{1em}\ant, w \Es u, w \Ea u \sar \suc_{i}}^{\gseq_{i}}\}_{i \in [n]}$}
\RightLabel{$\ruii$}
\UnaryInfC{$\underbrace{\ant,\, w \Es u,\, w \Ea u \sar \suc_{n{+}1}}_{\gseq_{n{+}1}}$\vspacer}
\DisplayProof \hspace{5mm}

&

\AxiomC{$\{\overbrace{\rule{0pt}{1em}\ant,\, w \Es u \sar \suc_{i}}^{\gseq_{i}'}\}_{i \in [n]}$}
\RightLabel{$\ruii$}
\UnaryInfC{$\underbrace{\ant,\, w \Es u \sar \suc_{n{+}1}}_{\gseq_{n{+}1}'}$\vspacer}
\DisplayProof
\end{tabular}
\end{center}
Let $\con = (\{v,z\},E,L)$ and $\gseq_{i} := (\vset_{i},\edgs_{i},\lfunc_{i})$ for $i \in [n{+}1]$. Furthermore, for $i \in [n{+}1]$, suppose that constraint maps $\conmap_{i} \colon \{v,z\} \to \vset_{i}$ exist such that (1) $\conmap_{i}(v) = \conmap_{j}(v)$ and $\conmap_{i}(z) = \conmap_{j}(z)$ for $i \neq j \in [n{+}1]$, (2) $\gseq_{i}$ satisfies $\con$ with $\conmap_{i}$, and (3) $\lfunc_{1}(\conmap_{1}(v)), \lfunc_{1}(\conmap_{1}(z)), \ldots, \lfunc_{n{+}1}(\conmap_{n{+}1}(v)), \lfunc_{n{+}1}(\conmap_{n{+}1}(z))$, and $\suc$ satisfy $\seqrel$. By \dfn~\ref{def:absorb}, observe that $\con \oplus \gram(\hrus) = (\set{v,z}, E, L')$ such that for any $(v',z') \in E$, $L'(v',z') = (\gram' \! \cup \gram(\hrus))(\etypa)$ \iffi $L(v',z') = \gram'(\etypa)$.

Let $\gseq_{i}' := (\vset_{i},\edgs_{i}',\lfunc_{i})$ for $i \in [n{+}1]$. It is straightforward to verify that conditions (1) and (3) hold in the above right instance of $\ruii$ with respect to the constraint maps $\conmap_{i}$. Regarding condition (2), since $\set{a \pto \stra, \conv{a} \pto \conv{\stra}} = \gram(\fhru) \subseteq \gram(\hrus)$, anytime $w \ppath{a} u$ or $u \ppath{\conv{a}} w$ is used in the propagation path witnessing the satisfaction of $L(v',z')$ in $\gseq_{i}$ with $\conmap_{i}$, we can use $w \ppath{s} u$ or $u \ppath{\conv{s}} w$ in $\gseq_{i}'$ instead. Thus, $\gseq_{i}'$ satisfies $\con$ with $\conmap_{i}$, meaning, condition (2) is satisfied as well. Therefore, $\fhru$ can be applied first, and $\ruii$ second.
\end{proof}


\subsection{Fracturing}

 We now introduce the fracture operation, which under certain conditions, functions as the inverse of the absorb operation, thus weakening constraints on initial and reachability rules. Subsequently, we define the notion of simulation, and show how weakened variants of initial and reachability rules can be simulated with the help of Horn rules.

\begin{definition}[Fracture]\label{def:fracture}
We define the \emph{fracture} operation between a constraint $\con = (V, E, L)$ and an $\etypset$-system $\gram$, denoted $\con \ominus \gram$, to be the constraint $(V, E, L')$ such that for each $(w,u) \in E$, $(\gram' \setminus \gram)(\etypa) = L'(w,u)$ \iffi $\gram'(\etypa) = L(w,u)$. For a constraint family $\confam = (\con_{1}, \ldots, \con_{n})$, we let $\confam \dg \gram = (\con_{1} \dg \gram, \ldots, \con_{n} \dg \gram)$. We lift the fracture operation from constraints to initial rules and reachability rules as follows:
$\id \ominus \gram = i(\con \ominus \gram, \seqrel)$ and
$\rru \ominus \gram = r(\con \ominus \gram, \seqrel)$.
\end{definition}

Observe that absorbing a grammar into a transmission rule (discussed in Sections~\ref{sec:overview} and~\ref{sec:abstract-calc}), yields a reachability rule, and that `fracturing' the grammar $\gram(\rru)$ from a reachability rule $\rru$, gives a transmission rule.

\begin{proposition} Let $\tru$ be a transmission rule, $\rru$ be a reachability rule, and $\hrus$ be a non-empty set of Horn rules. Then, (1) $\tru \ug \gram(\hrus)$ is a reachability rule, and (2) $\rru \dg \gram(\rru)$ is a transmission rule.
\end{proposition}

Moreover, one can confirm that under certain conditions, the absorb and fracture operations are inverses of one another, and exhibit the following properties. Note, the following serves both as a technical lemma, and a showcase of the duality between fracture and absorb operations.

\begin{lemma}\label{lem:abs-frc-inverse} Let $\rui \in \{\id, \rru\}$ and $\hrus$ be a set of Horn rules. Then,
\begin{enumerate}

\item $(\rui \ug \gram(\hrus)) \dg \gram(\hrus) = \rui \dg \gram(\hrus)$;

\item if $\gram(\hrus) \cap \gram(\rui) = \emptyset$, then $(\rui \ug \gram(\hrus)) \dg \gram(\hrus) = \rui$;

\item $(\rui \dg \gram(\hrus)) \ug \gram(\hrus) = \rui \ug \gram(\hrus)$;

\item if $\gram(\hrus) \subseteq \gram(\rui)$, then $(\rui \dg \gram(\hrus)) \ug \gram(\hrus) = \rui$.

\end{enumerate}
\end{lemma}

\subsection{Simulation}

 We now define the simulation and bi-simulation relation between rule sets and abstract calculi. In the sequel, we state a variety of useful properties concerning such relations.

\begin{definition}[Simulation]\label{def:simulaton} Let $\etypset$ be a set of edge types, and $\rus_{1}$ and $\rus_{2}$ two sets of rules. We say that $\rus_{2}$ \emph{simulates} $\rus_{1}$, written $\rus_{1} \simul \rus_{2}$, \iffi for any g-sequents $\gseq, \gseq_{1}, \ldots, \gseq_{n} \in \gspace{\etypset}$, if $\gseq$ is derivable from $\gseq_{1}, \ldots, \gseq_{n}$ with $\rus_{1}$, then $\gseq$ derivable from $\gseq_{1}, \ldots, \gseq_{n}$ with $\rus_{2}$. If $\rus_{1} \simul \rus_{2}$ and $\rus_{2} \simul \rus_{1}$, then we say that $\rus_{1}$ and $\rus_{2}$ \emph{bi-simulate} each other and write $\rus_{1} \bisimul \rus_{2}$. Let $\calc_{1} = (\gspace{\etypset}, \ops_{1})$ and $\calc_{2} = (\gspace{\etypset}, \ops_{2})$ be two abstract calculi. We say that $\calc_{2}$ \emph{simulates} $\calc_{1}$, and write $\calc_{1} \simul \calc_{2}$, \iffi $\ops_{1} \simul \ops_{2}$. We say that $\calc_{1}$ \emph{bi-simulates} $\calc_{2}$, and write $\calc_{1} \bisimul \calc_{2}$, \iffi $\calc_{1} \simul \calc_{2}$ and $\calc_{2} \simul \calc_{1}$.
\end{definition}


 It is a basic exercise to establish the following properties:

\begin{lemma}\label{lem:simulation-props} Let $\etypset$ and $\etypset'$ be sets of edge types, $\calc \in \calcspace{\etypset}$ with $\calc = (\gspace{\etypset}, \ops)$, and $\calcii \in \calcspace{\etypset'}$. Then, 
\begin{enumerate}

\item 
 $\simul$ is a pre-order over $\calcspace{\etypset}$;\vspacer

\item $\bisimul$ is an equivalence relation over $\calcspace{\etypset}$;\vspacer

\item if $\rus_{1} \subseteq \ops$ and $\rus_{1} \simul \rus_{2}$ with $\rus_{2}$ a set of rules, then $\calc \simul (\gspace{\etypset}, (\ops \setminus \rus_{1}) \cup \rus_{2})$;\vspacer

\item if $\calc \exby \calcii$, then $\calc \simul \calcii$.\vspacer

\end{enumerate}
\end{lemma}

\subsubsection{Ordered Rule Sets} To discuss simulations and properties thereof, we require the use of \emph{ordered rule sets}. In essence, an ordered rule set is a set $\rus = \rus_{1} \cup \cdots \cup \rus_{n}$ of rules such that any derivation constructed with $\rus$ must proceed in a certain order, being obtained by applying at least $0$ or $1$ rule applications from $\rus_{1}$, followed by at least $0$ or $1$ rule applications from $\rus_{2}$, etc.

\begin{definition}[Ordered Rule Sets] Let $\rus_{1}, \ldots, \rus_{n}$ be sets of rules. We define $\fby^{i_{1}} \rus_{1} \cdots \fby^{i_{n}} \rus_{n}$ with $i_{j} \in \{0,1\}$ to be an \emph{ordered rule set} such that any derivation constructed with the rules in $\rus_{1} \cup \cdots \cup \rus_{n}$ must proceed by first applying $i_{1}$ or more applications of rules from $\rus_{1}$, followed by $i_{2}$ or more applications of rules from $\rus_{2}$, etc. When a set of rules in an ordered rule set is a singleton $\{\rui\}$, we will simply write $\rui$. 
\end{definition}

Having introduced ordered rule sets, we now pursue two immediate goals. First, we establish a pair of technical lemmas that clarify how ordered rule sets interact with simulation and with the absorb and fracture operations. These lemmas also highlight the purpose of ordering in the context of rule interactions. Second, we use this machinery to clearly formulate \thm~\ref{thm:id-fby-H-sim-id}.

\thm~\ref{thm:id-fby-H-sim-id} is closely connected to \thm~\ref{thm:local-horn-perm}, \thm~\ref{thm:expansion-horn-perm}, and \thm~\ref{thm:reach-horn-perm}. Taken together, these results show that---by allowing constraints to increase in expressivity---Horn rules can be systematically ``permuted away.'' In other words, the combined effect of these four theorems is to demonstrate that the inferential content of Horn rules can be internalized into constraints, enabling derivations to proceed without explicit applications of those rules.



\begin{lemma}\label{lem:basic-ordered-facts} Let $\rus_{1}$ and $\rus_{2}$ be two sets of rules. Then,
\begin{enumerate}

\item if $\rus_{1} \permabove \rus_{2}$, then $(\fby^{1} \rus_{2} \fby^{1} \rus_{1}) \simul (\fby^{1} \rus_{1} \fby^{1} \rus_{2})$;

\item for $i \in \{0,1\}$, $\fby^{1} \rus_{1} \fby^{i} \rus_{2} \simul \fby^{0} \rus_{1} \fby^{i} \rus_{2}$;

\item for $i \in \{0,1\}$, $\fby^{i} \rus_{1} \fby^{1} \rus_{2} \simul \fby^{i} \rus_{1} \fby^{0} \rus_{2}$.

\end{enumerate}
\end{lemma}

 As stated in the lemma below, we find that applying the absorb operation to an initial or reachability rule $\rui$ \emph{strengthens} the rule in the sense that $\rui \ug \gram(\hrus)$ can simulate $\rui$ for a set $\hrus$ of Horn rules, and conversely, we find that the fracture operation \emph{weakens} an initial or reachability rule. Moreover, the absorb operation satisfies a monotonicity property relative to the subset relation over Horn rules, while the fracture operation satisfies an antitonicity property, as expressed by the fourth claim of the following lemma.

\begin{lemma}\label{lem:ud-dg-properties} Let $\hrus$ and $\hrus'$ be two sets of Horn rules, $\hrus \subseteq \hrus'$, $\rui \in \{\id, \rru\}$, and $i, j \in \{0,1\}$. Then,
\begin{enumerate}

\item $\rui \simul \rui \ug \gram(\hrus)$ and $\rui \dg \gram(\hrus) \simul \rui$;

\item $(\fby^{i} \rui \fby^{j} \hrus) \simul (\fby^{i} \rui \ug \gram(\hrus) \fby^{j} \hrus)$;

\item $( \fby^{i} \rui \dg \gram(\hrus) \fby^{j} \hrus) \simul (\fby^{i} \rui \fby^{j} \hrus)$;

\item $\rui \ug \gram(\hrus) \simul \rui \ug \gram(\hrus')$ and $\rui \dg \gram(\hrus') \simul \rui \dg \gram(\hrus)$.

\end{enumerate}
\end{lemma}

 The following theorem is crucial for our generic algorithms in \sect~\ref{sec:generic-algs}. The theorem states that any derivation consisting of initial rules followed by applications of Horn rules can be simulated by initial rules under absorption.

\begin{theorem}\label{thm:id-fby-H-sim-id} If $\id$ is an initial rule and $\hrus$ is a set of Horn rules, then $(\fby^{1} \id \fby^{0} \hrus) \simul \id \ug \gram(\hrus)$.
\end{theorem}

\begin{proof} We consider w.l.o.g. a forward Horn rule $\fhru \in \hrus$ as the case for a backward Horn rule is similar. Let us assume that we have an instance of $\id$ with $\con = (V,E,L)$, followed by an application of $\fhru$, as shown below.
\begin{center}
\AxiomC{}
\RightLabel{$\id$}
\UnaryInfC{$\ant,\, w \Es u,\, w \Ea u \sar \suc$\vspacer}
\RightLabel{$\fhru$}
\UnaryInfC{$\ant,\, w \Es u \sar \suc$\vspacer}
\DisplayProof
\end{center}

We let $\gseq = \ant, w \Es u, w \Ea u \sar \suc = (\vset, \edgs, \lfunc)$ and let $\gseq' = \ant, w \Es u \sar \suc = (\vset, \edgs', \lfunc)$. Furthermore, let $i(\inc{\con}, R) = \id \ug \gram(\hrus)$ and $\inc{\con} = \con \ug \gram(\hrus) = (V,E,L')$ be as in \dfn~\ref{def:absorb}. Since $\id$ is an initial rule, we know that $\gseq$ satisfies $\con$ with a constraint map $\conmap$. We aim to show that $\gseq'$ is an instance of $i(\inc{\con}, R)$, thus demonstrating that $i(\inc{\con}, R)$ can simulate $\fby^{1} \id \fby^{0} \hrus$. First, observe that the only difference between $\edgs$ and $\edgs'$ is that $\Ea' = \Ea \setminus \{(w,u)\}$ for $\etypa \in \etypset$, $\Ea \in \edgs$, and $\Ea' \in \edgs'$. It is trivial to confirm that $\gseq'$ satisfies condition (2) of an initial rule with $\conmap$. Therefore, we focus on showing that condition (1) is satisfied, i.e. $\gseq'$ satisfies $\inc{\con}$ with $\conmap$.

We must show for any $(v,z) \in E$, if $L'(v,z) = \gram'(\etypb)$, then $\exists \strc \in \gram'(b)$ and $\gseq' \models \conmap(v) \ppath{\strc} \conmap(z)$. Therefore, let $(v,z) \in E$ and suppose that $L'(v,z) = \gram'(\etypb)$ with $\gram' = \gram \cup \gram(\hrus)$. Due to the fact that $\gseq$ satisfies $\con$ with $\conmap$, we know that $\exists \strb \in \gram(b)$ such that $\gseq \models \conmap(v) \ppath{\strb} \conmap(z)$. We now use the path $\conmap(v) \ppath{\strb} \conmap(z)$ to find a new path $\conmap(v) \ppath{\strc} \conmap(z)$ in $\gseq'$. Since $w \Es u$ occurs in $\gseq'$, there exists a path $w \ppath{\stra} u$ and its converse $u \ppath{\conv{\stra}} w$ in $\gseq'$. To find $\conmap(v) \ppath{\strc} \conmap(z)$, we take $\conmap(v) \ppath{\strb} \conmap(z)$ and replace each occurrence of $w \ppath{\etypa}u$ by $w \ppath{\stra} u$, as well as replace each occurrence of $u \ppath{\conv{a}} w$ by $u \ppath{\conv{\stra}} w$. Hence, $\gseq' \models \conmap(v) \ppath{\strc} \conmap(z)$. Last, we need to show that $\strc \in (\gram \cup \gram(\hrus))(\etypb)$. To do this, we first recognize that $\{a \pto \stra, \conv{a} \pto \conv{\stra}\} = \gram(\fhru) \subseteq \gram(\hrus)$. Because $\strb \in \gram(\etypb)$ and $\gram(\etypb) \subseteq (\gram \cup \gram(\hrus))(\etypb)$, there exists a derivation $\etypb \pto^{*}_{\gram \cup \gram(\hrus)} \strb$. By applying the production rule $a \pto \stra$ to each occurrence of $a$ in $\strb$ and the production rule $\conv{a} \pto \conv{\stra}$ to each occurrence of $\conv{a}$ in $\strb$, we obtain a derivation $\etypb \pto^{*}_{\gram \cup \gram(\hrus)} \strc$, showing the desired claim. As a consequence, we have verified that $\gseq'$ is an instance of $i(\inc{\con}, R)$.
\end{proof}

\subsubsection{Dependency Graphs and Fracturable Sets}

Recall that our overarching goal is to trade Horn rules for increased complexity of constraints. In particular, one may wish to ``permute away'' or ``reintroduce'' only a subset of Horn rules; however, this cannot be done arbitrarily, and a certain stratification of Horn rules is required. Our goal in this subsection is to prove \thm~\ref{thm:frac-horn-perm}.

Let us turn our attention toward investigating simulations in the presence of fracturing. When fracturing an initial or reachability rule $\rui$ with an $\etypset$-system $\gram(\hrus)$ for $\hrus$ a set of Horn rules, we find that $\rui$ can be simulated by $\rui \dg \gram(\hrus)$ along with applications of other inference rules. Yet, it so happens that \emph{dependencies} between Horn rules in $\hrus$ are of importance when considering simulations in this context. Intuitively, one Horn rule $h_{1}$ depends on another Horn rule $h_{2}$ when an application of $h_{2}$ produces a g-sequent $\gseq$ such that $h_{1}$ becomes applicable to it. We have provided an example of Horn rules and dependencies (which are captured by the following notion of a \emph{dependency graph}) in Example~\ref{ex:exp-imp-example} of \sect~\ref{sec:generic-algs}. We now define these dependencies:

\begin{definition}[Dependency Graph]\label{def:dep-graph} Let $\gram$ be an $\etypset$-system with $(\pru, \conv{\pru})$ and $(\pru',\conv{\pru}')$ distinct propagation pairs in $P(\gram)$ such that $\pru = \charx \pto \stra$ and $\pru' = \chary \pto \strb$ with $\charx, \chary \in \etypset \cup \conv{\etypset}$. We say that $(\pru', \conv{\pru}')$ \emph{depends on} $(\pru,\conv{\pru})$, written $(\pru,\conv{\pru}) \sqsubset (\pru', \conv{\pru}')$, \iffi $\stra$ or $\conv{\stra}$ is of the form $\stra_{1} \chary \stra_{2}$.  We define the \emph{dependency graph} of $\gram$ to be the pair $\depgr(\gram) = (\depgrv, \depgre)$ such that $\depgrv = P(\gram)$ and $\depgre$ is the reflexive-transitive closure of $\sqsubset$. 

For a set $\hrus$ of Horn rules, we define $\depgr(\hrus) = (\hrus,\depgre')$ such that for $h,h' \in \hrus$, $h \depgre' h'$ \iffi for $(\pru, \conv{\pru}) \in P(h)$ and $(\pru', \conv{\pru}') \in P(h')$, $(\pru, \conv{\pru}) \depgre (\pru', \conv{\pru}')$ in $\depgr(\gram(\hrus)) = (\depgrv,\depgre)$. To capture dependency graphs over $\etypset$-systems and Horn rules in a uniform notation, we may denote them by $\depgr = (V,\depgre)$.
\end{definition}

Of critical importance in dependency graphs is the notion of a \emph{fracturable set}. In essence, for a dependency graph $\depgr = (V,\depgre)$, a fracturable set is a set $V' \subseteq V$ of vertices such that every vertex $v \in V'$ `sees' only vertices in $V'$.

\begin{definition}[Fracturable Set]\label{def:fracturable-set}
Given a dependency graph $\depgr = (V, \depgre)$ we say that a subset $V'$ of $V$ is \emph{fracturable} when there are no $\depgre$-edges from $V'$ to $V \setminus V'$, and we define a subset $V''$ of $V$ to be \emph{anti-fracturable} \iffi there exists a fracturable subset $V'$ and $V'' = V \setminus V'$. Given any $V' \subseteq V$ of a dependency graph $\depgr = (V, \depgre)$ of an $\etypset$-system or a set of Horn rules $\hrus$, we let $\gram(V')$ denote $\bigcup_{(\pru,\conv{\pru}) \in V'} \{\pru, \conv{\pru}\}$ and $\bigcup_{h \in V'}\gram(h)$, respectively. For a set $\hrus$ of Horn rules we say that a subset $\hrus'$ is \emph{(anti-)fracturable} \iffi $\hrus'$ is (anti-)fracturable in the dependency graph $\depgr(\hrus) = (\hrus, \depgre)$.
\end{definition}

The following properties of (anti-)fracturable subsets are useful and follow from the above definition.

\begin{lemma}\label{lem:(anti)-frac-set-props}
Let $\depgr = (V, \depgre)$ be a dependency graph. Then,
\begin{enumerate}

\item $V$ and $\emptyset$ are both (anti-)fracturable subsets of $V$; 

\item if both $V'$ and $V''$ are (anti-)fracturable subsets of $V$ and $V \setminus V'$ respectively, then $V' \cup V''$ is (anti-)fracturable.

\end{enumerate}
\end{lemma}

\begin{theorem}\label{thm:frac-horn-perm}
If $\hrus$ is a set of Horn rules with $\hrus'$ a fracturable subset of $\depgr(\hrus)$, then $\hrus \setminus \hrus' \permabove \hrus'$.
\end{theorem}

\begin{proof} We consider the case of a backward Horn rule $\bhru \in \hrus \setminus \hrus'$ as the forward case is analogous, and argue by contraposition. Suppose that $\bhru$ cannot be permuted above a rule $h' \in \hrus'$ (we assume w.l.o.g. that $h'$ is a forward Horn rule) in a derivation, which occurs \iffi a derivation of the form shown below exists, where $w\; \ppatha{s\conv{b}s'} \; u, w \Ea u$ and $v \ppatha{\strb} v', v' \ppatha{\etypb} v$ are `active' in $h'$ and $\bhru$, respectively.
\begin{center}
    \AxiomC{$\ant,\, v \ppatha{\strb} v',\; \overbrace{w \ppatha{\stra} v,\, v' \ppatha{\etypb} v,\, v' \ppatha{s'} u}^{\let\scriptstyle\textstyle\substack{w\; \ppatha{s\conv{b}s'} \; u}},\; w \ppatha{\etypa} u \sar \suc$}
    \RightLabel{$h'$}
    \UnaryInfC{$\ant,\, w \ppatha{\stra} v,\, v' \ppatha{s'} u,\, v \ppatha{\strb} v',\, v' \ppatha{\etypb} v \sar \suc$\vspacer}
    \RightLabel{$\bhru$}
    \UnaryInfC{$\ant,\, w \ppatha{\stra} v,\, v' \ppatha{s'} u,\, v \ppatha{\strb} v'\sar \suc$\vspacer}
    \DisplayProof
\end{center}
However, from the above, we can conclude that $h' \depgre \bhru$ in $\depgr(\hrus)$, showing that $\hrus'$ is not a fracturable subset.
\end{proof}

\subsubsection{Reintroduction of Horn Rules}

So far, we have primarily focused on permutations of Horn rules (\thm~\ref{thm:local-horn-perm}, \thm~\ref{thm:expansion-horn-perm}, \thm~\ref{thm:reach-horn-perm}, and \thm~\ref{thm:frac-horn-perm}) and their removal (\thm~\ref{thm:id-fby-H-sim-id}), typically at the cost of increasing the expressive complexity of constraints. In this section, our goal is to identify when Horn rules can be reintroduced while reducing the complexity of constraints (\thm~\ref{thm:id-to-id-and-horn} and \thm~\ref{thm:rru-to-rru-and-horn}).

\begin{definition}[Saturation]\label{def:satured}
Let $\hrus$ be a set of Horn rules with $h \in \hrus$. If $h$ is of the form shown below left, we define the \emph{inverse} $\invh{h}$ of $h$ to be the rule of the form shown below right:
\begin{center}
\begin{tabular}{c @{\hskip 3em} c}
\AxiomC{$\gseq$}
\RightLabel{$h$}
\UnaryInfC{$\gseq'$}
\DisplayProof

&

\AxiomC{$\gseq'$}
\RightLabel{$\conv{h}$}
\UnaryInfC{$\gseq$}
\DisplayProof
\end{tabular}
\end{center}
 We write $\conv{h}(\gseq) = \gseq'$ to mean that an application of $\conv{h}$ to $\gseq$ produces $\gseq'$. A g-sequent $\gseq$ is defined to be \emph{$\conv{h}$-saturated} \iffi every application of $\conv{h}$ to $\gseq$ produces $\gseq$. We say that $\invh{h}(\gseq)$ is \emph{permissible} \iffi $\invh{h}(\gseq)$ produces a g-sequent $\gseq' \neq \gseq$. We define $\conv{\hrus} = \{\conv{h} \ | \ h \in \hrus\}$ and define a g-sequent $\gseq$ to be $\conv{\hrus}$-saturated \iffi it is $\conv{h}$-saturated for every $\conv{h} \in \conv{\hrus}$.
\end{definition}

 We now provide a sequence of results that will ultimately be used to show under what conditions initial and reachability rules can be simulated with `weaker' variants along with applications of Horn rules. In what follows, we let $\invh{\hrus}(\gseq)$ denote the $\invh{\hrus}$-saturated g-sequent obtained by repeatedly applying all permissible applications of rules in $\invh{\hrus}$ to $\gseq$.

\begin{lemma}\label{lem:invhorn-props}
If $\gseq$ is a g-sequent and $\hrus$ is a finite set of Horn rules, then $\invh{\hrus}(\gseq)$ is computable in $\mathrm{PTIME}$ and is $\invh{\hrus}$-saturated.
\end{lemma}

\begin{proof} It should be clear, from the definition of $\invh{\hrus}(\gseq)$, that saturating $\gseq$ under applications of rules from $\invh{\hrus}$ builds a bottom-up derivation of the following form:
\begin{center}
    \AxiomC{$\invh{\hrus}(\gseq)$}
    \RightLabel{$h_1$}
    \UnaryInfC{$\hspace{5mm}\vdots\hspace{5mm}$}
    \RightLabel{$h_n$}
    \UnaryInfC{$\gseq$}
    \DisplayProof
\end{center}
where $h_1, \ldots, h_n \in \hrus$. Observe the following: (1) any bottom-up application of a Horn rule builds the upper g-sequent from the lower by adding a single edge. (2) As only permissible rules can be applied, the height of this derivation is bound by the maximal number of edges in a graph, which is quadratic in the number of its vertices. (3) Given a Horn rule $h$, verifying if a g-sequent $\gseq$ is $\invh{h}$-saturated can be done in $\mathrm{PTIME}$, as checking the existence of a path labeled with a given string between any two vertices of a graph can be done in $\mathrm{PTIME}$ by standard algorithmic techniques. (4) $\invh{\hrus}(\gseq)$ is obviously $\invh{\hrus}$-saturated by its definition.
\end{proof}

\begin{observation}\label{obs:fracseq-VE} Let $\hrus$ be a set of Horn rules, $\gseq = \ant \sar \suc$ be a g-sequent, and $\invh{\hrus}(\gseq) = \ant' \sar \suc'$. Then, $\univ(\gseq) = \univ(\invh{\hrus}(\gseq))$, $\ant \subseteq \ant'$, and $\suc = \suc'$.
\end{observation}

\begin{lemma}\label{lem:fracturing-constraints-aux}
Let $\hrus$ be a set of Horn rules, $\gseq$ be a g-sequent, and $\gram = \gram(\hrus)$. For any $\stra \in (\etypset \cup \conv{\etypset})^{*}$ and $u,w \in \univ(\gseq)$ if $\gseq \models w \ppath{s} u$, then for every string $\strb$ such that $\strb \pto^{*}_{\gram} \stra$, we have $\invh{\hrus}(\gseq) \models w \ppath{\strb} u$.
\end{lemma}

\begin{proof} Assume toward a contradiction that the lemma does not hold. Let $\hrus$, $\gseq$, $w$, $u$, $\stra$, and $\strb$ form a counterexample. Note that w.l.o.g. we can assume $\strb \pto_{\gram} \stra$, meaning $\stra$ is derivable from $\strb$ in one step. As $\strb \pto_{\gram} \stra$ we know that (1) there exists a rule $h\in\hrus$ (we may assume w.l.o.g. that $h = \fhru$) such that $\gram(h) = \set{ \etypa \pto \strc, \conv{\etypa} \pto \conv{\strc}}$, (2) $\strb$ is of the form $\strb_1 \etypa \strb_2$ (or $\strb_1 \conv{\etypa} \strb_2$), and (3) $\stra$ is of the form $\strb_1 \strc \strb_2$ (or $\strb_1 \conv{\strc} \strb_2$). We will consider the case of $\strb = \strb_1 \etypa \strb_2$, as the latter is analogous. Let $w', u' \in \univ(\gseq)$ be such that: $\invh{\hrus}(\gseq) \models w \ppath{\strb_1} w', w' \ppath{\strc} u', u' \ppath{\strb_2} u$, but not $\invh{\hrus}(\gseq) \models w \ppath{\strb} u$. Let $\invh{\hrus}(\gseq) = \ant \sar \suc$ and consider the following:
\begin{center}
\AxiomC{$\ant,\,  w' \mathcal{E}_{\etypa} u' \sar \suc$}
\RightLabel{$h$}
\UnaryInfC{$\ant \sar \suc$}
\DisplayProof
\end{center} 
When viewed bottom-up, this is a permissible application of $\overline{h}$, which leads to a contradiction as $\invh{\hrus}(\gseq)$ is $\invh{\hrus}$-saturated.
\end{proof}

We introduce the \emph{path weakening} rule (denoted $\wk$) in the lemma below, which serves as a restricted form of the conventional weakening rule (cf. WrL in~\cite{Vig00}), only permitting the introduction of single edges between vertices in a g-sequent. By standard arguments, one can establish that $\wk$ is \emph{admissible} in any abstract calculus, i.e. if a g-sequent $\gseq$ has a proof in an abstract calculus $(\gspace{\etypset}, \ops \cup \{\wk\})$, then it has a proof in the abstract calculus $(\gspace{\etypset}, \ops)$.

\begin{lemma}\label{lem:wk-admiss}
 The path weakening rule $\wk$ (shown below) is admissible in any abstract calculus: 
\begin{center}
\AxiomC{$\ant \sar \suc$}
\RightLabel{$\wk$}
\UnaryInfC{$\ant, \antii \sar \suc$}
\DisplayProof
\end{center}
where $\antii \in \{w \Ea u \ | \ \etypa \in \etypset \text{ and } w,u \in \univ(\ant \sar \suc) \}$.
\end{lemma}

\begin{proof} We will show that: (1) $\id$ followed by $\wk$ can be simulated by $\id$ alone, (2) $\wk \permabove \lru$, (3) $\wk \permabove \eru$, (4) $\wk \permabove \rru$, and (5) that for any derivation consisting of a Horn rule $h$ followed by a weakening rule $\rui$ one of the following holds: (5a) $\rui$ can be permuted above $h$, or (5b) both applications can be removed. Using these facts, the result follows by induction on the quantity of the given proof.

(1) If we have a derivation of the form shown below left, then we have a derivation of the form shown below right, as the satisfaction of $\con$ and $\seqrel$ cannot be falsified by the addition of edges.
\begin{center}
    \AxiomC{}
    \RightLabel{$\id$}
    \UnaryInfC{$\ant \sar \suc$}
    \RightLabel{$\wk$}
    \UnaryInfC{$\ant, \Sigma \sar \suc$}
    \DisplayProof
\hspace{5mm}
    \AxiomC{}
    \RightLabel{$\id$}
    \UnaryInfC{$\ant, \Sigma \sar \suc$}
    \DisplayProof
\end{center}

\noindent
(2) Assume we have a derivation of the form shown below left. As applicability of a local rule does not depend on the set of the edges of a g-sequent, any derivation of the form shown below right is valid as well.
\begin{center}
    \AxiomC{$\set{\ant \sar \suc_i}_{i \in [n]}$}
    \RightLabel{$\lru$}
    \UnaryInfC{$\ant \sar \suc$}
    \RightLabel{$\wk$}
    \UnaryInfC{$\ant, \Sigma \sar \suc$}
    \DisplayProof
\hspace{5mm}
    \AxiomC{$\set{\ant \sar \suc_i}_{i \in [n]}$}
    \RightLabel{$\wk \times n$}
    \UnaryInfC{$\set{\ant, \Sigma \sar \suc_i}_{i \in [n]}$}
    \RightLabel{$\lru$}
    \UnaryInfC{$\ant, \Sigma \sar \suc$}
    \DisplayProof
\end{center}

\noindent
(3) Assume we have a derivation of the form shown below left. Furthermore, assume w.l.o.g. that $\univ(\ant \sar \suc') \cap \{w \Eb u \} = \{w\}$. Note as $u' \neq u$ and $w' \neq u$ in the derivation below left, we may permute $\wk$ above $\eru$, obtaining the derivation shown below right. 
\begin{center}
    \AxiomC{$\ant, w \Eb u \sar \suc$}
    \RightLabel{$\eru$}
    \UnaryInfC{$\ant \sar \suc'$}
    \RightLabel{$\wk$}
    \UnaryInfC{$\ant, w'\Ea u' \sar \suc'$}
    \DisplayProof
\hspace{1mm}
    \AxiomC{$\ant, w \Eb u \sar \suc$}
    \RightLabel{$\wk$}
    \UnaryInfC{$\ant, w \Eb u, w'\Ea u' \sar \suc$}
    \RightLabel{$\eru$}
    \UnaryInfC{$\ant', w'\Ea u' \sar \suc'$}
    \DisplayProof
\end{center}

\noindent
(4) Assume we have a derivation of the form shown below left. As satisfaction of $\con$ and $\seqrel$ cannot be falsified by the addition of edges, we may permute $\wk$ above $\rru$, as shown below right.
\begin{center}
    \AxiomC{$\set{\ant \sar \suc_i}_{i \in [n]}$}
    \RightLabel{$\rru$}
    \UnaryInfC{$\ant \sar \suc$}
    \RightLabel{$\wk$}
    \UnaryInfC{$\ant, \Sigma \sar \suc$}
    \DisplayProof
\hspace{5mm}
    \AxiomC{$\set{\ant \sar \suc_i}_{i \in [n]}$}
    \RightLabel{$\wk \times n$}
    \UnaryInfC{$\set{\ant, \Sigma \sar \suc_i}_{i \in [n]}$}
    \RightLabel{$\rru$}
    \UnaryInfC{$\ant, \Sigma \sar \suc$}
    \DisplayProof
\end{center}

\noindent
(5) We will consider only the $h_f$ case as the $\bhru$ case is analogous. Assume we have a derivation of the form shown below left. If $\Sigma = \Sigma'$ then both applications can be removed from any derivation; otherwise, we may permute $\wk$ above $\fhru$ as shown below right.
\begin{center}
    \AxiomC{$\ant, \Sigma \sar \suc$}
    \RightLabel{$h_f$}
    \UnaryInfC{$\ant \sar \suc$}
    \RightLabel{$\wk$}
    \UnaryInfC{$\ant, \Sigma' \sar \suc$}
    \DisplayProof
\hspace{5mm}
    \AxiomC{$\ant, \Sigma \sar \suc$}
    \RightLabel{$\wk$}
    \UnaryInfC{$\ant, \Sigma, \Sigma' \sar \suc$}
    \RightLabel{$h_f$}
    \UnaryInfC{$\ant, \Sigma' \sar \suc$}
    \DisplayProof
\end{center}
This concludes the proof.
\end{proof}


The two lemmas below follow from Observation~\ref{obs:fracseq-VE} with the latter lemma also relying on Lemma~\ref{lem:fracturing-constraints-aux}. We use both lemmas in the proofs of the two subsequent theorems.

\begin{lemma}\label{obs:fracture-weakening} If $\hrus$ is a set of Horn rules, then $\invh{\hrus} \simul \wk$.
\end{lemma}

\begin{lemma}\label{lem:fracturing-constraints}
Let $\gseq$ be a g-sequent, $\con$ be a constraint, and $\hrus := \hrus(\gram(\con))$. If $\hrus'$ is a fracturable subset of $\hrus$ and $\gseq$ satisfies $\con$ with a constraint map $\conmap$, then $\invh{\hrus'}(\gseq)$ satisfies $\con \ominus \gram(\hrus')$ with $\conmap$.
\end{lemma}

\begin{theorem}\label{thm:id-to-id-and-horn} Let $\id$ be an initial rule with the set $\hrus := \hrus(\gram(\id))$. If $\hrus'$ is a fracturable subset of $\hrus$, then $\id \simul \fby^{1} \id \dg \gram(\hrus') \fby^{0} \hrus'$.
\end{theorem}

\begin{proof} Suppose we have an application of $\id$, yielding the g-sequent $\gseq$. By Lemma~\ref{lem:fracturing-constraints}, we know that the g-sequent $\gseq' = \invh{\hrus'}(\gseq)$ satisfies $\con \dg \gram(\hrus')$, and thus, serves as an instance of $\id \dg \gram(\hrus')$. As $\invh{\hrus'}(\gseq)$ is $\invh{\hrus'}$-saturated, we may apply the rules from $\hrus'$ to $\gseq'$ to derive $\gseq$.
\end{proof}

\begin{theorem}\label{thm:rru-to-rru-and-horn} Let $\rru$ be a reachability rule and let $\hrus := \hrus(\gram(\rru))$. If $\hrus'$ is a fracturable subset of $\hrus$, then $\rru \simul \fby^{0} \wk \fby^{1} \rru \dg \gram(\hrus') \fby^{0} \hrus'$.
\end{theorem}

\begin{proof} Suppose that we have an application of $\rru$ as shown below left. 
By \lem~\ref{obs:fracture-weakening}, we can derive the g-sequent $\gseq_{i}' = \invh{\hrus'}(\gseq_i)$ from $\gseq_{i}$ for each $i \in [n]$ via some number $k$ of applications of $\wk$. Moreover, one can verify that conditions (1)--(3) of a reachability rule are satisfied in the application of $\rru \ominus \gram(\hrus')$ shown below right; in particular, by Lemma~\ref{lem:fracturing-constraints}, we know that for $i \in [n{+}1]$, each g-sequent $\gseq_{i}'$ satisfies the constraint $\con_{i} \dg \gram(\hrus')$ with a constraint map $\conmap_{i}$. Thus, we may derive $\invh{\hrus'}(\gseq_{n{+}1})$ using $\rru \ominus \gram(\hrus')$, and since $\invh{\hrus'}(\gseq_{n{+}1})$ is $\invh{\hrus'}$-saturated, we can derive $\gseq_{n{+}1}$ by subsequent applications of $\hrus'$, as shown below right.
\begin{center}
    \AxiomC{$\set{\,\gseq_i\,}_{i \in [n]}$}
    \RightLabel{$\rru$}
    \UnaryInfC{$\gseq_{n{+}1}$}
    \DisplayProof
\hspace{3mm}
    \AxiomC{$\set{\,\gseq_i\,}_{i \in [n]}$}
    \RightLabel{$\wk \times k$}
    \UnaryInfC{$\set{\,\invh{\hrus'}(\gseq_i)\,}_{i \in [n]}$}
    \RightLabel{$\rru \ominus \gram(\hrus')$}
    \UnaryInfC{$\invh{\hrus'}(\gseq_{n{+}1})$}
    \RightLabel{$\hrus'$}
    \UnaryInfC{$\gseq_{n{+}1}$}
    \DisplayProof
\end{center}
This concludes the proof of the theorem.
\end{proof}


\section{Generic Transformation Algorithms}\label{sec:generic-algs}


\begin{wrapfigure}[11]{R}{0.3\textwidth}
\centering
\includegraphics[width=3.5cm]{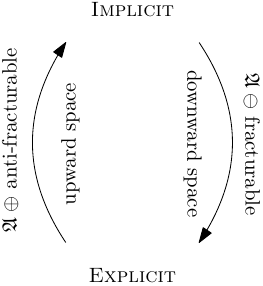}
\caption{\label{fig:section-5-summary}}
\end{wrapfigure}

As promised in the former section, we now put our permutation and simulation results to work, yielding generic transformations that establish the polynomial equivalence between abstract calculi within certain spaces. What we discover is rather remarkable: we find that certain abstract calculi sit within two isomorphic lattices (dubbed \emph{upward} and \emph{downward} spaces), with the top and bottom elements exhibiting unique properties. We call calculi that sit at that the top of a lattice \emph{implicit} and call calculi that sit at the bottom of a lattice \emph{explicit}. Implicit calculi perform Horn reasoning with constraints only, whereas explicit calculi perform such reasoning with Horn rules. In \sect~\ref{subsec:lab-nest-dual}, we make the observation that many known (cut-free) reachability nested systems and labeled sequent systems serve as examples of implicit and explicit calculi, respectively, for a diverse number of logics. 
Then, in Section~\ref{sec:example} we show an instantiation of our formalism.

As we will be constantly shifting back and forth between different types of calculi in this section, and since the objects we work with always have a dual counterpart, many arguments may become difficult to navigate. We have therefore included a ``map'' in Figure \ref{fig:section-5-summary} that gives a high level depiction of the transformations and concepts discussed in this section.

\subsection{Upward Spaces/Lattices and Explicit Calculi}

 We first investigate the \emph{upward space} of an arbitrary abstract calculus, which is defined by taking an abstract calculus $\calc$ and strengthening its rules with $\gram(\hrus)$ for $\hrus$ a specific set of Horn rules, while simultaneously removing $\hrus$ from $\calc$. This operation permits us to define a partial order over the produced abstract calculi, which can be viewed as a complete lattice (see \thm~\ref{thm:upspace-is-lattice} below). We will utilize these lattices later to identify spaces of polynomially equivalent calculi. 
 First, let us lift the absorption operation $\oplus$ and fracturing operation $\ominus$ to the level of calculi.

\begin{definition}[Calculus Absorption and Fracturing] Let $\calc = (\gspace{\etypset},\ops)$ be an abstract calculus, $\gram$ be an $\etypset$-system, and $\udg \in \{\ug, \dg\}$. We define $\calc \udg \gram$ to be the same as $\calc$, but where each initial rule $\id$ and reachability rule $\rru$ in $\ops$ is replaced by $\id \udg \gram$ and $\rru \udg \gram$, respectively.
\end{definition}

\begin{definition}[Upward Space]\label{def:upward-space} Let $\calc = (\gspace{\etypset},\ops)$ be an abstract calculus. We define the \emph{upward space} $\upspace{\calc} = (\cs,\uprel)$ inductively: (1) $\calc \in \cs$, and (2) if $\calcii \in \cs$ with $\depgr(\hrus) = (\hrus, \depgre)$ the dependency graph of $\hrus := \hrus(\calcii)$, then for an anti-fracturable subset $\hrus' \subseteq \hrus$, we have that $\calcii \uprel \calciii$ with $\calciii = \big( (\calcii  \ug \gram(\hrus')) \setminus \hrus' \big) \in \cs$.
\end{definition}

\begin{theorem}\label{thm:upspace-is-lattice}
Let $\calc = (\gspace{\etypset},\ops)$ be an abstract calculus with $\upspace{\calc} = (\cs,\uprel)$ its upward space. Then, the $\uprel$ relation is a connected partial-order. Moreover, for $\calcii, \calciii \in \cs$, if we take $\calcii \wedge \calciii = \inf\{\calcii,\calciii\}$ and $\calcii \vee \calciii = \sup\{\calcii,\calciii\}$ under $\uprel$, then $(\cs,\wedge,\vee)$ is a complete lattice with $\calc = \bot$.
\end{theorem}

\begin{proof} Let $\calc = (\gspace{\etypset},\ops)$ be an abstract calculus with $\upspace{\calc} = (\cs,\uprel)$ its upward space. We first prove that $\uprel$ is a partial order. By \lem~\ref{lem:(anti)-frac-set-props}, we know that $\emptyset$ is an anti-fracturable subset in any dependency graph. Observe that for any $\calcii \in \cs$, $\calcii = (\calcii \ug \gram(\emptyset)) \setminus \emptyset$, showing that $\calcii \uprel \calcii$, i.e. $\uprel$ is reflexive. 

 Second, assume that $\calcii \uprel \calciii$ and $\calciii \uprel \calciv$ for $\calcii, \calciii, \calciv \in \cs$. Then, there exist anti-fracturable subsets $\hrus_{1}'$ of $\hrus(\calcii)$ and $\hrus_{2}'$ of $\hrus(\calciii)$ such that $\calciii = (\calcii \ug \gram(\hrus_{1}')) \setminus \hrus_{1}'$ and $\calciv = (\calciii \ug \gram(\hrus_{2}')) \setminus \hrus_{2}'$. Observe that $\calciv = (\calcii \ug \gram(\hrus_{1}' \cup \hrus_{2}')) \setminus (\hrus_{1}' \cup \hrus_{2}')$. Therefore, to complete the proof of the case, we need to show that $\hrus_{1}' \cup \hrus_{2}'$ is anti-fracturable. Observe that the sets $\hrus(\calcii) \setminus \hrus_{1}' = \hrus(\calciii)$ and $\hrus(\calciii) \setminus \hrus_{2}'$ are fracturable, meaning $(\hrus(\calcii) \setminus \hrus_{1}') \setminus \hrus_{2}' = \hrus(\calcii) \setminus (\hrus_{1}' \cup \hrus_{2}')$ is fracturable; consequently, $\hrus_{1}' \cup \hrus_{2}'$ is anti-fracturable, 
 showing that $\uprel$ is transitive.
 
Third, let us assume that $\calcii \uprel \calciii$ and $\calciii \uprel \calcii$ for $\calcii, \calciii \in \cs$. Then, there exist anti-fracturable subsets $\hrus_{1}'$ of $\hrus(\calcii)$ and $\hrus_{2}'$ of $\hrus(\calciii)$ such that $\calciii = (\calcii \ug \gram(\hrus_{1}')) \setminus \hrus_{1}'$ and $\calcii = (\calciii \ug \gram(\hrus_{2}')) \setminus \hrus_{2}'$. Plugging either equation into the other shows that $\hrus_{1}' = \emptyset = \hrus_{2}'$, implying that $\calciii = \calcii$.
 
Last, one can show that for any $\calcii \in \cs$, $\calc \uprel \calcii$, that is, $\upspace{\calc}$  is connected. Moreover, it is straightforward to argue that $(\cs,\wedge,\vee)$ forms a complete lattice with $\calc$ the bottom element.
\end{proof}

Given an abstract calculus $\calc$ and its upward space $\upspace{\calc}$, we find that any proof in $\calc$ may be transformed along the partial order of $\upspace{\calc}$ in $\mathrm{PTIME}$. We also find that proofs may shrink when transformed along the partial order of $\upspace{\calc}$.

\begin{theorem}[Up the Upward Space]\label{thm:up-upward-lattice}
Let $\calc = (\gspace{\etypset},\ops)$ be an abstract calculus with $\upspace{\calc} = (\cs,\uprel)$ its upward space. For any abstract calculus $\calcii \in \cs$ and proof $\prf$ of a g-sequent $\gseq$ in $\calc$, there exists a proof $\prf'$ of $\gseq$ in $\calcii$ such that $\prf'$ is computable from $\prf$ in $\mathrm{PTIME}$ with $\dsize{\prf'} \leq \dsize{\prf}$.
\end{theorem}

\begin{proof} Let $\calc$ be an abstract calculus, $\upspace{\calc} = (\cs,\uprel)$ its upward space, and $\calcii \in \cs$. By \thm~\ref{thm:upspace-is-lattice}, we know that $\calc \uprel \calcii$, and hence, there exists a set $\hrus'$ of Horn rules such that $\hrus' \subseteq \hrus = \hrus(\calc)$ is an anti-fracturable subset in $\depgr(\hrus)$ which witnesses that the relation $\calc \uprel \calcii$ holds. Let us take a proof $\prf$ of a g-sequent $\gseq$ in $\calc$ and consider the topmost application of a Horn rule in $\hrus'$. We assume w.l.o.g. that the Horn rule is $\fhru$ and argue by induction on the quantity of the proof that $\fhru$ can be permuted upward and eliminated entirely.

\textit{Base case.} Suppose we have an application of $\id$ followed by an application of $\fhru$. By \lem~\ref{lem:basic-ordered-facts}, we know that $\fby^{1} \id \fby^{1} \fhru \simul \fby^{1} \id \fby^{0} \fhru$, and by \thm~\ref{thm:id-fby-H-sim-id}, we know that $\fby^{1} \id \fby^{0} \fhru \simul \id \oplus \gram(\fhru)$. By \lem~\ref{lem:ud-dg-properties}, $\id \oplus \gram(\fhru) \simul \id \ug \gram(\hrus')$ since $\fhru \in \hrus'$. As simulations are transitive (\lem~\ref{lem:simulation-props}), we know that $\fby^{1} \id \fby^{1} \fhru \simul \id \ug \gram(\hrus')$. We can therefore replace the application of $\id$ followed by $\fhru$ by a single application of $\id \ug \gram(\hrus')$.

\textit{Inductive step.} Suppose we have an application of a local, expansion, reachability, or Horn rule $\rui$ in $\calc$ followed by an application of $\fhru$. By \thm~\ref{thm:local-horn-perm} (local case), \thm~\ref{thm:expansion-horn-perm} (expansion case), and \thm~\ref{thm:frac-horn-perm} (Horn rule case) we know that $\fhru$ can be permuted above $\rui$. In the case where $\rui$ is a reachability rule, we note that $\rui$ can be replaced by $\rui \ug \gram(\hrus')$ (which is a rule in $\calcii$) by Lemma~\ref{lem:ud-dg-properties} and $\fhru$ can be permuted above $\rui \ug \gram(\hrus')$ by \thm~\ref{thm:reach-horn-perm}. Therefore, by the induction hypothesis, $\fhru$ is eliminable from the proof altogether. Note that by definition $\hrus'$ is an anti-fracturable subset of $\hrus(\calc)$, and thus, \thm~\ref{thm:frac-horn-perm} is applicable.

We repeat the above algorithm by successively considering each topmost occurrence of a Horn rule in $\hrus'$ until all such rules are eliminated, giving a $\mathrm{PTIME}$ procedure for computing the output proof $\prf'$ in $\calcii$. Moreover, we have that $\max\{\seqsize{\gseq} \ | \ \gseq \in \prf'\} \leq \max\{\seqsize{\gseq} \ | \ \gseq \in \prf\}$ since the elimination of Horn rules removes edges from g-sequents in $\prf$, giving potentially smaller g-sequents in $\prf'$. Additionally, $|\{\gseq \in \gspace{\etypset} \ | \ \gseq \in \prf'\}| \leq |\{\gseq \in \gspace{\etypset} \ | \ \gseq \in \prf\}|$ since Horn rules are `absorbed' into the initial rules. Thus, we have that $\dsize{\prf'} \leq \dsize{\prf}$.
\end{proof}

To transform proofs `down' the partial order of an upward space $\upspace{\calc}$ into a proof of $\calc$ requires an additional condition, namely, the initial and reachability rules of $\calc$ must satisfy a certain set of equations. This gives rise to the notion of an \emph{explicit calculus}. Intuitively, an explicit calculus is one where all initial and reachability rules are parameterized with minimal constraints, i.e. constraints $\con$ such that $\csize{\con} = 0$. This has the effect that if a proof utilizes Horn rules, then such rules cannot be eliminated 
as they cannot be `mimicked' by other rules of the calculus.

\begin{definition}[Explicit Calculus]\label{def:explicit-calc} Let $\calc = (\gspace{\etypset},\ops)$ be an abstract calculus. We define $\calc$ to be \emph{explicit} \iffi for every initial rule $\id$ and reachability rule $\rru$ in $\ops$: $\id \dg \gram(\calc) = \id$ and $\rru \dg \gram(\calc) = \rru.$
\end{definition}

\begin{theorem}[Down the Upward Space]\label{thm:down-upward-lattice}
Let $\calc = (\gspace{\etypset},\ops)$ be an explicit calculus with $\upspace{\calc} = (\cs,\uprel)$ its upward space. For any abstract calculus $\calcii \in \cs$ and proof $\prf$ of a g-sequent $\gseq$ in $\calcii$, there exists a proof $\prf'$ of $\gseq$ in $\calc$ such that $\prf'$ is computable from $\prf$ in $\mathrm{PTIME}$ with $\dsize{\prf'} = \mathcal{O}(\dsize{\prf}^{2})$.
\end{theorem}

\begin{proof} Let $\calc$ be an explicit calculus, $\upspace{\calc} = (\cs,\uprel)$ its upward space, and $\calcii \in \cs$. By \thm~\ref{thm:upspace-is-lattice}, we know that $\calc \uprel \calcii$, and hence, there exists a set $\hrus'$ of Horn rules such that $\hrus' \subseteq \hrus = \hrus(\calc)$ is an anti-fracturable subset in $\depgr(\hrus)$ and which witnesses that the relation $\calc \uprel \calcii$ holds.

Let us first consider any initial rule $i(\inc{\con},\seqrel)$ or reachability rule $r(\inc{\con}',\seqrel)$ in $\calcii$. We know that each rule was obtained from an initial rule $\id$ and reachability rule $r(\con',\seqrel)$ in $\calc$ such that $\inc{\con} = \con \ug \gram(\hrus')$ and $\inc{\con}' = \con' \ug \gram(\hrus')$. Since $\calc$ is explicit, we know that $\gram(\rui) \cap \gram(\hrus') = \emptyset$ for $\rui \in \{\id,r(\con',\seqrel)\}$, and thus, by \lem~\ref{lem:abs-frc-inverse} $(\rui \ug \gram(\hrus')) \dg \gram(\hrus') = \rui$. This fact is required to complete the remainder of the proof.

Let us now suppose that we have a proof $\prf$ of $\gseq$ in $\calcii$. By \thm~\ref{thm:id-to-id-and-horn} and~\ref{thm:rru-to-rru-and-horn}, we can replace every application of an initial rule $i(\inc{\con},\seqrel)$ and reachability rule $r(\inc{\con}',\seqrel)$ from $\calcii$ by a derivation from $\fby^{1} i(\inc{\con},\seqrel) \dg \gram(\hrus') \fby^{0} \hrus'$ and $\fby^{0} \wk \fby^{1} r(\inc{\con}',\seqrel) \dg \gram(\hrus') \fby^{0} \hrus'$ as $\hrus'$ is a fracturable subset of $\hrus'$ by \lem~\ref{lem:(anti)-frac-set-props}. As mentioned above, $i(\inc{\con},\seqrel) \dg \gram(\hrus')$ and $r(\inc{\con}',\seqrel) \dg \gram(\hrus')$ are initial and reachability rules in $\calc$. Furthermore, by \lem ~\ref{lem:wk-admiss}, we know that $\wk$ is 
 admissible, and thus, by eliminating all occurrences of $\wk$, we obtain a proof $\prf'$ that is a proof in $\calc$.

The above yields a $\mathrm{PTIME}$ algorithm for transforming $\prf$ into $\prf'$, and in the worst-case
$$
\max\{\seqsize{\gseq} \ | \ \gseq \in \prf'\} = \mathcal{O}(\max\{\seqsize{\gseq} \ | \ \gseq \in \prf\}^{2})
$$
 as applications of $\wk$ could transform g-sequents from $\prf$ into at most complete graphs with quadratically more edges. Moreover, for each application of a rule in $\prf$, we could have (in the worst-case) at most quadratically many applications of rules in $\prf'$ as Horn rules can only be applied at most quadratically many times (by what was said above) until the edges from a g-sequent are completely removed. Therefore, $\dsize{\prf'} = \mathcal{O}(\dsize{\prf}^{2})$.
\end{proof}

\subsection{Downward Spaces/Lattices and Implicit Calculi}

Above, we investigated the upward spaces of calculi obtained by strengthening initial and reachability rules via absorption. Conversely, we obtain \emph{downward spaces} by weakening initial and reachability rules via fracturing.

\begin{definition}[Downward Space]\label{def:downward-space} Let $\calc = (\gspace{\etypset},\ops)$ be an abstract calculus. We define the \emph{downward space} $\downspace{\calc} = (\cs,\downrel)$ inductively as follows: (1) $\calc \in \cs$, and (2) if $\calcii \in \cs$ with $\hrus := \hrus(\calcii)$ and $\depgr(\gram) = (\depgrv, \depgre)$ the dependency graph of $\gram := \gram(\calcii \setminus \hrus)$, then for a fracturable subset $\depgrv' \subseteq \depgrv$, we have that $\calciii \downrel \calcii$ with $\calciii = \big(( \calcii \dg \gram(\depgrv')) \cup \hrus(\gram(\depgrv'))\big) \in \cs$.
\end{definition}

 As with upward spaces, we obtain that downward spaces are partially ordered sets, which can be viewed as lattices.

\begin{theorem}\label{thm:downspace-is-lattice}
Let $\calc = (\gspace{\etypset},\ops)$ be an abstract calculus with $\downspace{\calc} = (\cs,\downrel)$ its downward space. Then, the $\downrel$ relation is a connected partial order. Moreover, for $\calcii, \calciii \in \cs$, if we take $\calcii \wedge \calciii = \inf\{\calcii,\calciii\}$ and $\calcii \vee \calciii = \sup\{\calcii,\calciii\}$ under $\downrel$, then $(\cs,\wedge,\vee)$ is a complete lattice with $\calc = \top$.
\end{theorem}

\begin{proof} Let $\calc = (\gspace{\etypset},\ops)$ be an abstract calculus with $\downspace{\calc} = (\cs,\downrel)$ its downward space. We first prove that $\downrel$ is a partial order. By \lem~\ref{lem:(anti)-frac-set-props}, we know that $\emptyset$ is a fracturable subset in $\depgr(\gram(\calcii \setminus \hrus))$ with $\calcii$ an abstract calculus and $\hrus := \hrus(\calcii)$, implying that for any $\calcii \in \cs$, $\calcii = (\calcii \dg \emptyset) \cup \hrus(\emptyset)$, showing that $\calcii \downrel \calcii$, i.e. $\downrel$ is reflexive. 

 Second, assume that $\calcii \downrel \calciii$ and $\calciii \downrel \calciv$ for $\calcii, \calciii, \calciv \in \cs$. Then, there exist fracturable subsets $P_{1}'$ of $\gram(\calciii \setminus \hrus_{1})$ with $\hrus_{1} := \hrus(\calciii)$ and $P_{2}'$ of $\gram(\calciv \setminus \hrus_{2})$ with $\hrus_{2} := \hrus(\calciv)$ such that $\calcii = (\calciii \dg \gram(P_{1}')) \cup \hrus(P_{1}')$ and $\calciii = (\calciv \dg \gram(P_{2}')) \cup \hrus(P_{2}')$. Observe that $\calcii = (\calciv \dg \gram(P_{1}' \cup P_{2}')) \cup \hrus(P_{1}' \cup P_{2}')$. Therefore, to complete the proof of the case, we need to show that $P_{1}' \cup P_{2}'$ is fracturable. We know that $P_{1}' \subseteq \gram(\calciii \setminus \hrus_{1}) \subseteq \gram(\calciv \setminus \hrus_{2})$ as $\gram(\calciii \setminus \hrus_{1}) = \gram(\calciv \setminus \hrus_{2}) \setminus \gram(P_{2}')$. Since $P_{2}'$ is fracturable in $\gram(\calciv \setminus \hrus_{2})$ and $P_{1}' \subseteq \gram(\calciv \setminus \hrus_{2})$, we know that $P_{2}'$ does not depend on any pair in $P_{1}'$ in the dependency graph of $\gram(\calciv \setminus \hrus_{2})$. Furthermore, a pair in $P_{1}'$ may depend on a pair in $P_{2}'$, but cannot depend on any pair in $\gram(\calciv \setminus \hrus_{2}) \setminus \gram(P_{2}')$. Thus, $P_{1}' \cup P_{2}'$ is a fracturable subset of $\gram(\calciv \setminus \hrus_{2})$, showing that $\calcii \downrel \calciv$, i.e. $\downrel$ is transitive.
 
Third, let us assume that $\calcii \downrel \calciii$ and $\calciii \downrel \calcii$ for $\calcii, \calciii \in \cs$. Then, there exist fracturable subsets $P_{1}'$ of $\gram(\calcii \setminus \hrus_{1})$ with $\hrus_{1} := \hrus(\calcii)$ and $P_{2}'$ of $\gram(\calciii \setminus \hrus_{2})$ with $\hrus_{2} := \hrus(\calciii)$ such that $\calciii = (\calcii \dg \gram(P_{1}')) \cup \hrus(P_{1}')$ and $\calcii = (\calciii \dg \gram(P_{2}')) \cup \hrus(P_{2}')$. Plugging either equation into the other shows that $P_{1}' = \emptyset = P_{2}'$, implying that $\calciii = \calcii$.
 
Finally, one can show that for any $\calcii \in \cs$, $\calcii \downrel \calc$, that is, $\downspace{\calc}$ is connected. Moreover, it is straightforward to argue that $(\cs,\wedge,\vee)$ forms a complete lattice with $\calc$ the top element.
\end{proof}

 Given an abstract calculus $\calc$, we can translate proofs `down' the partial order of $\downspace{\calc}$ in $\mathrm{PTIME}$, similar to \thm~\ref{thm:down-upward-lattice}.
 
\begin{theorem}[Down the Downward Space]\label{thm:down-downward-lattice}
Let $\calc = (\gspace{\etypset},\ops)$ be a calculus with $\downspace{\calc} = (\cs,\downrel)$ its downward space. For any abstract calculus $\calcii \in \cs$ and proof $\prf$ of a g-sequent $\gseq$ in $\calc$, there exists a proof $\prf'$ of $\gseq$ in $\calcii$ such that $\prf'$ is computable from $\prf$ in $\mathrm{PTIME}$ with $\dsize{\prf'} = \mathcal{O}(\dsize{\prf}^{2})$.
\end{theorem}

\begin{proof} Let $\calc$ be an abstract calculus, $\downspace{\calc} = (\cs,\downrel)$ its downward space, and $\calcii \in \cs$. By \thm~\ref{thm:downspace-is-lattice}, we know that $\calcii \downrel \calc$ and hence there exists a fracturable subset $\depgrv'$ in the dependency graph $\depgr(\gram(\calc \setminus \hrus)) = (\depgrv, \depgre)$ with $\hrus := \hrus(\calc)$, witnessing that the relation $\calcii \downrel \calc$ holds. 

Let us suppose that we have a proof $\prf$ of $\gseq$ in $\calc$. By \thm~\ref{thm:id-to-id-and-horn} and~\ref{thm:rru-to-rru-and-horn}, we can replace every application of an initial rule $\id$ and reachability rule $\rru$ from $\calc$ by a derivation from $\fby^{1} \id \dg \gram(P') \fby^{0} \hrus(\gram(P'))$ and $\fby^{0} \wk \fby^{1} \rru \dg \gram(P') \fby^{0} \hrus(\gram(P'))$. By \lem ~\ref{lem:wk-admiss}, we know that $\wk$ is 
 admissible, and thus, by eliminating all occurrences of $\wk$, we obtain a proof $\prf'$ that is a proof in $\calcii$.

The above yields a $\mathrm{PTIME}$ algorithm for transforming $\prf$ into $\prf'$, and a complexity analysis similar to the one in \thm~\ref{thm:down-upward-lattice} shows that $\dsize{\prf'} = \mathcal{O}(\dsize{\prf}^{2})$.
\end{proof}
 
 We find that transforming proofs `up' the partial order of a downward space $\downspace{\calc}$ requires that $\calc$ is of a specific form. Namely, we find that such proofs can be transformed when $\calc$ is an \emph{implicit calculus}, defined below.


\begin{definition}[Implicit Calculus]\label{def:implicit-calc} Let $\calc = (\gspace{\etypset},\ops)$ be an abstract calculus. We define $\calc$ to be \emph{implicit} \iffi $\calc$ does not contain any Horn rules, and every initial rule $\id$ and reachability rule $\rru$ in $\ops$, satisfies the following equations: $\id \ug \gram(\calc) = \id$ and
$\rru \ug \gram(\calc) = \rru$.
\end{definition}

\begin{theorem}[Up the Downward Space]\label{thm:up-downward-lattice}
Let $\calc = (\gspace{\etypset},\ops)$ be an implicit calculus with $\downspace{\calc} = (\cs,\downrel)$ its downward space. For any abstract calculus $\calcii \in \cs$ and proof $\prf$ of a g-sequent $\gseq$ in $\calcii$, there exists a proof $\prf'$ of $\gseq$ in $\calc$ such that $\prf'$ is computable from $\prf$ in $\mathrm{PTIME}$ with $\dsize{\prf'} \leq \dsize{\prf}$.
\end{theorem}

\begin{proof} Let $\calc$ be an implicit calculus, $\downspace{\calc} = (\cs,\downrel)$ its downward space, and $\calcii \in \cs$. By \thm~\ref{thm:downspace-is-lattice}, we know that $\calcii \downrel \calc$, and hence, there exists a fracturable subset $P'$ of $\depgr(\gram(\calc))$ (note that $\hrus(\calc) = \emptyset$ as $\calc$ is implicit), which witnesses that the relation $\calcii \downrel \calc$ holds. 

Let us first consider any initial rule $i(\dec{\con},\seqrel)$ or reachability rule $r(\dec{\con}',\seqrel)$ in $\calcii$. We know that each rule was obtained from an initial rule $\id$ and reachability rule $r(\con',\seqrel)$ in $\calc$ such that $\dec{\con} = \con \dg \gram(P')$ and $\dec{\con}' = \con' \dg \gram(P')$. Since $\calc$ is implicit, we know that $\gram(P') \subseteq \gram(\rui)$ for $\rui \in \{\id,r(\con',\seqrel)\}$, and thus, by \lem~\ref{lem:abs-frc-inverse} and the fact that $\gram(P') = \gram(\hrus(P'))$, we have $(\rui \dg \gram(P')) \ug \gram(P') = \rui$. This fact is required to complete the remainder of the proof.

Let us take a proof $\prf$ of a g-sequent $\gseq$ in $\calcii$ and consider the topmost application of a Horn rule in $\hrus' := \hrus(\gram(P'))$. We assume w.l.o.g. that the Horn rule is $\fhru$ and argue by induction on the quantity of the proof that $\fhru$ can be permuted upward and eliminated entirely.

\textit{Base case.} Suppose we have an application of $i(\dec{\con},\seqrel)$ followed by an application of $\fhru$. By \lem~\ref{lem:basic-ordered-facts}, we know that $\fby^{1} i(\dec{\con},\seqrel) \fby^{1} \fhru \simul \fby^{1} i(\dec{\con},\seqrel) \fby^{0} \fhru$, and by \thm~\ref{thm:id-fby-H-sim-id}, we know that $\fby^{1} i(\dec{\con},\seqrel) \fby^{0} \fhru \simul i(\dec{\con},\seqrel) \oplus \gram(\fhru)$. By \lem~\ref{lem:ud-dg-properties}, $i(\dec{\con},\seqrel) \oplus \gram(\fhru) \simul i(\dec{\con},\seqrel) \ug \gram(\hrus')$ since $\fhru \in \hrus'$. 
As simulations are transitive (\lem~\ref{lem:simulation-props}), we know that $\fby^{1} i(\dec{\con},\seqrel) \fby^{1} \fhru \simul i(\dec{\con},\seqrel) \ug \gram(\hrus')$. We can therefore replace the application of $i(\dec{\con},\seqrel)$ followed by $\fhru$ by a single application of $i(\dec{\con},\seqrel) \ug \gram(\hrus')$, which is an initial rule in $\calc$ since $(\rui \dg \gram(P')) \ug \gram(P') = \rui$ for every initial rule $\rui$ in $\calc$.

\textit{Inductive step.} Suppose we have an application of a local, expansion, or reachability $\rui$ in $\calcii$ followed by an application of $\fhru$. In the case that $\rui$ is a reachability rule, replace $r(\dec{\con},\seqrel)$ by $r(\dec{\con},\seqrel) \ug \gram(\hrus') = \rru$, which is a reachability rule in $\calc$ since $(\rui \dg \gram(P')) \ug \gram(P') = \rui$ for every reachability rule $\rui$ in $\calc$. By \thm~\ref{thm:local-horn-perm} (local case), \thm~\ref{thm:expansion-horn-perm} (expansion case), and \thm~\ref{thm:reach-horn-perm} (reachability case), we know that $\fhru$ can be permuted above $\rui$, and thus by the induction hypothesis, $\fhru$ is eliminable from the proof altogether. Note that we need not consider permutations of $\fhru$ above Horn rules as $\fhru$ is a topmost application of a Horn rule. 

We repeat the above algorithm by successively considering each topmost occurrence of a Horn rule in $\hrus'$ until all such rules are eliminated, giving a $\mathrm{PTIME}$ procedure for computing the output proof $\prf'$ of $\calc$. Moreover, we have that $\max\{\seqsize{\gseq} \ | \ \gseq \in \prf'\} \leq \max\{\seqsize{\gseq} \ | \ \gseq \in \prf\}$ since the elimination of Horn rules removes edges from g-sequents in $\prf$, giving potentially smaller g-sequents in $\prf'$. Additionally, $|\{\gseq \in \gspace{\etypset} \ | \ \gseq \in \prf'\}| \leq |\{\gseq \in \gspace{\etypset} \ | \ \gseq \in \prf\}|$ since Horn rules are `absorbed' into the initial rules. Thus, we have that $\dsize{\prf'} \leq \dsize{\prf}$.
\end{proof}

Beyond their use in the theorem above, another interesting feature of implicit calculi is that every \emph{complete proof} employs only g-sequents of a polytree shape.\footnote{See \sect~\ref{sec:abstract-calc} for the definition of complete proofs and polytree g-sequents.} Such calculi are reminiscent of nested sequent systems~\cite{Bul92,Kas94}, and later on, we will identify a number of reachability nested systems appearing in the literature with implicit calculi.
 
\begin{theorem}\label{thm:implicit-polytrees}
If $\calc = (\gspace{\etypset},\ops)$ is an implicit calculus, then every complete proof is a polytree proof.
\end{theorem}

\begin{proof} Consider a complete proof of a g-sequent $\sar w : \seq$ in an implicit calculus $\calc$. Observe that $\sar w : \seq$ is a polytree g-sequent, and any bottom-up application of $\lru$, $\eru$, or $\rru$ will yield polytree g-sequents as the premises, so the entire proof will consist of polytree g-sequents.
\end{proof}

\subsection{Generic Calculus Transformations}

\begin{figure}[t]
\small

\begin{center}
\begin{minipage}{.49\textwidth}
\probbox{
	\textbf{Algorithm:} $\ialg(\calc)$\\
	\textbf{Input:} An explicit calculus $\calc$.\\
	\textbf{Output:} The upward space $\upspace{\calc}$.\\[1ex]
\textbf{Set} $\upspace{\calc} := (\cs,\uprel)$, $\cs := \{\calc\}$, and $\uprel \ := \emptyset$;\\
\textbf{While} $\upspace{\calc}$ grows;\\
\hbox{}\hspace{2mm}\textbf{For each} $\uprel$-maximal element $\calcii$ of $\upspace{\calc}$; \\
\hbox{}\hspace{4mm}\textbf{For each} anti-frac. 
subset $\hrus'$ in $\depgr(\hrus(\calcii))$;\\
\hbox{}\hspace{6mm}\textbf{Add} $f(\calcii,\hrus')$ \textbf{to} $\cs$ and $\calcii \uprel f(\calcii,\hrus')$ \textbf{to} $\uprel$;\\
\textbf{Return} $\upspace{\calc}$.
}
\caption{The $\ialg$ algorithm takes an explicit calculus as input and computes its upward space.\label{fig:ialg}}

\end{minipage} \hfill
\begin{minipage}{0.49\textwidth}
\small
\probbox{
	\textbf{Algorithm:} $\ealg(\calc)$\\
	\textbf{Input:} An implicit calculus $\calc$.\\
	\textbf{Output:} The downward space $\downspace{\calc}$.\\[1ex]
\textbf{Set} $\downspace{\calc} := (\cs,\downrel)$, $\cs := \{\calc\}$, and $\downrel \ := \emptyset$;\\
\textbf{While} $\downspace{\calc}$ grows;\\
\hbox{}\hspace{2mm}\textbf{For each} $\downrel$-minimal element $\calcii$ of $\downspace{\calc}$;\\
\hbox{}\hspace{4mm}\textbf{For each} fracturable $P$ in $\depgr$$\big(\gram(\calcii \setminus \hrus(\calcii))\big)$;\\
\hbox{}\hspace{6mm}\textbf{Add} $g(\calcii,P)$ \textbf{to} $\cs$ and $g(\calcii,P) \downrel \calcii$ \textbf{to} $\downrel$;\\
\textbf{Return} $\downspace{\calc}$.
}
\caption{The $\ealg$ algorithm takes an implicit calculus as input and computes its downward space.\label{fig:ealg}}
\end{minipage}
\end{center}
\end{figure}

Our framework yields a new discovery, namely, 
certain abstract calculi participate in lattices of polynomially equivalent calculi. These lattices can be identified by transforming an explicit calculus into its upward space, or by transforming an implicit calculus into its downward space. We provide two calculus transformation algorithms $\ialg$ and $\ealg$, which take an abstract calculus as input, and compute its upward or downward space, effectively generating a lattice of polynomially equivalent calculi. To state these algorithms we employ the following notation:
 
\begin{definition}\label{def:f-and-g-functions} Let $\calc$ be an abstract calculus with $\hrus$ a set of Horn rules and $P$ a set of production pairs. We define: 
$$
f(\calc,\hrus) := (\calc \ug \gram(\hrus)) \setminus \hrus
 \quad
g(\calc,P) := (\calc \dg \gram(P)) \cup \hrus(P)
$$
\end{definition}
 
Our first calculus transformation algorithm $\ialg$ is presented in \fig~\ref{fig:ialg}. The algorithm is named `$\ialg$' as it successively computes better approximations of the implicit calculus $\calcii$ that is polynomially equivalent to the input. It is straightforward to verify that $\ialg$ terminates as every execution of the while-loop strictly reduces the finite set of Horn rules associated with all $\uprel$-maximal calculi. The following relies on Theorems~\ref{thm:up-upward-lattice} and~\ref{thm:down-upward-lattice}, and we remark that $\top$ can be obtained from $\calc$ in $\mathrm{PTIME}$ by computing $\top := f(\calc,\hrus(\calc))$.

\begin{theorem}\label{thm:main-explicit-thm}
Let $\calc = (\gspace{\etypset},\ops)$ be an explicit calculus with $\upspace{\calc} = (\cs,\uprel)$ its upward space. Then,
\begin{enumerate}
\item \label{itm:i-main-explicit-thm} $\upspace{\calc} = \ialg(\calc)$;
\item \label{itm:ii-main-explicit-thm} $\upspace{\calc}$ is computable from $\calc$ in $\mathrm{EXPTIME}$;

\item \label{itm:iii-main-explicit-thm} If $\calcii, \calciii \in \cs$, then $\calcii \pdeq \calciii$;\footnote{Recall that the relation $\pdeq$ denotes the polynomial equivalence between two abstract calculi as defined 
on p.~\pageref{def:poly-equiv}.}

\item \label{itm:iv-main-explicit-thm} $\top$ is computable from $\calc$ in $\mathrm{PTIME}$;

\item \label{itm:v-main-explicit-thm} $\top$ is the only implicit calculus in $\upspace{\calc}$.


\end{enumerate}
\end{theorem}

\begin{proof} As (1) is straightforward to confirm, we argue claims (\ref{itm:ii-main-explicit-thm})--(\ref{itm:v-main-explicit-thm}) in turn:

\smallskip
\noindent
(\ref{itm:ii-main-explicit-thm}) During the execution of $\ialg$ the algorithm considers all anti-fracturable subsets of any calculus $\calcii \in \cs$, which may be exponential in the size of the input. Therefore, $\ialg$ has a worst-case complexity of $\mathrm{EXPTIME}$.

\smallskip
\noindent
(\ref{itm:iii-main-explicit-thm}) Suppose $\calcii, \calciii \in \cs$. Then, by Theorems~\ref{thm:up-upward-lattice} and~\ref{thm:down-upward-lattice}, we know that $\calcii \pdeq \calc$ and $\calc \pdeq \calciii$, which implies that $\calcii \pdeq \calciii$.

\smallskip
\noindent
(\ref{itm:iv-main-explicit-thm})
By taking $\hrus := \hrus(\calc)$ and defining $\calcii := f(\calc,\hrus)$, we obtain the calculus $\top$. This procedure can be performed in $\mathrm{PTIME}$ in the size of $\calc$.

\smallskip
\noindent
(\ref{itm:v-main-explicit-thm})
Since $\top$ is free of Horn rules and $\calc$ is explicit, $\top$ satisfies the properties of an implicit calculus by \dfn~\ref{def:implicit-calc}. Moreover, for every calculus $\calcii \in \cs$ such that $\calcii \neq \top$ and $\calcii \uprel \top$, it must be the case that $\calcii$ contains Horn rules by the definition of $\uprel$.
\end{proof}

 Our second calculus transformation algorithm $\ealg$ is displayed in \fig~\ref{fig:ealg}. The algorithm is named `$\ealg$' since it successively computes better approximations of the explicit calculus polynomially equivalent to the input. We note that $\ealg$ depends on sets of production pairs, and observe that $\ealg$ terminates since each grammar $\gram(\calcii \setminus \hrus(\calcii))$ strictly decreases for each $\downrel$-minimal calculus $\calcii$ after each execution of the while-loop.

 The following theorem is similar to \thm~\ref{thm:main-explicit-thm}, but relies on \thm~\ref{thm:down-downward-lattice} and~\ref{thm:up-downward-lattice} to establish the polynomial equivalence of all abstract calculi in the downward space. Moreover, we remark that $\bot$ can be obtained from the input $\calc$ in $\mathrm{PTIME}$ by computing $\bot := g(\calc,P(\calc))$. 

\begin{theorem}\label{thm:main-implicit-thm}
Let $\calc = (\gspace{\etypset},\ops)$ be an implicit calculus with $\downspace{\calc} = (\cs,\downrel)$ its downward space. Then,
\begin{enumerate}
\item $\downspace{\calc} = \ealg(\calc)$;

\item $\downspace{\calc}$ is computable from $\calc$ in $\mathrm{EXPTIME}$;

\item If $\calcii, \calciii \in \cs$, then $\calcii \pdeq \calciii$;

\item $\bot$ is computable from $\calc$ in $\mathrm{PTIME}$;

\item $\bot$ is the only explicit calculus in $\downspace{\calc}$.


\end{enumerate}
\end{theorem}

\begin{figure}[t]
\begin{center}
\begin{tikzpicture}[
		Dotted/.style={
			dash pattern=on 0.1\pgflinewidth off #1\pgflinewidth,line cap=round,
			shorten >=#1\pgflinewidth/2,shorten <=#1\pgflinewidth/2},
			box/.style = {draw,inner sep=.5pt,rounded corners=5pt},
		Dotted/.default=3]

\node[] [] (ntop) [] {$\calcii$};

\node[] [] (n1) [below=of ntop, xshift=-1.25cm,yshift=.25cm] {$g(\calcii,\!P_{2})$};
\node[] [] (n2) [below=of ntop, xshift=1.25cm,yshift=.25cm] {$g(\calcii,\!P_{1})$};
\draw[-,color=black] (ntop) -- (n1) node [] { };
\draw[-,color=black] (ntop) -- (n2) node [] { }; 

\node[] [] (n11) [below=of ntop,yshift=-1.00cm] {$g(\calcii,\!P_{1} {\cup} P_{2})$};
\draw[-,color=black] (n11) -- (n1) node [] { };
\draw[-,color=black] (n11) -- (n2) node [] { }; 

\node[] [] (n) [below=of n11,yshift=.25cm] {$g(\calcii,\!P_{1} {\cup} P_{2} {\cup} P_{3})$};
\draw[-,color=black] (n11) -- (n) node [] { };

\node[] [] (ntopx) [left=of ntop,xshift=-4cm] {$f(\calc,\!\{h_{1},\!h_{2},\!h_{3}\})$};

\node[] [] (n1x) [below=of ntopx, xshift=-1.25cm,yshift=.25cm] {$f(\calc,\!\{h_{1},\!h_{3}\})$};
\node[] [] (n2x) [below=of ntopx, xshift=1.25cm,yshift=.25cm] {$f(\calc,\!\{h_{2},\!h_{3}\})$};
\draw[-,color=black] (ntopx) -- (n1x) node [] { };
\draw[-,color=black] (ntopx) -- (n2x) node [] { }; 

\node[] [] (n11x) [below=of ntopx,yshift=-1.00cm] {$f(\calc,\!\{h_{3}\})$};
\draw[-,color=black] (n11x) -- (n1x) node [] { };
\draw[-,color=black] (n11x) -- (n2x) node [] { }; 

\node[] [] (nx) [below=of n11x,yshift=.25cm] {$\calc$};
\draw[-,color=black] (n11x) -- (nx) node [] { };


\end{tikzpicture}
\caption{The lattice above left corresponds to 
$\ialg(\calc)$, where $\calc$ is defined as in Example~\ref{ex:exp-imp-example}. The lattice shown above right corresponds to a computation of $\ealg(\calcii)$, where $\calcii = f(\calc,\!\{h_{1},\!h_{2},\!h_{3}\})$.\label{fig:lattice}}
\end{center}
\end{figure}

\begin{example}\label{ex:exp-imp-example} To demonstrate the functionality of our calculus transformation algorithms, we consider an example with an explicit calculus $\calc$ consisting of the following rules, and where $\csize{\con} = 0$.
\begin{center}
\begin{tabular}{c @{\hskip .5em} c @{\hskip .5em} c @{\hskip .5em} c}
\AxiomC{$\phantom{\gseq}$}
\RightLabel{$\id$}
\UnaryInfC{$\ant \sar \suc$}
\DisplayProof

&

\AxiomC{$\ant,\, w \Ec u,\, u \Ec v,\, w \Ea v \sar \suc$}
\RightLabel{$h_{1}$}
\UnaryInfC{$\ant,\, w \Ec u,\, u \Ec v \sar \suc$\vspacer}
\DisplayProof

&

\AxiomC{$\ant,\, w \Ea w \sar \suc$}
\RightLabel{$h_{2}$}
\UnaryInfC{$\ant\sar \suc$}
\DisplayProof

&

\AxiomC{$\ant,\, w \Ea u,\, u \Eb w \sar \suc$}
\RightLabel{$h_{3}$}
\UnaryInfC{$\ant,\, w \Ea u \sar \suc$}
\DisplayProof
\end{tabular}
\end{center}
Let $\hrus = \{h_{1},h_{2},h_{3}\}$ and observe that in the dependency graph $\depgr(\hrus) = (\hrus,\depgre)$, $h_{3} \depgre h_{1}$ and $h_{3} \depgre h_{2}$. In this setting, $\ialg(\calc)$ constructs the lattice shown on the left of \fig~\ref{fig:lattice}, eventually yielding the implicit calculus $f(\calc,\{h_{1},h_{2},h_{3}\}) = \calcii$ at the top. Furthermore, if we let $P(\hrus) = P_{1} \cup P_{2} \cup P_{3}$ such that $P_{i} = P(h_{i}) = \{(\pru_{i},\conv{\pru}_{i})\}$, then in $\depgr(P) = (P,\depgre')$, $(\pru_{3},\conv{\pru}_{3}) \depgre' (\pru_{1},\conv{\pru}_{1})$ and $(\pru_{3},\conv{\pru}_{3}) \depgre' (\pru_{2},\conv{\pru}_{2})$. If we run $\ialg(\calcii)$, we obtain the lattice shown on the right of \fig~\ref{fig:lattice}.
\end{example}

\begin{theorem}\label{thm:up-down-iso}
Let $\calc$ be an explicit calculus, $\upspace{\calc} = (\cs,\uprel)$, $\calcii$ be an implicit calculus, and $\downspace{\calcii} = (\cs',\downrel)$. If either $\calcii \in \cs$ or $\calc \in \cs'$, then $\cs = \cs'$, $\upspace{\calc} \cong \downspace{\calcii}$, and for any $\calciii, \calciv \in \cs = \cs'$, $\gram(\calciii) = \gram(\calciv)$.
\end{theorem}

\begin{proof} Assume w.l.o.g. that $\calcii \in \cs$. As $\upspace{\calc}$ contains only one implicit calculus (\thm~\ref{thm:main-explicit-thm}) we know that $\calcii$ is the top element of $\upspace{\calc}$. Let $\hrus := \hrus(\calc)$, and observe that for any anti-fracturable subset $\hrus'$ of $\hrus$, $f(\calc,\hrus') = g(\calcii,P(\hrus \setminus \hrus'))$. We define $h(f(\calc, \hrus')) = g(\calcii,P(\hrus \setminus \hrus'))$, and note that $h$ serves as an isomorphism between $\upspace{\calc}$ and $\downspace{\calcii}$, meaning $\upspace{\calc} \cong \downspace{\calcii}$. The fact that for any $\calciii, \calciv \in \cs = \cs'$, $\gram(\calciii) = \gram(\calciv)$ follows from the fact that all Horn rules of $\calc$ either occur explicitly in any calculus of $\upspace{\calc}$ or were absorbed into initial and reachability rules.
\end{proof}

\subsection{Discussion: Labeled Systems, Reachability Nested Systems, and Complexity}\label{subsec:lab-nest-dual}

Our abstract framework for the study of multisequent systems yields insights into the relationship between labeled sequent systems and reachability nested systems. A wide variety of labeled sequent systems, such as those mentioned in \fig~\ref{fig:logics-and-calculi}, serve as instances of explicit calculi. The reason being, if one formulates a `typical' labeled sequent system as an abstract calculus in our framework, then the initial and reachability rules will be parameterized with minimal constraints (i.e. constraints $\con$ such that $\csize{\con} = 0$), meaning, such rules will satisfy the equations stated in \dfn~\ref{def:explicit-calc}. This is due to the fact that labeled sequent systems are usually formulated so that structural/Horn reasoning is carried out with \emph{structural rules} that manipulate the edges of labeled sequents. This has the effect that initial and reachability rules do not require complex constraints that encode such reasoning as evidenced by Theorems~\ref{thm:id-to-id-and-horn} and~\ref{thm:rru-to-rru-and-horn}.

Alternatively, a large number of reachability nested systems 
(see \fig~\ref{fig:logics-and-calculi}) can be identified as implicit calculi. This arises from the fact that such systems employ nested sequents, which have the form of (poly)trees. As a consequence, such systems cannot include Horn rules since bottom-up applications of such rules will break the (poly)tree structure of nested sequents. This means that such systems must encode Horn reasoning in the constraints of their initial and reachability rules to ensure the superfluity of Horn rules as evidenced by \thm~\ref{thm:id-fby-H-sim-id}, Theorems~\ref{thm:local-horn-perm}--\ref{thm:reach-horn-perm}, and \thm~\ref{thm:frac-horn-perm}. This ensures that such calculi are sound and complete despite only allowing (poly)tree proofs (see \thm~\ref{thm:implicit-polytrees}).

Since many labeled sequent calculi can be identified as explicit calculi and many reachability nested calculi can be identified as implicit calculi, the $\ialg$ and $\ealg$ algorithms can be used to compute polynomially equivalent reachability nested and labeled sequent systems. This suggests that labeled sequent systems and reachability nested systems tend to come in pairs and that systems of each type tend to yield a `dual' system of the other type with labeled systems serving as bottom elements in lattices and reachability nested systems serving as top elements. 

Nevertheless, we do note that the classes of explicit calculi and implicit calculi are \emph{not} disjoint from one another. This can arise, e.g. when a calculus omits Horn rules and all of its initial and reachability rules utilize minimal constraints. As concrete examples, labeled sequent calculi can be formulated for the classical modal logic $\mathsf{K}$ and the intuitionistic modal logic $\mathsf{IK}$, which omit structural rules altogether (cf.~\cite{CiaLyoRamTiu21,Lyo21a}). As discussed previously~\cite{CiaLyoRamTiu21,Lyo21a}, such systems can be simultaneously viewed as nested sequent systems since they remain sound and complete when labeled sequents are restricted to a tree shape in proofs. Thus, systems sitting at the intersection of the explicit and implicit classes can be appropriately viewed as labeled sequent systems and reachability nested systems; the difference is purely based on the shape of multisequents allowed in proofs.

Moreover, notice that the proof transformation algorithms described in Theorems \ref{thm:up-upward-lattice}, \ref{thm:down-upward-lattice}, \ref{thm:down-downward-lattice}, and \ref{thm:up-downward-lattice} do not introduce any version of a cut rule when transforming proofs between different abstract calculi in a lattice. This shows that for any cut-free labeled or reachability nested system (which can be viewed as an abstract calculus in our framework) its respective dual reachability nested or labeled sequent system will also be cut-free, i.e. cut-admissibility is preserved via our generic proof transformation algorithms.

Regarding the proof complexity of labeled and reachability nested systems, our results establish a key point. Labeled proofs can be quadratically larger than their nested sequent counterparts (Theorems~\ref{thm:down-upward-lattice} and~\ref{thm:down-downward-lattice} ). That is, nested systems generally admit shorter proofs with syntactically simpler sequents, as shown in Theorems~\ref{thm:up-upward-lattice} and~\ref{thm:up-downward-lattice}. The underlying reason is that labeled calculi perform explicit Horn reasoning: relational properties must be established through Horn rule applications, increasing both the number of rule applications and the size of sequents. Consequently, nested proofs provide more compact witnesses of validity. Nevertheless, from the standpoint of proof‑search, this difference in proof size is not significant. The polynomial equivalence of labeled and nested calculi (Theorems~\ref{thm:main-explicit-thm} and~\ref{thm:main-implicit-thm}) ensures that proofs can be translated between the two systems with only polynomial overhead. Since the logics to which these systems are typically applied have decision problems beyond $\mathrm{PTIME}$, such polynomial factors do not alter their overall computational complexity.

\section{Instantiating the Abstract Formalism}\label{sec:example} 


In this section, we illustrate our generic calculus and proof transformation algorithms in a concrete setting. Our aim is twofold. First, we show how our abstract framework can be instantiated to obtain an ordinary labeled sequent calculus, making explicit how its rules, structural constraints, and sequent constraints arise as instances of our generic definitions. Second, we demonstrate how a reachability nested calculus can be automatically obtained via the $\ialg$ algorithm. For this purpose we work with the labeled sequent calculus $\gtsfour$ for the modal logic $\sfour$ (as formulated in~\cite{CiaLyoRamTiu21}). This calculus provides representatives of all inference rule types while remaining comparatively simple, making it an ideal example.

We will formalize $\gtsfour$ as a specific abstract calculus $(\gspace{\etypset}, \ops)$. This will be accomplished in a two-step process: first, we define the set $\gspace{\etypset}$ of labeled sequents that appear in $\gtsfour$. This is achieved by specifying the set $\etypset$ of edge types and the set $\seqset$ of sequents used to label vertices in a g-sequent. Second, we show how the various parameters used in rule types, namely, sequent constraints and structural constraints, can be instantiated to produce the set $\ops$ of rules for $\gtsfour$. Once $\gtsfour$ has been properly formalized, we show how the $\ialg$ algorithm can be used to compute a dual nested sequent calculus $\nsfour$ and give an example of inter-translatable proofs.

The modal language $\langsfour$ is the set of all formulae generated by the following grammar in BNF:
$$
\phi ::= p \mid \neg p \mid \phi \lor \phi \mid \phi \land \phi \mid \Box \phi \mid \dia \phi
$$
such that $p$ ranges over a set of (propositional) atoms. We use $p$, $q$, $r$, $\ldots$ to denote atoms and $\phi$, $\psi$, $\chi$, $\ldots$ to denote formulae. Furthermore, as this section is entirely concerned with exemplifying our abstract formalism and algorithms, we omit the semantics of $\sfour$ and assume that the reader is familiar; the interested reader can consult~\cite{BlaRijVen01} for background material on the logic.

Let us now define the set $\gspace{\etypset}$ of labeled sequents that appear in $\gtsfour$. Recall that a g-sequent is a labeled graph $\gseq = (\vset, \edgs, \lfunc)$ with $\edgs$ a family of typed edges and $\lfunc$ a labeling function mapping vertices in $\vset$ to sequents in $\seqset$ (see \dfn~\ref{def:g-sequent}). In the current setting, we take the set $\etypset := \{\etypa\}$ of edge types to be a singleton for two reasons: first, $\sfour$ admits a relational semantics with a \emph{single} accessibility relation, and second, the edges used in labeled sequents \emph{encode} accessibility relations (cf.~\cite{Sim94,Vig00}). Therefore, only a single type of edge is needed in our labeled sequents, which we will denote by $\E$ rather than $\Ea$ for simplicity.

Next, we define a \emph{sequent} to be an expression of the form $X$ such that $X \subseteq \langsfour$ is a (potentially empty) finite set of formulae. We take the set $\seqset := \set{X, Y, Z, \ldots}$ to be the set of all such sequents. Therefore, in the current example, each g-sequent $\gseq = (\vset, \edgs, \lfunc)$ is a labeled graph such that $\vset$ is a finite set of vertices, $\edgs \subseteq \vset \times \vset$, and $\lfunc$ maps each vertex to a set of formulae. We may equivalently write each such g-sequent $\gseq = (\vset, \edgs, \lfunc)$ in the notation $\ant \sar \suc$ such that $w \E u \in \ant$ \iffi $(w,u) \in \E$, and $w : X \in \suc$ \iffi $\lfunc(w) =  X$. We formally define a \emph{labeled sequent} to be a g-sequent of the above form and let $\gspace{\etypset}$ be the set of all such labeled sequents.


Let us now define the set $\ops$ of rules for the labeled sequent calculus $\gtsfour$. As with labeled sequents, we recast the rules of $\gtsfour$ in the format of our abstract framework to show how such rules can be viewed as instances of inference rule types. For a more traditional presentation of $\gtsfour$, the reader may consult~\cite{CiaLyoRamTiu21}. We will define the various rules of $\gtsfour$ in sequence, starting with the initial rules, then presenting the local rules, expansion rules, transmission rules, and finally, the Horn rules. Such rules will be obtained by using specific sequent constraints and structural constraints to instantiate inference rule types into concrete rules from $\gtsfour$.

\smallskip
\noindent
\textbf{Initial Rules.} The labeled sequent calculus $\gtsfour$ contains one initial rule $\init(\con,\seqrel)$, shown below right, which can be obtained by instantiating the inference rule type $\id$, shown below left. This is accomplished by taking the constraint $\con := (\{w\},\emptyset,\emptyset)$ and defining the sequent constraint $\seqrel \subseteq \seqset \times 2^{\pseqset}$ as: $(X_{1},\suc') \in \seqrel$ \iffi $p, \neg p \in X_{1}$ for some atom $p$. The context $\suc' = \suc \setminus \{w : X_{1}\}$ does not play a role in the sequent constraint, i.e. the rule is context independent. For an example of an instantiation of our framework that includes context dependent rules, see~\cite[Section~6]{LyoOstarxiv}.
\begin{center}
\begin{tabular}{c c}
\AxiomC{}
\RightLabel{$\id$}
\UnaryInfC{$\ant \sar \suc$}
\DisplayProof

&

\AxiomC{}
\RightLabel{$\init$}
\UnaryInfC{$\Sigma \sar \Pi, w : (X, p, \neg p)$}
\DisplayProof
\end{tabular}
\end{center}

\smallskip
\noindent
\textbf{Local Rules.} $\gtsfour$ contains two local rules: one for the $\lor$ connective and one for the $\land$ connective. We first explain how the conjunction rule $\conr$, shown below middle, can be obtained as an instance of the local rule type $\lru$, shown below left. First, we fix the number of premises to two, let the structural constraint $\con := (\{w\},\emptyset,\emptyset)$ and define the sequent constraint $\seqrel \subseteq \seqset \times \seqset \times \seqset \times 2^{\pseqset}$ as follows: $(X_{1},X_{2},X_{3},\suc') \in \seqrel$ \iffi (1) $\phi_{1} \in X_{1}$, (2) $\phi_{2} \in X_{2}$, (3) $X_{3} = (X_{1} \setminus \phi_{1}), (X_{2} \setminus \phi_{2}), \phi_{1} \land \phi_{2}$, and (4) $X_{1} \setminus \phi_{1} = X_{2} \setminus \phi_{2} = X_{3} \setminus \phi_{1} \land \phi_{2}$ for some $\phi_{1}, \phi_{2} \in \langsfour$. It is simple to verify that $R$ holds for $X_{1} = X, \phi_{1}$, $X_{2} = X, \phi_{2}$, $X_{3} = X, \phi_{1} \land \phi_{2}$, and arbitrary $\Delta$. The disjunction rule, shown below right, can be obtained from $\lru$ analogously.
\begin{center}
\begin{tabular}{c c c}
\AxiomC{$\{\,\ant \sar \suc,\, w : \seq_{i}\,\}_{i \in [n]}$}
\RightLabel{$\lru$}
\UnaryInfC{$\ant \sar \suc,\, w : \seq_{n+1}$}
\DisplayProof

&

\AxiomC{$\set{\Sigma \sar \Pi, w : (X, \phi_{i})}_{i \in [2]}$}
\RightLabel{$\conr$}
\UnaryInfC{$\Sigma \sar \Pi, w : (X, \phi_{1} \land \phi_{2})$}
\DisplayProof

&

\AxiomC{$\Sigma \sar \Pi, w : (X, \phi, \psi)$}
\RightLabel{$\disr$}
\UnaryInfC{$\Sigma \sar \Pi, w : (X, \phi \lor \psi)$}
\DisplayProof
\end{tabular}
\end{center}

\smallskip
\noindent
\textbf{Expansion Rules.} $\gtsfour$ contains one expansion rule $\boxr$, shown below right. The constraint family is determined by the definition of an expansion rule (see p.~\pageref{def:expansion-rule}). The $\boxr$ rule can be obtained as an instance of the expansion rule type $\eru$ shown below left by defining the sequent constraint $\seqrel \subseteq \seqset \times \seqset \times \seqset \times 2^{\pseqset}$ as: $(X_{1},X_{2},X_{3},\Delta') \in \seqrel$ \iffi $X_{3} = X_{1}, \Box \phi$ and $X_{2} = \phi$ for some $\phi \in \langsfour$.
\begin{center}
\begin{tabular}{c c}
\AxiomC{$\ant, \antii \sar \suc, w : \seq_{1}, u : \seq_{2}$}
\RightLabel{$\eru$}
\UnaryInfC{$\ant \sar \suc,\, w : \seq_{3}$}
\DisplayProof

&

\AxiomC{$\Sigma, w \E u \sar \Pi, w : X, u : \phi$}
\RightLabel{$\boxr$}
\UnaryInfC{$\Sigma \sar \Pi, w : (X, \Box \phi)$}
\DisplayProof
\end{tabular}
\end{center}

\smallskip
\noindent
\textbf{Transmission Rules.} The labeled sequent calculus $\gtsfour$ contains one transmission rule $\diar$, shown below right. (NB. Recall that transmission rules form a proper subclass of reachability rules.) This rule can be obtained as an instance of the transmission rule type $\tru$ shown below left by specifying the constraint $\con$ and sequent constraint $\seqrel$. To obtain the $\diar$ rule, we fix the number of premises to one and set $\con := (\{w,u\},\{(w,u)\},L)$ with $L(w,u) = \{a\}$. The sequent constraint $\seqrel \subseteq \seqset \times \seqset \times \seqset \times \seqset \times 2^{\pseqset}$ is defined as follows: $(X_{1}, Y_{1}, X_{2}, Y_{2},\Delta') \in \seqrel$ \iffi $\dia \phi \in X_{1} = X_{2}$ and $Y_{1} = Y_{2},\phi$ for some $\phi \in \langsfour$. Observe that $\seqrel(X_{1}, Y_{1}, X_{2}, Y_{2},\Delta')$ holds when (1) $X_{1} = X_{2} = X, \dia \phi$, (2) $Y_{1} = Y, \phi$, and (3) $Y_{2} = Y$, which is what is explicitly indicated in the rule below right.
\begin{center}
\begin{tabular}{c c}
\AxiomC{$\{\,\ant \sar \suc,\, w : \seq_{i},\, u : \seq_{i}'\,\}_{i \in [n]}$}
\RightLabel{$\tru$}
\UnaryInfC{$\ant \sar \suc,\, w : \seq_{n+1},\, u : \seq_{n+1}'$}
\DisplayProof

&

\AxiomC{$\Sigma \sar \Pi, w : (X, \dia \phi), u : (Y,\phi)$}
\RightLabel{$\diar$}
\UnaryInfC{$\Sigma \sar \Pi, w : (X, \dia \phi), u : Y$}
\DisplayProof
\end{tabular}
\end{center}

\smallskip
\noindent
\textbf{Horn Rules.} There are two Horn rules included in $\gtsfour$, namely, the reflexivity rule $\refl$ and the transitivity rule $\trans$, shown below left and right, respectively. The $\refl$ rule encodes the reflexivity condition and the $\trans$ rule encodes the transitivity condition imposed on $\sfour$ relational models. One can verify that $\refl$ serves as an instance of either a forward or backward Horn rule, while $\trans$ is an instance of a forward Horn rule.
\begin{center}
\begin{tabular}{c c}
\AxiomC{$\Sigma, w \E w \sar \Pi$}
\RightLabel{$\refl$}
\UnaryInfC{$\Sigma \sar \Pi$}
\DisplayProof

&

\AxiomC{$\Sigma, w \E u,\, u \E v,\, w \E v, \sar \Pi$}
\RightLabel{$\trans$}
\UnaryInfC{$\Sigma, w \E u,\, u \E v\sar \Pi$}
\DisplayProof
\end{tabular}
\end{center}

\begin{figure}[t]
\centering

\begin{minipage}{.48\columnwidth}
\AxiomC{}
\RightLabel{$\init$}
\UnaryInfC{$w \E u, u \E v, w \E v \sar w : \dia p, v : \neg p, v : p$}
\RightLabel{$\diar$}
\UnaryInfC{$w \E u, u \E v, w \E v \sar w : \dia p, v : \neg p$}
\RightLabel{$\trans$}
\UnaryInfC{$w \E u, u \E v \sar w : \dia p, v : \neg p$}
\RightLabel{$\boxr$}
\UnaryInfC{$w \E u \sar w : \dia p, u : \Box \neg p$}
\RightLabel{$\boxr$}
\UnaryInfC{$\sar w : (\Box \Box \neg p, \dia p)$}
\RightLabel{$\disr$}
\UnaryInfC{$\sar w : \Box \Box \neg p \lor \dia p$}
\DisplayProof
\end{minipage}
\hfill 
\begin{minipage}{.48\columnwidth}
\begin{tikzpicture}[
		Dotted/.style={
			dash pattern=on 0.1\pgflinewidth off #1\pgflinewidth,line cap=round,
			shorten >=#1\pgflinewidth/2,shorten <=#1\pgflinewidth/2},
			box/.style = {draw,inner sep=.5pt,rounded corners=5pt},
		Dotted/.default=3]

\node[] [] (ntop) [] {$\nsfour$};

\node[] [] (n1) [below=of ntop, xshift=0cm,yshift=.5cm] {$g(\nsfour, P(\{\refl\}))$};
\draw[-,color=black] (ntop) -- (n1) node [] { };

\node[] [] (n11) [below=of n1,yshift=.5cm] {$g(\nsfour, P(\{\refl,\trans\})) = \gtsfour$};
\draw[-,color=black] (n11) -- (n1) node [] { };


\node[] [] (ntopx) [left=of ntop,xshift=-.25cm] {$f(\gtsfour,\!\{\refl,\!\trans\}) = \nsfour$};

\node[] [] (n1x) [below=of ntopx, xshift=0cm,yshift=.5cm] {$f(\gtsfour, \{\trans\})$};
\draw[-,color=black] (ntopx) -- (n1x) node [] { };

\node[] [] (nx) [below=of n1x,yshift=.5cm] {$\gtsfour$};
\draw[-,color=black] (nx) -- (n1x) node [] { };


\end{tikzpicture}
\end{minipage}

\caption{An example proof in $\gtsfour$ of the $\mathsf{4}$ axiom $\Box \Box \neg p \lor \dia p$ is shown above left. The left lattice above corresponds to $\ialg(\gtsfour)$ and the right lattice above corresponds to the computation of $\ealg(\nsfour)$.\label{fig:lattice-g3iq}}
\end{figure}

We define the labeled sequent calculus $\gtsfour := (\gspace{\etypset}, \ops)$ with $\gspace{\etypset}$ the set of labeled sequents defined above and $\ops$ the set of rules defined above. This demonstrates that $\gtsfour$ can be viewed as a specific abstract calculus within our framework, and thus, our results in Sections~\ref{sec:perm-sim} and~\ref{sec:generic-algs} are applicable to $\gtsfour$. An example of a proof in $\gtsfour$ of the $\mathsf{4}$ axiom $\Box \Box \neg p \lor \dia p$ is displayed to the left in \fig~\ref{fig:lattice-g3iq} for the interested reader.

Observe that $\gram(\gtsfour) = \{a \pto \empstr, \conv{a} \pto \empstr, a \pto a a, \conv{a} \pto \conv{a} \conv{a}\}$ due to the inclusion of the rules $\refl$ and $\trans$ in $\gtsfour$. $\gtsfour$ is an explicit calculus because its initial and transmission rules satisfy the equations of an explicit calculus (see \dfn~\ref{def:explicit-calc}), namely, the following holds:
$$
\init \dg \gram(\gtsfour) = \init
\text{ and }
\diar \dg \gram(\gtsfour) = \diar
$$
By running the algorithm $\ialg(\gtsfour)$, we obtain the upward space $\upspace{\gtsfour}$, which is the left of the two lattices shown in \fig~\ref{fig:lattice-g3iq}. Since $\depgr(\hrus(\gtsfour)) = (\{\refl,\trans\},\depgre)$ with $\trans \depgre \refl$, the anti-fracturable subsets of $\hrus(\gtsfour) = \{\refl,\trans\}$ are $\emptyset$, $\{\trans\}$, and $\{\refl,\trans\}$, which gives rise to the structure of the upward space. By \thm~\ref{thm:main-explicit-thm}, we know that the top element $\top = f(\gtsfour,\!\{\refl,\!\trans\})$ is an implicit calculus. We name this implicit calculus $\nsfour$ and now discuss its definition.

By making use of absorption, we compute the reachability rule $\diarii := \diar \ug \gram(\gtsfour)$. The constraint associated with the $\diarii$ rule is of the form $\con := (\{w,u\},\{(w,u)\},L)$ with $L(w,u) = \gram'(a)$ and $\gram' = \gram(\gtsfour)$. The initial rule is unaffected by absorption due to the fact that the structural constraint $(\{w\},\emptyset,\emptyset)$ does not have any edges; consequently, $\init = \init \ug \gram(\gtsfour)$. The calculus $\nsfour := f(\gtsfour,\!\{\refl,\!\trans\})$ therefore uses the same rules as $\gtsfour$, but omits the $\dia$ rule, the Horn rules $\refl$ and $\trans$, and includes $\diarii$ instead, that is: $\nsfour = (\gtsfour \setminus \set{\diar,\refl,\trans}) \cup \set{\diarii}$.

By \thm~\ref{thm:implicit-polytrees} and the fact that $\nsfour$ is an implicit calculus, we know that every complete proof in the calculus is a polytree proof. In the case of $\nsfour$ however, it turns out that we obtain an even stronger claim, that is, every complete proof is a \emph{tree proof}. A tree proof is proof $\prf$ such that for every g-sequent $\gseq = (\vset, \edgs, \lfunc)$ occurring in the proof, the graph $(\vset, \edgs)$ is a tree. In the context of our example, we refer to a labeled sequent in the shape of a tree as a \emph{labeled tree sequent}. Recall that a complete proof is a proof ending with a g-sequent of the form $\sar w : \seq$; hence, in the setting of $\nsfour$, a complete proof ends with a labeled sequent of the form $\sar w : X$. If we consider bottom-up applications of rules from $\nsfour$, all rules either preserve the edge atoms in an inference or, in the case of the $\boxr$, introduce a single edge atom protruding forward to a fresh vertex. Therefore, only labeled sequents of a tree shape will participate in a complete proof. It is well known that labeled (poly)tree sequents are notational variants of nested sequents, and thus, if we restrict $\nsfour$ to only using labeled tree sequents in proofs, $\nsfour$ can be viewed as a reachability nested calculus for $\sfour$.

Because $\nsfour$ is an implicit calculus (see \dfn~\ref{def:implicit-calc}), we can run the algorithm $\ealg(\nsfour)$ with $\nsfour$ as input to compute the downward space $\downspace{\nsfour}$, which is the right of the two lattices in \fig~\ref{fig:lattice-g3iq}. Observe that $P(\nsfour) = P(\{\refl,\trans\}) = \{(a \pto \empstr, \conv{a} \pto \empstr), (a \pto a a, \conv{a} \pto \conv{a} \conv{a})\}$ due to the fact that $P(\{\refl\}) = \{(a \pto \empstr, \conv{a} \pto \empstr)\}$ and $P(\{\trans\}) = \{(a \pto a a, \conv{a} \pto \conv{a} \conv{a})\}$. Thus, in $\depgr(P(\nsfour)) = (P(\{\refl,\trans\}),\depgre')$, we have that $(a \pto a a, \conv{a} \pto \conv{a} \conv{a}) \depgre' (a \pto \empstr, \conv{a} \pto \empstr)$, showing that the fracturable subsets of $P(\{\refl,\trans\})$ are $\emptyset$, $P(\{\refl\})$, and $P(\{\refl,\trans\})$. This gives rise to the structure of the downward space. One may verify that the bottom element $\bot  = g(\nsfour, P(\{\refl,\trans\}))$ in $\downspace{\nsfour}$ is indeed the labeled sequent calculus $\gtsfour$. 

By making use of the generic proof transformation described in \thm~\ref{thm:up-upward-lattice} or~\ref{thm:up-downward-lattice}, the labeled sequent proof shown in \fig~\ref{fig:lattice-g3iq} can be transformed into the tree proof in $\nsfour$ shown below left. By making use of the proof transformation described in \thm~\ref{thm:down-upward-lattice} or~\ref{thm:down-downward-lattice}, we obtain the converse transformation. For those familiar with the nested sequent formalism (cf.~\cite{Bul92,Kas94}), observe that the tree proof shown below left can indeed be recast as the nested sequent proof as shown below right.
\begin{center}
\begin{tabular}{c @{\hskip 1cm} c}
\AxiomC{}
\RightLabel{$\init$}
\UnaryInfC{$w \E u, u \E v \sar w : \dia p, v : \neg p, v : p$}
\RightLabel{$\diarii$}
\UnaryInfC{$w \E u, u \E v \sar w : \dia p, v : \neg p$}
\RightLabel{$\boxr$}
\UnaryInfC{$w \E u \sar w : \dia p, u : \Box \neg p$}
\RightLabel{$\boxr$}
\UnaryInfC{$\sar w : (\Box \Box \neg p, \dia p)$}
\RightLabel{$\disr$}
\UnaryInfC{$\sar w : \Box \Box \neg p \lor \dia p$}
\DisplayProof

&

\AxiomC{}
\RightLabel{$\init$}
\UnaryInfC{$\dia p, [[\neg p, p]]$}
\RightLabel{$\diarii$}
\UnaryInfC{$\dia p, [[\neg p]]$}
\RightLabel{$\boxr$}
\UnaryInfC{$\dia p, [\Box \neg p]$}
\RightLabel{$\boxr$}
\UnaryInfC{$\Box \Box \neg p, \dia p$}
\RightLabel{$\disr$}
\UnaryInfC{$\Box \Box \neg p \lor \dia p$}
\DisplayProof
\end{tabular}
\end{center}

Finally, we note that a second, more elaborate instantiation example is available in an online version of this work~\cite{LyoOstarxiv}, where the framework is applied to first‑order intuitionistic logic.

\section{Concluding Remarks}\label{sec:conclusion}


In this paper, we introduced (the foundations of) an abstract framework permitting the study of multisequent systems and inference rules in a logic-independent setting. This has the advantage that our results hold generally for any proof system that can be viewed as an object in our framework, covering multisequent systems for diverse classes of logics of import in computer science, mathematics, and philosophy (e.g. see \fig~\ref{fig:logics-and-calculi}). We identified a number of inference rule types subsuming concrete rules that typically appear in multisequent systems, established permutation and simulation relationships between them, and showed how the functionality of certain rules could be strengthened or weakened via the novel operations of absorption and fracturing. We utilized these relationships and operations on inference rule types to specify generic proof and calculus transformation algorithms, which led to the discovery that abstract calculi sit within lattices of polynomially equivalent systems.

Furthermore, we made the observation that many labeled sequent systems can be identified as explicit calculi, which serve as bottom elements in lattices and perform Horn reasoning in an explicit manner via Horn structural rules. Conversely, many reachability nested calculi can be identified as implicit calculi, which serve as top elements in lattices and perform Horn reasoning implicitly via structural constraints in initial and reachability rules. This observation led to the finding that (Horn) labeled sequent systems and reachability nested systems tend to come in pairs, expressing a duality between them. That is to say, labeled and reachability nested systems are not fundamentally distinct, but occupy different positions along a common spectrum---they differ in terms of the degree to which they make structural reasoning implicit or explicit.

We expect the approach given in this paper to be generalizable and adaptable to alternative settings, enabling the identification of generic proof transformations between calculi within distinct formalisms, e.g. hypersequents, linear nested sequents, or non-wellfounded systems. Such transformations would yield broadly applicable proof-complexity results, which, by extension, would give information on the relative computational complexity of automated reasoning within distinct proof formalisms. Moreover, we anticipate that generic calculus transformation algorithms will be useful in extracting new multisequent systems for non-classical logics, e.g. multi-modal, temporal, or intuitionistic logics and first-order variants thereof, with potential applications to decidability, complexity, and interpolation.

There are various avenues for future research. 
First, we could generalize the types of structural rules considered in our framework, moving beyond Horn rules. In fact, as discovered in~\cite{LyoBer19,Lyo21thesis}, certain proof systems utilizing disjunctive properties admit transformations similar to those in \sect~\ref{sec:generic-algs}. Examining these cases and incorporating them into our framework seems promising. Second, we could investigate an even broader class of g-sequents, e.g. hypergraphs of sequents, similar to those used for relevance logics~\cite{Vig00}. Third, we could consider a larger set of inference rule types; for example, rules that introduce fresh edges between edges in a g-sequent (similar to the $\rightarrow^{2}_{R}$ rule in~\cite{KuzLel18}) or display rules like those used in display calculi~\cite{Bel82,Wan02} that switch the designated vertex of a g-sequent (i.e. the `root') to another vertex. Fourth, we could investigate 
other types of proof systems (e.g. linear nested sequents) through the lens of our framework, identifying the spaces these calculi exist within and uncovering transformations that navigate them. 
 

\begin{acks}
Work supported by the European Research Council (ERC) Consolidator Grant 771779 
 (DeciGUT).
\end{acks}

\bibliographystyle{ACM-Reference-Format}
\bibliography{bibliography}


\end{document}
\endinput